\documentclass[aps,superscriptaddress,tightenlines,nofootinbib,notitlepage,floatfix,longbibliography]{revtex4-1}
\usepackage[left=23mm,right=30mm,top=23mm,bottom=16mm]{geometry}

\usepackage[utf8x]{inputenc}
\usepackage[T1]{fontenc}
\usepackage[english]{babel}

\usepackage{amsmath,amssymb}
\usepackage{bm}
\usepackage{dsfont}
\usepackage[mathscr]{euscript}
\usepackage{comment}
\usepackage{graphicx}
\usepackage[dvipsnames]{xcolor}
\usepackage{slashed}
\usepackage{bbold}
\usepackage{pgf}
\usepackage{placeins}
\usepackage{hyperref}

\newcommand{\ikp}{
    Institut f\"{u}r Kernphysik,
    Forschungszentrum J\"{u}lich, 54245 J\"{u}lich Germany
}

\newcommand{\ias}{
    Institute for Advanced Simulation,
    Forschungszentrum J\"{u}lich, 54245 J\"{u}lich Germany
}

\newcommand{\bonn}{
    Helmholtz-Institut f\"{u}r Strahlen- und Kernphysik,
    Rheinische Friedrich-Williams-Universit\"{a}t Bonn, 53012 Bonn Germany
}

\newcommand{\berkeley}{
    Department of Physics, 
    University of California, Berkeley, CA 94720, USA
}

\usepackage{xspace}
\usepackage{bbm}
\usepackage{afterpage}
\usepackage{rotating}

\newcommand{\Secref}[1]{Section~\ref{sect:#1}}

\newcommand{\Appref}[1]{Appendix~\ref{sect:#1}}

\newcommand{\Tabref}[1]{Table~\ref{tab:#1}\xspace}

\newcommand{\Figref}[1]{Figure~\ref{fig:#1}\xspace}

\newcommand{\Reference}[1]{Reference~\cite{#1}}

\newcommand{\goesto}{\ensuremath{\rightarrow}}
\newcommand{\one}{\ensuremath{\mathbbm{1}}}
\newcommand{\up}{\ensuremath{\uparrow}\xspace}
\newcommand{\down}{\ensuremath{\downarrow}\xspace}
\newcommand{\Z}{\ensuremath{\mathcal{Z}}}
\newcommand{\transpose}{\ensuremath{{}^{\top}}}
\newcommand{\Reals}{\ensuremath{\mathds{R}}}

\newcommand{\nt}{{\ensuremath{N_t}}\xspace}
\newcommand{\nx}{{\ensuremath{N_x}}\xspace}
\newcommand{\Mexp}{\ensuremath{M^e}}
\newcommand{\Mdia}{\ensuremath{M^d}}
\newcommand{\kappatilde}{{\ensuremath{\tilde{\kappa}}}}
\newcommand{\Utilde}{{\ensuremath{\tilde{U}}}}
\newcommand{\inverse}{\ensuremath{{}^{-1}}}
\newcommand{\with}{\ensuremath{\leftrightarrow}}
\newcommand{\infinity}{\ensuremath{\infty}}
\newcommand{\permutation}{\ensuremath{\mathcal{P}}\xspace}

\renewcommand{\mod}[1]{\ensuremath{\ \left(\text{mod }#1\right)}}

\newcommand{\red}{\color[rgb]{0.8149019607843137, 0.18117647058823538, 0.18745098039215682}}

\newcommand{\purple}{\color[rgb]{0.571372549019608, 0.3392156862745098, 0.6058823529411764}}

\renewcommand{\vec}[1]{\boldsymbol{#1}}

\begin{document}

\title{Avoiding Ergodicity Problems in Lattice Discretizations of the Hubbard Model}

\author{Jan-Lukas Wynen}        \affiliation{\ias}
\author{Evan Berkowitz}         \affiliation{\ias} \affiliation{\ikp}
\author{Christopher K\"orber}   \affiliation{\ias} \affiliation{\ikp} \affiliation{\bonn} \affiliation{\berkeley}
\author{Timo A. L\"{a}hde}         \affiliation{\ias} \affiliation{\ikp}
\author{Thomas Luu}             \affiliation{\ias} \affiliation{\ikp} \affiliation{\bonn}

\date{\today}

\begin{abstract}
The Hubbard model arises naturally when electron-electron interactions are added to the tight-binding descriptions of many condensed matter systems.
For instance, the two-dimensional Hubbard model on the honeycomb lattice is central to the \textit{ab initio} description of the electronic structure of carbon nanomaterials, such as graphene.
Such low-dimensional Hubbard models are advantageously studied with Markov chain Monte Carlo methods, such as Hybrid Monte Carlo (HMC).
HMC is the standard algorithm of the lattice gauge theory community, as it is well suited to theories of dynamical fermions.
As HMC performs continuous, global updates of the lattice degrees of freedom, it provides superior scaling with system size relative to local updating methods.
A potential drawback of HMC is its susceptibility to ergodicity problems due to so-called exceptional configurations, for which the fermion operator cannot be inverted.
Recently, ergodicity problems were found in some formulations of HMC simulations of the Hubbard model.
Here, we address this issue directly and clarify under what conditions ergodicity is maintained or violated in HMC simulations of the Hubbard model.
We study different lattice formulations of the fermion operator and provide explicit, representative calculations for small systems, often comparing to exact results.
We show that a fermion operator can be found which is both computationally convenient and free of ergodicity problems.
\end{abstract}

\maketitle

\tableofcontents

\clearpage


\section{Introduction}

Modern Lattice QCD simulations are performed using the Hybrid Monte Carlo (HMC) algorithm~\cite{Duane:1987de, Kennedy:2006ax}.
While the basic structure of the HMC algorithm has remained unchanged since it was introduced, much effort has been directed toward the development of efficient iterative solvers, with accelerated convergence~\cite{saad2003iterative}.
Such highly optimized solvers have been instrumental in tearing down the so-called computational ``Berlin Wall'', which for a long time prevented simulations with dynamical fermions at the physical pion mass, for  systems of a realistic size~\cite{Bernard:2002pd,Jansen:2008vs,Urbach:2005ji,Orginos:2006en,Clark:2006wq}.
An interesting aspect of Lattice QCD is that a large freedom of choice exists for the formulation of the lattice action, provided that the proper continuum limit is recovered as the lattice spacing $a \to 0$.
This freedom has been exploited to formulate lattice actions that conserve more symmetries at finite $a$, or sacrifice exact symmetries in order to gain computational performance or scaling with system size $V$.
This is particularly true for chiral symmetry in Lattice QCD, which is intimately connected to the fermion ``doubling problem'' through the Nielsen-Ninomiya theorem~\cite{Nielsen:1980rz,Nielsen:1981xu}.
While lattice actions have been formulated which circumvent the doubling problem, and maintain exact or near-exact chiral symmetry at non-zero $a$ (such as overlap~\cite{Narayanan:1993sk,Narayanan:1994gw,Neuberger:1997fp,Neuberger:1998wv} or domain wall~\cite{Kaplan:1992bt,Shamir:1993zy,Furman:1994ky} fermions), they typically come at a high computational cost compared with lattice actions where exact chiral symmetry needs to be recovered by extrapolation.
For this reason, the development of Lattice QCD has tended to emphasize computational simplicity and efficiency, together with Symanzik improvement~\cite{Symanzik:1982dy,Symanzik:1982dz} of the lattice action, which accelerates the approach to the continuum limit by the systematic removal of lattice artifacts.

Remarkably, HMC has rarely been applied to problems in condensed matter physics.
In part, this can be traced back to the higher dimensionality of QCD---lattice QCD researchers have few options besides relying on the expected $\sim V^{5/4}$ scaling of HMC~\cite{Creutz:1988wv}.
In contrast, Monte Carlo (MC) simulations of Hamiltonian theories with electron-electron interactions, such as the Hubbard model~\cite{Hubbard_paper, Hubbard_paper2, Hubbard_paper3}, are usually performed using one of the many possible formulations of Auxiliary Field Quantum Monte Carlo (AFQMC).
In the Blankenbecler-Sugar-Scalapino (BSS) algorithm~\cite{Blankenbecler:1981jt, Hirsch:1983zza, White:1989zz, Bercx:2017pit}, the four-fermion interactions are split by means of a Hubbard-Stratonovich (HS) or ``auxiliary'' field $\phi$, which is sometimes taken to be discrete~\cite{Hirsch:1983zza}, in analogy with lattice models of spin systems.
Unlike the case of Lattice QCD, early attempts at combining BSS with HMC updates~\cite{Scalettar:1987zz, Hirsch:1988} were not pursued further, largely because the configuration space of auxiliary fields becomes increasingly fragmented into regions with positive and negative fermion determinant at low temperatures.
The HMC algorithm requires a continuous (rather than discrete) auxiliary field, and furthermore $M[\phi]$ should be invertible at every point during the HMC Hamiltonian update, or \emph{trajectory}, for a new configuration proposal.
At the boundaries between regions of different sign $\det M[\phi]$ vanishes, which causes HMC to become trapped in the region of the starting configuration.
As a result, algorithms which combine BSS with HMC updates are in general not ergodic\footnote{We stress that the BSS algorithm with discrete auxiliary fields is ergodic, and we do not consider such algorithms further in this paper.}.
Note that similar ergodicity problems appear for overlap fermions in Lattice QCD, which has prompted the development of a class of HMC algorithms which can reflect from or refract through boundaries where $\det M[\phi]$ changes sign~\cite{Fodor:2003bh, Fodor:2004wx, Cundy:2005pi}.
However, such algorithms are rather costly computationally, which suggests that a more amenable approach is to find a lattice action free from ergodicity problems.

The Hubbard and Hubbard/Coulomb models are key components of the {\it ab initio} description of electron-electron interactions in low-dimensional materials~\cite{Wehling2011, Tang2015} (e.g. graphene), so the problem of combining HMC with the Hubbard Hamiltonian has recently been revisited.\footnote{We restrict our presentation for simplicity to the Hubbard model, but our conclusions hold for the Hubbard/Coulomb model because, as we will see, the ergodicity problems we address here depend only on the fermion operator, and not on the form of the gaussian part of the action of the auxiliary field, where fermion-fermion interactions ultimately appear.}
The dimensionality of such systems is between that of QCD and problems of great interest in applied condensed matter physics and materials science, such as quantum dots and nanoribbons~\cite{CastroNeto2009, Kotov2012}.
For any 2-d lattice, one therefore expects that the (expected) superior scaling with $V$ of HMC over BSS would be advantageous in the study of critical phenomena, such as high $T_c$ superconductivity for the square lattice or the anti-ferromagnetic Mott insulating (AFMI) phase for the honeycomb lattice (and possibly other types of spin-liquid phases) which appears at moderate values of the on-site electron-electron coupling $U$~\cite{Paiva2005, Meng2010, Otsuka2016}.
Recently, new attempts have been made by Beyl {\it et al.} in Ref.~\cite{Beyl:2017kwp} to circumvent the ergodicity problem of the BSS+HMC method by making the auxilliary field $\phi$ complex-valued.
While this doubles the number of lattice degrees of freedom, it would represent an acceptable trade-off, if the computational scaling would be dramatically improved.
However, it was found that HMC simulations of the complexified theory failed to deliver the expected $\sim V^{5/4}$ scaling.
For the Su-Schrieffer-Heeger (SSH) electron-phonon Hamiltonian~\cite{SSH:1988}, which lacks the ergodicity problem of the Hubbard case, this scaling was realized.
The recent studies of Ref.~\cite{2018arXiv180407195K} also found favorable scaling with HMC when using Hasenbusch preconditioning.

From the perspective of the Lattice QCD community, MC simulations of the Hubbard model have several attractive features.
Some of these features are the apparent simplicity of the Hubbard model in comparison to the QCD Lagrangian, the possibility to identify the spatial lattice discretization $a$ with the physical lattice spacing, and the potential of applying the versatile and sophisticated toolbox of numerical methods developed for Lattice QCD to a problem with many promising applications.
The seminal work of Ref.~\cite{Brower:2012zd} introduced the Brower-Rebbi-Schaich (BRS) algorithm for the Hubbard model, which is inspired by Lattice QCD methods.
The BRS algorithm has recently been applied not only to graphene~\cite{Ulybyshev:2013swa, Smith:2014tha}, but also to carbon nanotubes~\cite{Luu:2015gpl, Berkowitz:2017bsn}.
The BSS and BRS algorithms are closely related.
The main differences are the treatment of the hopping matrix $h$ in the fermion operator $M$ and that BRS uses a purely imaginary auxiliary field.
Specifically, BSS uses a ``compact'' operator $M[\phi] \sim \mathbb{1} - \exp(h)\exp(\phi)$ (as in AFQMC) where both the hopping term and the auxiliary field appear as arguments of exponentials, while BRS uses the ``half-compact'' operator $M[\phi] \sim \mathbb{1} - h - \exp(i\phi)$ (as in Lattice QCD, where the phase is replaced by the parallel transporter, or gauge link).
Ref.~\cite{Smith:2014tha} found that ``non-compact'' formulations with $M[\phi] \sim \mathbb{1} - h - i\phi$ are numerically unstable due to round-off error.
These studies in terms of BRS found no indication of an ergodicity problem.

Still, the ergodicity of BRS remains controversial.
It has been noted~\cite{Beyl:2017kwp, Ulybyshev:2017hbs, Buividovich:2018yar} that the ergodicity problem related to the use of HMC with BSS cannot be eliminated simply by switching to an imaginary auxiliary field, as in BRS.
Irrespective of whether the auxiliary field is purely real or purely imaginary, the BSS fermion determinant $\det M[\phi]$ factorizes, becoming proportional to real-valued function which is not positive definite.
Hence, the configuration space of BSS should remain fragmented into regions separated by boundaries of \emph{exceptional configurations}, which have vanishing determinant, that HMC cannot cross.
However, Ref.~\cite{Beyl:2017kwp} found that using a complex auxiliary field solved the ergodicity problem of BSS.
The argument for factorization of Ref.~\cite{Ulybyshev:2017hbs} only applies to the compact version (BSS) of $M[\phi]$, which may explain why no ergodicity problem was found in previous BRS simulations.
Here, we show in detail that BRS avoids the ergodicity problem associated with factorization of $\det M[\phi]$, and we also reproduce the ergodicity problems reported in earlier work using BSS+HMC (with both real-valued and imaginary-valued auxiliary fields).
We note that the compact and half-compact versions of $M[\phi]$ are equivalent, up to terms of higher order in the Euclidean time step $\delta$.
We also discuss the relative merits of each formulation including the extrapolation of observables to the temporal continuum limit $\delta \to 0$.
It can be argued that BRS represents a case where symmetries of the continuum theory are sacrificed on the lattice, in order to improve computational scaling and gain the applicability of the computational toolbox of Lattice QCD.\@
Unlike the complexified BSS formulation, we find no indication of adverse computational scaling with BRS.\@
A thorough analysis of the computational scaling will be given in an upcoming publication.

This paper is organized as follows.
We describe HMC and the Hubbard model in \Secref{formalism}, including different choices of basis and discretization, the associated symmetries, and the properties of the fermion matrix $M$ and its eigenvalues.
We study the ergodicity problem in \Secref{ergo}, and show its connection to the eigenvalues of $M$.
We also give explicit examples for small numbers of lattice sites, which demonstrates how and when ergodicity issues appear.
We explore possible ways to circumvent such issues in \Secref{solution}.
By taking into account the symmetries of the system, we propose novel ways to effect large jumps between configurations, thereby crossing regions of low or zero probability.
We recapitulate and conclude in \Secref{conclusion}.



\section{Formalism\label{sect:formalism}}

We start by giving a cursory description of the HMC algorithm and describing the different discretizations of the Hubbard model in the literature.
Throughout, we assume the system is at half-filling (with zero chemical potential) and when we discretize the number of time slices \nt is even.


\subsection{Hybrid Monte Carlo}\label{sect:hmc}

The Hybrid Monte Carlo (HMC) algorithm is a Markov chain Monte Carlo (MCMC) method which can be used to estimate
multi-dimensional integrals
\begin{equation}
\label{eqn:mc integral}
\left( \prod\limits_{d=1}^{N_d} \int d \phi_d \right) W[\phi] O[\phi]
\simeq \frac{1}{N_c} \sum _{i=1}^{N_c} O[\phi^{(i)}],
\quad \phi^{(i)} \sim W,
\end{equation}
using importance sampling according to $W[\phi]$.
Each HMC step generates a new \emph{configuration}, or integration point $\phi^{(i)}$ in the $N_d$-dimensional space, based on the previous configuration $\phi^{(i-1)}$ in the Markov chain.
The larger the integration probability density $W[\phi^{(i)}]$ for a configuration $\phi^{(i)}$, the higher the probability that $\phi^{(i)}$ will be generated during the MC evolution.
Once an \emph{ensemble} consisting of $N_c$ configurations $\{ \phi^{(1)}, \cdots, \phi^{(N_c)} \}$ has been generated, operator expectation values can be estimated stochastically, by performing the sum over $i$ in~\eqref{eqn:mc integral}.

HMC is a global algorithm: all $N_d$ field components of a configuration $\phi$ are updated simultaneously.
Each field component $\phi_d$ is assigned a canonically conjugate momentum component $\pi_d^{}$, and the resulting $(\phi,\pi)$ system is evolved in a fictitious time by numerical integration of the Hamiltonian equations of motion.
This is done using the Hybrid Molecular Dynamics (HMD) algorithm, which combines the stochastic Langevin and deterministic Molecular Dynamics (MD) methods.
Specifically, each Langevin update (where the conjugate momenta are refreshed from a random Gaussian distribution) is interspersed with a number of MD integration steps, where the field $\phi$ follows a \emph{trajectory} through the field space.
The key advantage of HMC is the treatment of the HMD update as the proposal machine for the Metropolis algorithm.
In principle, energy is conserved during an MD trajectory, but as numerical integration schemes have finite truncation errors energy conservation is violated.
This violation is incorporated into the acceptance criterion of the Metropolis test---if the energy were exactly conserved, every proposed configuration would be accepted.
Unlike HMD and similar algorithms, HMC does not require extrapolation of the step size of the MD integration rule to the continuum.
The computational scaling of HMC as a function of system size $V$ is expected to be $\sim V^{5/4}$~\cite{Creutz:1988wv, 2018arXiv180407195K}, superior to the cubic (or nearly cubic) scaling of local updates in theories of dynamical fermions.
For this reason, HMC is the method of choice for computing ensembles of configurations in high-dimensional theories, such as Lattice QCD.\@

Viewed as a Markov process, HMC converges to the desired equilibrium probability distribution $W[\phi]$ if:
\begin{enumerate}\label{enum:mcmc requirements}

\item The detailed balance condition $W[\phi] \Omega(\phi \to \phi^\prime) = W[\phi^\prime] \Omega(\phi^\prime \to \phi)$ is satisfied, where $W[\phi]$ is the normalized Boltzmann
distribution $\exp(-S[\phi])/\mathcal{Z}$, and $S[\phi]$ the Euclidean action of the theory.
Also, $\Omega(\phi \to \phi^\prime)$ is the transition probability from configuration $\phi$ to $\phi^\prime$.

\item The Markov chain is ergodic meaning that the equilibrium distribution $W[\phi]$ is unique and independent of the starting configuration of the chain.
In other words, given a configuration $\phi$ for which $W[\phi] \neq 0$, every other configuration $\phi^\prime$ for which $W[\phi^\prime] \neq 0$ should be reachable from $\phi$ in a finite number of steps (or amount of MC time).

\end{enumerate}
For detailed balance to be satisfied, the MD integration should be performed with an integration rule which is reversible and symplectic (such as the leapfrog and Omelyan integrators).
Such integrators also ensure that the acceptance rate of HMC only depends weakly on $V$, as there is \emph{a priori} no guarantee that HMC can perform large global updates with significant decorrelation between successive configurations.

The second criterion is much harder to enforce, especially for multi-dimensional probability densities.
While indicators for ergodicity issues can be monitored during the generation of configurations (for example one can monitor the force, watch for large changes in the acceptance rate, or the freezing of observables), a formal proof that HMC is ergodic for a particular system is usually not available.
In some cases, a physical understanding of ergodicity problems is possible, such as the difficulty of tunneling between different topological sectors in Lattice QCD.\@
Note that the ergodicity problems referred to here should not be confused with the lack of ergodicity in algorithms that effect updates in terms of pure MD trajectories, with no periodic refreshment of the conjugate momenta or Metropolis accept/reject step.

The violation of energy conservation during the MD trajectory should remain small if an HMC update should be accepted with high likelihood.
The classical MD evolution is driven by the functional derivative $F[\phi] = -\delta S[\phi]/\delta\phi$, the HMC {\it force term}.
HMC is susceptible to barriers or discontinuities in the landscape of $W[\phi]$.
Such barriers can occur when $W[\phi_0^{}] = 0$ and \emph{exceptional configurations} $\phi_0^{}$ can separate the integration domain into disconnected regions.
As exceptional configurations correspond to singularities in the force term, attempts to cross the barrier generate a large energy violation, and the HMC update is rejected with a high probability.
The inability of the standard HMC algorithm to cross barriers between disconnected regions of $W[\phi]$ leads to an ergodicity problem.
In other words, the HMC Markov chain becomes locked in a region with boundaries $\phi_0$, where $W[\phi_0] = 0$.
It should be noted that HMC can cross such boundaries infrequently if the numerical integration of the Hamiltonian equations of motion is sufficiently coarse.
In general too coarse MD updates cannot maintain a high acceptance rate as $V$ is increased.

As we have already noted, an ergodicity problem appears whenever the fermion determinant $\det M[\phi]$ becomes proportional to a real-valued function $f[\phi]$ which is not positive definite~\cite{Ulybyshev:2017hbs}.
While such a problem can appear also for complex-valued $\phi$ and $\det M[\phi]$, it is usually more severe in theories where $\phi$ and $\det M[\phi]$ are real-valued,
such as the overlap formulation of chiral fermions in Lattice QCD.\@
Another example is the AFQMC treatment of Ref.~\cite{Scalettar:1987zz}, which combined the HMC algorithm with the BSS formulation of the Hubbard model.
There, the fermion matrix $M[\phi]$ is not guaranteed to satisfy $\det M[\phi] > 0$, which becomes apparent at low $T$ and at strong on-site coupling $U$,  where the energy landscape fragments into multiple regions of the positive and negative $\det M[\phi]$.
For overlap fermions, specialized HMC algorithms have been developed which can tunnel through or reflect from infinite-force barriers in a reversible manner, maintaining ergodicity and a high acceptance rate\cite{Fodor:2003bh, Fodor:2004wx, Cundy:2005pi}.
Here, we take a different approach and instead seek an optimal representation of $M[\phi]$, which minimizes or eliminates ergodicity problems altogether.



\subsection{Choice of basis}

The nearest-neighbor tight-binding Hamiltonian $H_0$,
\begin{equation}\label{eqn:tight-binding}
    H_0 = -\kappa\sum_{\langle x,y \rangle}\left(a^\dag_{x,\uparrow}a_{y,\uparrow}^{} +a^\dag_{x,\downarrow}a_{y,\downarrow}^{}\right)
\end{equation}
contains a kinetic term only, which describes free electrons of spin \up and spin \down hopping between different lattice sites with hopping parameter $\kappa$.
The bracket $\langle x,y \rangle$ denotes pairs of nearest neighbors.
The Hubbard model adds on-site interactions,
\begin{equation}
    H = H_0 -\frac{U}{2}\sum_{x}{\left(n_{x,\uparrow}-n_{x,\downarrow}\right)}^2\label{eqn:alpha 0 case}
\end{equation}
where number operator $n_{x,s} \equiv a^\dag_{x,s}a_{x,s}^{}$ counts electrons of spin $s$ at position~$x$.

We can change basis via a particle-hole transformation on the spin-$\downarrow$ electrons
\begin{align}
  b^\dag_{x,\downarrow}\equiv a_{x,\downarrow},\qquad b_{x,\downarrow}\equiv a^\dag_{x,\downarrow} \ .\label{eqn:particle hole transform}
\end{align}
Up to an overall irrelevant constant, the Hamiltonian under this transformation is
\begin{equation}
  H = -\kappa\sum_{\langle x,y \rangle}\left(a^\dag_{x}a_{y}^{} - b^\dag_{x}b_{y}^{}\right)
  + \frac{U}{2}\sum_{x}{\left(n_{x}-\tilde{n}_{x}\right)}^2
  \ ,
\end{equation}
where the number operator $\tilde{n}_{x} \equiv b^\dag_{x}b_{x}^{}$ counts spin-$\downarrow$ holes at position~$x$.
The degrees of freedom here are electrons of spin $\uparrow$ and holes with spin $\downarrow$.
This lets us drop the spin indices.
Both bases describe the same system and give the same relative spectrum.

In the case of bipartite lattices (like the honeycomb lattice of graphene and carbon nanotubes) it is possible to modify transformation~\eqref{eqn:particle hole transform} to switch signs on one sublattice,
\begin{align}
b^\dag_{x,\downarrow}\equiv \mathcal{P}_x a_{x,\downarrow},\qquad b_{x,\downarrow}\equiv \mathcal{P}_x a^\dag_{x,\downarrow} \ ,\label{eqn:bipartite particle hole transform}
\end{align}
where $\mathcal{P}_x$ is $+1$ if $x$ is on one sublattice and $-1$ if it is on the other.
This keeps $H_0$ invariant under the particle-hole transformation, but still flips the sign of the on-site interaction compared to \eqref{eqn:alpha 0 case},
\begin{equation}\label{eqn:alpha 1 case}
    H = -\kappa \sum_{\langle x,y \rangle}\left(a^\dag_{x}a_{y}^{} + b^\dag_{x}b_{y}^{}\right) + \frac{U}{2}\sum_{x}{\left(n_{x}-\tilde{n}_{x}\right)}^2;
\end{equation}
we recognize the first term as simply the tight binding Hamiltonian $H_0$ given in~\eqref{eqn:tight-binding} with $b$ in lieu of $a_\down$ and spin labels dropped.
We henceforth specialize to bipartite lattices.
We say that this Hamiltonian is written in the \emph{particle/hole basis}, while~\eqref{eqn:alpha 0 case} is in the \emph{spin basis}.
The only difference between~\eqref{eqn:alpha 0 case} and~\eqref{eqn:alpha 1 case} is the sign in front of the on-site interaction term.

It is possible to write down a Hamiltonian that includes both types of interactions, parameterized via $\alpha \in [0, 1]$ as
\begin{align}
  H = H_0
  +\alpha\frac{\mathcal{U}}{2}\sum_{x}{\left(n_{x,\uparrow}-n_{x,\downarrow}\right)}^2
  -(1-\alpha)\frac{U}{2}\sum_{x}{\left(n_{x,\uparrow}-n_{x,\downarrow}\right)}^2
  \ .\label{eqn:alpha case}
\end{align}
Ignoring the superficial difference in labelling of spin-\down electrons or spin-\down holes, when $\alpha=0$ one recovers the Hamiltonian of the spin basis \eqref{eqn:alpha 0 case} while $\alpha=1$ yields the Hamiltonian in the particle/hole basis \eqref{eqn:alpha 1 case} if $\mathcal{U}=U$.
For arbitrary $\alpha\in(0,1)$, Hubbard-Stratonovich transformations will introduce auxiliary fields with both real and imaginary components, as thoroughly investigated in Ref.~\cite{Beyl:2017kwp}, which found ergodicity problems for the extreme values 0 and 1.

As our investigations revolve around issues related to ergodicity, in what follows we concentrate only on the extreme values $\alpha=0$, the spin basis, and $\alpha=1$, the particle/hole basis.
Sometimes we use $\alpha$ to label the different bases for brevity.


\subsection{Discretization}

Discretizing the Hubbard model path integral, and the introduction of auxiliary fields $\phi$ by Hubbard-Stratonovich transformation, has been discussed before (see~\cite{Brower:2012zd, Ulybyshev:2013swa, Smith:2014tha, Luu:2015gpl}, for example).
After discretization, the partition function in the spin basis, up to an overall normalization, can be written
\begin{align}
  \mathcal{Z}_{\uparrow\downarrow}
     =  \int        \left[\prod_{x,t}\mathrm{d}\phi_{xt}\right] W[\phi]
    &=  \int        \left[\prod_{x,t}\mathrm{d}\phi_{xt}\right]
                    \det M[\phi] \det M[-\phi] \exp{\Big(-\frac{1}{2\Utilde}\sum_{x,t}\phi_{xt}^2\Big)}
\\
  &=    \int        \left[\prod_{x,t}\mathrm{d}\phi_{xt}\right]
                    \exp{\Big(-\frac{1}{2\Utilde}\sum_{x,t}\phi_{xt}^2+\log\det M[\phi] +\log \det M[-\phi] \Big)}
\ ,\label{eqn:spin basis partition fn}
\end{align}
where $W[\phi]$ is the probability weight of a configuration $\phi$ and $M[\phi]$ is the fermion matrix.
$M$ is also a function of the hopping matrix $h_{x',x} = \kappatilde\delta_{\langle x',x \rangle}$ which corresponds to the nearest-neighbor connections in the tight-binding Hamiltonian $H_0$ with hopping strength $\kappatilde = \kappa\delta$, with $\delta=\beta/\nt$ the discretization of the inverse temperature into $\nt$ evenly-spaced slices.
The interaction strength is $\Utilde=U\delta$.

The partition function in the particle/hole basis uses imaginary fields in the fermion matrix which is a consequence of the different sign in front of the on-site interaction term in~\eqref{eqn:alpha 1 case} compared to~\eqref{eqn:alpha 0 case}.
The partition function is otherwise identical to the on in the spin basis
\begin{equation}
\mathcal{Z}_{ph} =\int\left[\prod_{x,t}\mathrm{d}\phi_{xt}\right] \exp{\Big(-\frac{1}{2\Utilde}\sum_{x,t}\phi_{xt}^2+\log\det M[i\phi] +\log \det M[-i\phi] \Big)}\ .\label{eqn:ph basis partition fn}
\end{equation}
A negative sign in front of the interaction requires completely real auxiliary fields, whereas a positive sign requires completely imaginary fields.

Differences in discretizations manifest themselves in the structure of the fermion matrix $M$.
In Refs.~\cite{Meng2010,Beyl:2017kwp,Ulybyshev:2017hbs}, for example, matrix elements of the fermion operator have the form
\begin{equation}\label{eqn:M1}
  {\Mexp[\phi]}_{x^\prime t^\prime, xt} = \delta_{x^\prime, x} \delta_{t^\prime, t} - {[e^{h}]}_{x^\prime, x} e^{\phi_{xt}} \mathcal{B}_{t'} \delta_{t^\prime, t + 1}\quad\quad\text{(\emph{exponential} discretization)}\ ,
\end{equation}
where space and time directions are combined and $x^\prime t^\prime$ denote the row and $xt$ the column index.
$\mathcal{B}_{t'} = +1$ for $0 < t' < \nt$ and $\mathcal{B}_0 = -1$ explicitly encodes the anti-periodic boundary condition in time.
In Ref.~\cite{Smith:2014tha}, on the other hand, the matrix elements are
\begin{equation}\label{eqn:M2}
  {M^{\ell}[\phi]}_{x^\prime t^\prime, xt} = \delta_{x^\prime,x} \delta_{t^\prime,t} - \left(h_{x',x} + e^{\phi_{xt}} \delta_{x',x}\right) \mathcal{B}_{t'} \delta_{t^\prime, t+1} \quad\quad\text{(\emph{linear} discretization)}\ ,
\end{equation}
Finally, in Refs.~\cite{Brower:2012zd,Luu:2015gpl,Berkowitz:2017bsn}, the hopping term is moved to the time diagonal\footnote{In addition to moving the hopping term to the time diagonal, in~\cite{Luu:2015gpl,Berkowitz:2017bsn} a mixed forward and backward differencing scheme was applied to the underlying sublattices.},
\begin{equation}\label{eqn:M3}
  {\Mdia[\phi]}_{x^\prime t^\prime, xt} = \left(\delta_{x^\prime,x} - h_{x',x}\right) \delta_{t^\prime,t} - e^{\phi_{xt}} \delta_{x',x} \mathcal{B}_{t'} \delta_{t^\prime, t+1}\quad\quad\text{(\emph{diagonal} discretization)} \ .
\end{equation}
This last discretization is more akin to what is done in lattice gauge theories, where the gauge links (parallel transporters) reside between discretization slices.
The exponential, linear, and diagonal discretizations in \eqref{eqn:M1},~\eqref{eqn:M2},~ and~\eqref{eqn:M3} formally agree up to $\mathcal{O}(\delta^2)$, and thus have the same continuum $\delta\to0$ ($\nt\to\infty$) limit.
Observables calculated with these different discretizations should only be compared after a continuum limit extrapolation.
In this work we will focus on the exponential and diagonal discretizations, only occasionally commenting on the linear discretization.

It will prove useful to consider the matrix $S=M-\one$ for the various discretizations with matrix elements
\begin{align}
    {S^e[\phi]}_{x't',xt} &= - {\left[e^h\right]}_{x',x} e^{\phi_{xt}} \mathcal{B}_{t'} \delta_{t',t+1}                  && \text{(\emph{exponential})}   \label{eqn:F exponential}\\
    {S^d[\phi]}_{x't',xt} &= - h_{x',x} \delta_{t',t} - e^{\phi_{xt}} \delta_{x',x} \mathcal{B}_{t'} \delta_{t',t+1}     && \text{(\emph{diagonal})}      \label{eqn:F diagonal}
\end{align}
which in the exponential case is entirely off-diagonal.
Each eigenvalue of $M$ differs from an eigenvalue of $S$ by 1.


\subsection{Symmetries, Fermion Determinants, and Fermion Matrix Eigenvalues}\label{sect:eigenvalues}

Understanding the symmetries and limits of the physical problem and the discretizations will prove valuable for later discussion and inspiration for how to alleviate some ergodicity problems.
The impatient reader may prefer to skip this detailed discussion, though we do rely on observations here throughout the rest of the paper.

It is useful to consider the probability weight of a field configuration $\phi$
\begin{equation}
    W[\phi] = \det M[\phi] \det M[-\phi] \exp\left(-\frac{1}{2\Utilde}\sum_{x,t}\phi_{xt}^2\right)
\end{equation}
that appears in the spin-basis partition function~\eqref{eqn:spin basis partition fn} and its analog in~\eqref{eqn:ph basis partition fn} where the arguments of the fermion matrices get an $i$.

\subsubsection{Charge Conjugation}\label{sect:charge conjugation}

The first, most obvious symmetry of the probability weight is the change of the sign of $\phi$.
When one sends $\phi \goesto -\phi$ the quadratic piece is invariant and the determinants change roles, so that
\begin{equation}
    W[\phi] = W[-\phi].
\end{equation}
The two determinants arise from the different spins or species, depending on the basis.
Thus, sending $\phi \goesto -\phi$ exchanges the spins, or exchanges particles and holes.
This symmetry is broken when away from half filling, or, put another way, with nonzero chemical potential.
Away from half filling one must also negate the chemical potential to achieve equality of $W$.
This is the analog to charge conjugation symmetry $C$.
Interestingly, for non-bipartite lattices the sign of the tight binding Hamiltonian differs between the two determinants, and $\phi\goesto-\phi$ fails to be a symmetry, even at half filling.

\subsubsection{Characteristic Polynomials}\label{sect:characteristic polynomials}

To go beyond this observation it will prove useful to have a firm understanding of the fermion matrix, its eigenvalues, and its determinant in the four different cases described in the previous section.
It is simpler to consider the eigenvalues of $S=M-\one$, given in \eqref{eqn:F exponential} and \eqref{eqn:F diagonal}.
We restrict our attention to even \nt for simplicity.

We will demonstrate equalities of the characteristic polynomial of $S$,
\begin{equation}
    P[\phi](s) = \det(S[\phi] - s \one).
\end{equation}
When $s$ is a root of $P[\phi]$ it is an eigenvalue of $S[\phi]$, and $\lambda = s+1$ is an eigenvalue of $M[\phi]$.
Note that since $M = S+\one$, we know $\det M[\phi] = P[\phi](-1)$.
In the fully general case $P$ also depends on the chemical potential and on the sign of the adjacency matrix.
We suppress these dependencies for clarity and focus on the bipartite, half-filling case and only comment when ignoring these assumptions invalidates a conclusion.

First let us consider the exponential case with even \nt.
We will use the identity~\eqref{eqn:det M 1} shown in Appendix~\ref{sect:determinants}, letting
\begin{align}
        D_{x',x}   &= -s \delta_{x',x}
    &
        {\left[T_t\right]}_{x',x} &= -{\left[e^h\right]}_{x',y} {F_t[\phi]}_{y,x}
\end{align}
where
\begin{equation}\label{eqn:F alpha}
    {F_t[\phi]}_{x',x} =
        \begin{cases}
            e^{-\phi_{x(t-1)}} \delta_{x',x}    &   \alpha=0    \\
            e^{-i\phi_{x(t-1)}} \delta_{x',x}   &   \alpha=1
        \end{cases}
\end{equation}
is a diagonal matrix of auxiliary fields on a given timeslice and the $t$ index is understood modulo \nt.\footnote{
Note that here we have explicitly included the $\alpha$-dependent factor of $i$ in $F$ so that we can always just think of $\phi$ as real.}
Note that $F$ has the property
\begin{equation}
    \label{eqn:F inverse}
    F_{t}[\phi]\inverse = F_{t}[-\phi]
\end{equation}
generally and
\begin{equation}
    \label{eqn:F star}
    F_{t}[\phi]^* =
        \begin{cases}
            F_{t}[\phi]     &   \alpha=0    \\
            F_{t}[-\phi]    &   \alpha=1
        \end{cases}\ .
\end{equation}
Then the characteristic polynomial is given by
\begin{align}
    P^e[\phi](s)
    &= {\det(-s \one_\nx)}^\nt \ \det\left(\one_\nx + \frac{\one_\nx}{s} e^h F_{\nt-1} \frac{\one_\nx}{s} e^h F_{\nt-2} \cdots \frac{\one_\nx}{s} e^h F_{0}\right)  \nonumber\\
    &= \det(s^\nt \one_\nx) \det\left(\one_\nx + s^{-\nt} e^h F_{\nt-1} e^h F_{\nt-2} \cdots e^h F_{0}\right)  \nonumber\\
    &= \det\left( s^\nt \one_\nx + e^h F_{\nt-1} e^h F_{\nt-2} \cdots e^h F_{0}\right)                          \label{eqn:P(S) exponential}
\end{align}
which is a polynomial in the variable $s^\nt$.
So, if $s$ is a root of $P^e[\phi]$, any other $s$ with the same $\nt^\text{th}$ power is also a root.
This establishes that, in the fully-interacting exponential case, if $s$ is an eigenvalue, so is $s \exp(2\pi i / \nt)$, which can be bootstrapped all the way around the circle.
That is, the eigenvalues come equally spaced around rings in the exponential case, independent of $\alpha$.

In the diagonal case we use~\eqref{eqn:det M 2}, an equivalent determinant identity also shown in Appendix~\ref{sect:determinants} but now with
\begin{align}
        D_{x',x}     &=  - s \delta_{x',x} - h_{x',x}
    &
        {\left[T_t\right]}_{x'x}   &=  - {F_t(\phi)}_{x',x}.
\end{align}
One finds, dropping \nt powers of minus signs,
\begin{align}
    P^d[\phi](s)
    &= \det(F_{\nt-1}\cdots F_{0})\det\left( \one + F_0\inverse (s\one_\nx+h) \cdots F_{\nt-2}\inverse (s\one_\nx+h) F_{\nt-1}\inverse (s\one_\nx+h) \right)    \nonumber\\
    &= \left(\prod_t \det(F_t)\right) \det\left( \one + F_0\inverse (s\one_\nx+h) \cdots F_{\nt-2}\inverse (s\one_\nx+h) F_{\nt-1}\inverse (s\one_\nx+h) \right).
       \label{eqn:P(S) diagonal}
\end{align}
This is not a polynomial in $s^\nt$ and we do not expect to find perfect rings---were $h=0$ we would.

In the linearized case \eqref{eqn:M2}, the diagonal of $S$ is again zero, $s$ can be gathered in the characteristic polynomial as for $P^e[\phi]$, and we expect perfect rings.
We have verified this expectation numerically.

\subsubsection{Field Shifts and Periodicity}\label{sect:shifts and periodicity}

From \eqref{eqn:F alpha} it is clear that in the $\alpha=1$ case, $F$ remains invariant if any field component changes by $2\pi$, and the determinant is therefore invariant under such a shift.
In fact, we can make a more generic transformation,
\begin{equation} \label{eqn:field shift}
    \phi_{xt} \goesto \phi_{xt}+\theta_t
\end{equation}
shifting the all the fields on timeslice $t$ by a constant $\theta_t$.
Under that transformation, $F$ changes by an overall phase
\begin{equation}
    F_t[\phi+\theta] = F_t[\phi]e^{-i\theta_{t-1}}.
\end{equation}
In the exponential case, the characteristic polynomial of the transformed configuration,
\begin{align}
    P^e[\phi_{xt}+\theta_t](s)
    &=
    \det\left(s^\nt \one + e^h F_{\nt-1}e^{-i\theta_{\nt-2}} \cdots e^h F_1 e^{-i\theta_0}e^h F_0 e^{-i\theta_{\nt-1}}\right)
    \nonumber\\
    &=
    \det\left(s^\nt \one + e^{-i \sum_t \theta_t} e^h F_{\nt-1} \cdots e^h F_0 \right)
\end{align}
which is equal to $P^e[\phi](s)$, so long as
\begin{equation}\label{eqn:periodicity condition}
    \sum_t \theta_t \equiv 0 \mod{2\pi}.
\end{equation}
The same condition holds for the diagonal discretization, analogously.
Of course, these shift transformations are only symmetries of the fermion determinant; the gaussian part of the action changes, in general.
These shifts leave each Polyakov loop, the time-ordered product of links around the temporal direction,
\begin{align}\label{eqn:polyakov loop}
    \mathscr{P}_x &= \prod_t e^{i\phi_{xt}} = e^{i\Phi_x}
    &
    \Phi_x &= \sum_t \phi_{xt}
\end{align}
invariant.

In the $\alpha=0$ case we can make analogous field shifts to each timeslice.
However, rather than finding a periodic requirement on the sum of the shifts, the requirement to get the same eigenvalues is that the sum of the shifts vanishes, $\sum_t\theta_t=0$, as the links in the Polyakov loop lose their factor of $i$.

We also see that if we send an even number of $F$s to minus themselves, the determinant is invariant.
When $\alpha=1$ we can flip a single $F$ by shifting all of the fields on a single timeslice by $\pm\pi$, making an independent choice for each,
\begin{equation}
    \phi_{x} \goesto \phi_{x} + \pi j_{x}
\end{equation}
where each $j_x$ is an odd integer.
In fact, we can think of this transformation as a composition of individual $2\pi$ jumps and the coordinated field shift \eqref{eqn:field shift}, setting $\theta_t=\pi$.
One choice that leaves the sum of all the field variables invariant on a bipartite lattice is $j_x = \pm\mathcal{P}_x$, where $\mathcal{P}_x$ is $+1$ on one sublattice and $-1$ on the other, as in the particle-hole transformation \eqref{eqn:bipartite particle hole transform}.
As long as an even number of timeslices get so transformed, the eigenvalues are invariant, though the gaussian part of the action may change.

\subsubsection{The Non-Interacting Case}\label{sect:noninteracting case}

In the non-interacting $U=0$ case we can solve the fermion matrix exactly.
The gaussian controlling the auxiliary field $\phi$ in equations \eqref{eqn:spin basis partition fn} and \eqref{eqn:ph basis partition fn} becomes infinitely narrow and we need only consider $\phi=0$, so all $F$s are the identity matrix.
In this case we can find the exact spectrum of the fermion matrix.
Independent of whether $\alpha$ is 0 or 1, the eigenvalues are given by
\begin{align}\label{eqn:M eigenvalues}
s_{i,n}
    &=
        \begin{cases}
            e^{\delta\epsilon_i+i\omega_n}      &   \text{exponential}\\
            -\delta\epsilon_i+e^{i\omega_n}     &   \text{diagonal}
        \end{cases}
    &
    \lambda_{i,n} = s_{i,n}+1
\end{align}
where the Matsubara frequencies $\omega_n=\frac{2\pi}{\nt}\left(n+\frac{1}{2}\right)$ for $n$ in the integers from 0 to $\nt-1$ and $\epsilon_i$ are the non-interacting eigenvalues of $H_0/\kappa$.

In the exponential case the eigenvalues of $M$ come in rings concentric around 1, as discussed following~\eqref{eqn:P(S) exponential}, with radii $r_i=\exp\delta\epsilon_i$.
When the hopping is bipartite, the eigenvalues $\epsilon$ come in additive-inverse pairs and the corresponding radii are multiplicative inverses.
That is, if one ring has a radius $r$, another has radius $1/r$---if $s$ is an eigenvalue of $S$, so is $1/s^*$.
We will show, below, that this remains true in the exponential $\alpha=1$ case and that generally, in the interacting exponential case, if $s$ is an eigenvalue for one species, $1/s$ is an eigenvalue for the other, a statement of exact chiral symmetry.

\begin{figure}[htbp]
    \centering
    \includegraphics[width=.8\textwidth]{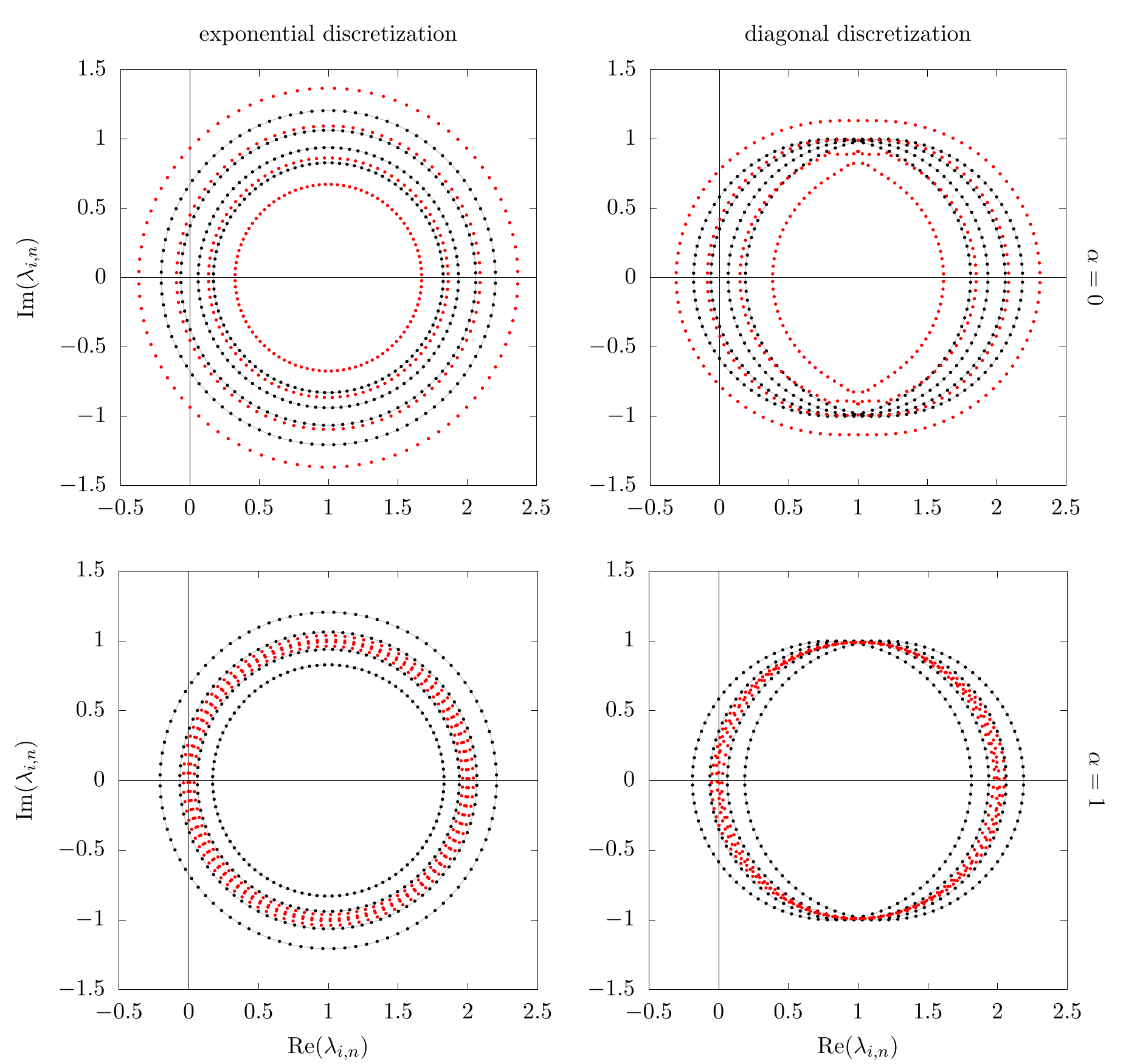}
    \caption{
		In the four panels we show as lighter red points the spectrum of a fermion matrix with a particular $\alpha$ and discretization, as indicated, for a four-site honeycomb lattice with $\nt=96$.
	    The darker, black points are the spectrum of the noninteracting case ($\phi=0$), which depends only on which discretization is used.
	    We show light gray rings to guide the eye for the eigenvalue rings as described by \eqref{eqn:M eigenvalues}.
	    There are four rings because there are four spatial sites; each ring has $\nt=96$ points.
	    In all four cases we use the same configuration $\phi$ which was randomly sampled from a gaussian with width $\sqrt{U\beta/\nt}$, where $U\beta=60$.
	    \label{fig:fermion-matrix-spectra}
    }
\end{figure}

In the diagonal case the eigenvalues of $M$ come in rings, all of radius 1, centered on $1-\delta\epsilon_i$.
In the continuum limit $\delta\goesto0$ both discretizations give the same $\nx$-degenerate ring of eigenvalues, as expected.
In both cases, the eigenvalues are evenly distributed around their respective rings, at angles given by the Matsubara frequencies; with interactions this perfect spacing is true for the exponential case, as already shown.
\autoref{fig:fermion-matrix-spectra} shows in black the non-interacting eigenvalues for the four-site honeycomb lattice, which has $\epsilon$ in $\{\pm1,\pm3\}$, with $\nt=96$.

In both discretizations, the eigenvalues in~\eqref{eqn:M eigenvalues} come in complex conjugate pairs.
However, this simple picture of perfect rings is broken when $\phi\neq0$, corresponding to $U\neq0$.
Moreover, once interactions are turned on the spectra differ depending on $\alpha$.
\autoref{fig:fermion-matrix-spectra} shows in red an example fermion matrix spectrum for the same example lattice, using the same field configuration for each discretization.

\subsubsection{Reality of the Probability Weight}\label{sect:reality}

Starting from \eqref{eqn:P(S) exponential} and using the fact that $\det(A^*) = \det(A)^*$, one finds
\begin{align}
    P[+\phi](s)   &=  \det\left( (s^*)^\nt \one + e^h F_{\nt-1}^* \cdots e^h F_0^* \right)^*
                    \nonumber\\
                    &=  \begin{cases}
                            P[+\phi](s^*)^*   &\alpha=0   \\
                            P[-\phi](s^*)^*   &\alpha=1
                        \end{cases} \label{eqn:P(S) complex conjugate}
\end{align}
where we used the properties of $F$ under complex conjugation in~\eqref{eqn:F star} and the fact that $h$ is real.
In the $\alpha=0$ case, the eigenvalues of a single fermion matrix come in complex-conjugate pairs or as real singletons, each determinant is independently real, and so the probability weight is real.
In the $\alpha=1$ case, the probability weight is guaranteed to be real and positive because, as~\eqref{eqn:P(S) complex conjugate} shows, the particle eigenvalues are the complex conjugates of the hole eigenvalues.
An analogous argument starting from~\eqref{eqn:P(S) diagonal} shows that~\eqref{eqn:P(S) complex conjugate} holds for the diagonal discretization as well.
We will return to the question of positivity in the $\alpha=0$ in Section~\ref{sect:eigenvalue summary}.

In the case of nonzero chemical potential~\eqref{eqn:P(S) complex conjugate} is not enough to show the reality of the probability, as the exchange of determinants also requires flipping the chemical potential and the right-hand side is not the characteristic polynomial for the other species.
Similarly, in the case of non-bipartite lattices, the sign of $h$ flips for the other species but is unchanged by this manipulation and we again find an identity between characteristic polynomials, but not the ones we need to demonstrate reality.
Thus, we expect a sign problem at nonzero chemical potential or on non-bipartite lattices; the difficulty of those sign problems is a question for future work.

\subsubsection{Particle-Hole Symmetry}\label{sect:particle-hole symmetry}

We can take advantage of the bipartite structure of the adjacency matrix.
Let $\Sigma$ be a diagonal matrix that is $+1$ on one sublattice and $-1$ on the other.
Then
\begin{align}
    \Sigma^2    &=  \one_\nx    &
    \Sigma h \Sigma &=  -h      &
    \Sigma F \Sigma &=  F.
    \label{eqn:bipartite sigma}
\end{align}
By repeatedly inserting $\Sigma^2$ we see that
\begin{equation}
    \label{eqn:sign of h}
    P[+h, \phi](s) = P[-h, \phi](s)
\end{equation}
independent of discretization or choice of $\alpha$.
In the general case without applying~\eqref{eqn:bipartite particle hole transform} one species' characteristic polynomial naturally occurs with a $-h$; in the bipartite case the sign of $h$ can be flipped without repercussion for the eigenvalues of the fermion matrix.

When \nt is even we see that the eigenvalues come in additive inverse pairs,
\begin{align}
    P^d[-h, \phi](s)
        &= \left(\prod_t \det(F_t)\right) \det\left( \one_{N_x} + F_0\inverse (s\one_\nx-h) \cdots F_{\nt-1}\inverse (s\one_\nx-h) \right)
        \nonumber\\
        &= \left(\prod_t \det(F_t)\right) \det\left( \one_{N_x} + {(-1)}^\nt F_0\inverse (-s\one_\nx+h) \cdots F_{\nt-1}\inverse (-s\one_\nx+h) \right)
        \nonumber\\
        &= P^d[+h, \phi](-s)
\end{align}
so that if $s$ is a root of $P^d[\phi]$, $-s$ is too.
In the exponential case this is already guaranteed by the fact that the eigenvalues come evenly spaced around rings.

\subsubsection{Temporal Shifts}\label{sect:temporal shifts}

By using Sylvester's determinant identity
\begin{equation}\label{eqn:determinant cyclicity}
    \det(\one + AB) = \det(\one + BA)
\end{equation}
we can cyclically permute matrices around the determinant.

In particular, from both \eqref{eqn:P(S) exponential} and \eqref{eqn:P(S) diagonal} we see immediately, by shifting both the hopping and the first $F$, that
\begin{equation}
    P[\phi_t](s) = P[\phi_{t+1}](s)
\end{equation}
where $\phi_{t+1}$ is the field configuration $\phi_t$ but shifted by one timeslice (modulo \nt).
Since $\phi$ is bosonic, we need not worry about antiperiodic boundary conditions; they are built into the fermion matrix directly through $\mathcal{B}_t$.
This shift can be repeatedly applied and ultimately guarantees the time-translation invariance of the probability weight, independent of discretization scheme and $\alpha$.

\subsubsection{Time Reversal}\label{sect:time reversal}

We also immediately see that $\phi$ has a time-reversal symmetry $T$.
Since the determinant is invariant under transposition, starting from~\eqref{eqn:P(S) exponential} one sees
\begin{align}
    P[\phi_{+t}](s) &=  \det\left( s^\nt \one\transpose + F_0\transpose e^h\transpose \cdots F_{\nt-1}\transpose e^h\transpose \right)
                    \nonumber\\
                    &=  \det\left( s^\nt \one + F_0 e^h \cdots F_{\nt-1} e^h \right)
                    \nonumber\\
                    &=  \det\left( s^\nt \one + e^h F_0 \cdots e^h F_{\nt-1} \right)
                    \nonumber\\
                    &=  P[\phi_{-t}](s) \label{eqn:time reversal}
\end{align}
where we used the fact that $h$ is symmetric, $F$ is diagonal, and Sylvester's identity \eqref{eqn:determinant cyclicity}.
This shows that the field configuration $\phi_{-t}$, which is time-reversed with respect to $\phi_{+t}$, so that $\phi_{-t}$ on the first timeslice is $\phi_{+t}$ on the last and so on, yields the same eigenvalues.
An analogous proof holds for the diagonal case, starting from \eqref{eqn:P(S) diagonal}.
Thus, time reversal $T$ holds independent of discretization scheme and $\alpha$.

\subsubsection{Spatial Symmetries}\label{sect:spatial symmetries}

The spatial lattice may have some rotational, translational, or parity or reflection symmetries.
An operation \permutation of these kinds permutes spatial sites,
\begin{equation}
    x \goesto \permutation x
\end{equation}
and is a symmetry if its application commutes with the Hamiltonian \eqref{eqn:alpha case}.\footnote{Strictly speaking, an operation commuting with the Hamiltonian is not enough to guarantee an invariance of the discretized action for a particular field configuration.
On the honeycomb lattice, for example, some parity symmetries of the lattice exchange sublattices.
In the mixed differencing scheme of Refs.~\cite{Luu:2015gpl,Berkowitz:2017bsn} those sublattices have different differencing operators and the weight is not guaranteed to be invariant under those parity operations.
However, those operations followed by a time reversal and charge conjugation keep the overall action invariant.}
Since every on-site interaction is of the same strength $U$ it is automatically invariant under any relabeling of sites; the symmetries of the lattice are those permutations that commute with the tight-binding Hamiltonian or $h$, $[\permutation, h]=0$.
Two configurations related by one of these symmetries have the same weight in the path integral.
They manifestly have the same gaussian factor.
Note that the action of a permutation \permutation on $F[\phi]$ is
\begin{equation}
    \permutation F[\phi_x] \permutation\inverse = F[\phi_{\permutation x}]
\end{equation}
where $\phi_{\permutation x}$ is configuration where the fields have changed sites according to the permutation.
Since \permutation commutes with $h$ (and also, obviously, $e^h$), we may insert $\one_\nx = \permutation\inverse \permutation$ everywhere, and find
\begin{equation}
    P[\phi_x](s) = P[\phi_{\permutation x}](s)
\end{equation}
independent of discretization scheme and $\alpha$.

\subsubsection{Exact Chiral Symmetry}\label{sect:chiral symmetry}

We can calculate the characteristic polynomial in the exponential case using the other identity,~\eqref{eqn:det M 2}, and massage it such that we can identify~\eqref{eqn:P(S) exponential},
\begin{align}
    P^e[+h, +\phi_{+t}](s)
        &=  \det\left(e^h F_{\nt-1} \cdots e^h F_{0}\right) \det\left(\one_\nx + F_0\inverse e^{-h} s \one_\nx \cdots F_{\nt-1}\inverse e^{-h} s \one_\nx\right)
        \nonumber\\
        &=  \left(\prod_t \det(F_t)\right) \det(s^\nt \one_\nx) \det\left( s^{-\nt} \one_\nx + F_0\inverse e^{-h} \cdots F_{\nt-1}\inverse e^{-h} \right)
        \nonumber\\
        &=  \left(\prod_t \det(F_t)\right) \det(s^\nt \one_\nx) P^e[-h, -\phi_{-t}](1/s)
        \nonumber\\
        &=  \left(\prod_t \det(F_t)\right) \det(s^\nt \one_\nx) P^e[-h, -\phi_{+t}](1/s)
        \label{eqn:exponential one over s}
\end{align}
where we used the fact that the determinant of $e^h$ is unity to simplify the leading factor, and we applied the time reversal identity for the characteristic polynomial~\eqref{eqn:time reversal}.
This identity shows that the eigenvalues $s$ for one species are the reciprocal of the eigenvalues for the other, even without the assumption of a bipartite lattice.
Using the ``other'' determinant identity for the diagonal case yields no useful identity, because $s$ does not appear alone, but rather in combination with the tight binding Hamiltonian, and cannot be gathered together and factored out.
In the linearized case of~\eqref{eqn:M2} we also find no useful identity, as suggested in \Reference{Buividovich:2018yar}.

Since, in the exponential case, the off-diagonal blocks in $S$ are the timeslice-to-timeslice transfer matrices,~\eqref{eqn:exponential one over s} shows that the transfer matrix for one species has the inverse eigenvalues of the transfer matrix for the other.
This is the exact chiral symmetry discussed in \Reference{Buividovich:2018yar}, and only appears in the exponential case.
Note that this identity does not rely on particle-hole symmetry.

Using~\eqref{eqn:sign of h} in the bipartite case and~\eqref{eqn:P(S) complex conjugate} in the $\alpha=1$ case, we arrive at
\begin{align}\label{eqn:conjugate reciprocity}
    P^e[\phi](s)    &=   \left(\prod_t \det(F_t)\right) \det(s^\nt \one_\nx) {P^e[\phi](1/s^*)}^*
    &
    (\alpha         &=  1).
\end{align}
As long as $s\neq0$, if $s$ is a root of $P^e[\phi]$, so is $1/s^*$.
This demonstrates that in the exponential $\alpha=1$ case each ring of eigenvalues has a partner ring with a reciprocal radius and the same angular alignment, as we observed in the non-interacting exponential case.
We say that $P^e[\phi]$ is proportional to its own \emph{conjugate reciprocal polynomial} ${P^e[\phi]}^\dagger$.
The conjugate reciprocal polynomial $p^\dagger$ of a polynomial $p$ is given by
\begin{equation}
    p^\dagger(z) = z^n \overline{p(\bar{z}\inverse)}
\end{equation}
here we use the overbar to indicate complex conjugation to avoid confusion with the asterisk that sometimes indicates the reciprocal polynomial.

Essential to demonstrating this conjugate reciprocity in the interacting case was~\eqref{eqn:P(S) complex conjugate} and therefore the properties of $F$~\eqref{eqn:F star}.
Although it is true for the noninteracting case and it is visually plausible for the interacting exponential $\alpha=0$ example in \Figref{fermion-matrix-spectra}, the radii are not, in fact, multiplicative inverses.

Also essential was the particle-hole symmetry that allowed us to flip the sign of $h$.
Without that symmetry, we find a relation between two characteristic polynomials, but not a relation that can be used to show this conjugate reciprocity.

Recall that setting $s=-1$ in a characteristic polynomial gives $\det(M[\phi])$.
It is now helpful to return the $\alpha$ dependence to the argument of $M$ rather than implicit in $F$ as in~\eqref{eqn:F alpha}.
Starting from~\eqref{eqn:exponential one over s}, using the particle-hole identity~\eqref{eqn:sign of h},  and plugging in $s=-1$ leads to the identity
\begin{align}
    \det M^e[+\phi] &= e^{+\Phi} \det M^e[-\phi] \nonumber\\
    f[\phi] = e^{-\Phi/2} \det M^e[+\phi]    &= e^{+\Phi/2} \det M^e[-\phi] = f[-\phi]  \label{eqn:exponential-factorization}
\end{align}
where we define $\Phi = \sum_{x,t} \phi_{x,t}$, $f[\phi] = e^{-\Phi/2} \det M^e[\phi]$ and immediately see that $f$ is even.
When $\alpha=0$, $f[\phi]$ must be real, because the determinant is real, as shown in \eqref{eqn:P(S) complex conjugate}.
When $\alpha=1$, we can use \eqref{eqn:P(S) complex conjugate} and find
\begin{equation}
    f[i\phi]    = e^{-i\Phi/2} \det M^e[+i\phi]
                = e^{+i\Phi/2} \det M^e[-i\phi]
                = e^{+i\Phi/2} \det M^e[+i\phi]^* = f[i\phi]^*
\end{equation}
so that $f$ is also real when $\alpha=1$.
Put another way, starting from \eqref{eqn:conjugate reciprocity} one finds
\begin{equation}
    \det(M^e[i\phi]) = e^{i\Phi} \det(M^e[i\phi])^*
\end{equation}
Writing, in radial coordinates, $ \det(M[i\phi]) = f[i\phi] e^{i\theta(\phi)}$, one finds
\begin{equation}\label{eqn:Q}
    \theta(\phi) = \frac{\Phi}{2} + Q\pi.
\end{equation}
where $Q$ is an integer.
So
\begin{equation}\label{eqn:alpha1 exp reality}
  e^{-i\Phi/2} \det M[i\phi] \in \Reals
\end{equation}
but of either sign, as discussed in (28) of \Reference{Ulybyshev:2017hbs}\footnote{Actually, \Reference{Ulybyshev:2017hbs} differs by a factor of two in the exponent.
For the one-site problem we later give the explicit form in \eqref{eqn:det M2}, confirming our result.
Note that one does get $\prod_t\det F_t = \exp(-\Phi)$ in forms like \eqref{eqn:P(S) diagonal} but no demonstration of a partitioning of the configuration space follows.}.
Since $f$ is continuous in $\phi$, changing $Q$ requires passing through $f=0$.
This is the origin of formal ergodicity problems, as we discuss later.

When $\alpha=0$ it is not guaranteed that $f$ must take both signs.
Indeed, as we later show for the one-site problem~\eqref{eqn:det M1} and a simple two-site problem~\eqref{eqn:f two site}, it may be that the sign of $f[\phi]$ is fixed in this case.
That is, at least for some examples $f[\phi]$ does not change sign and there are no formal ergodicity problems, but we stress that this is not necessarily generic.
We will see, one way or the other, that there are also in-practice problems in the exponential $\alpha=0$ case.

In the diagonal discretization there is no analog of~\eqref{eqn:exponential one over s}, no natural factorization of the determinant emerges, and there are no sectors that are separated by a vanishing determinant, even when $\alpha=1$.
Thus, the diagonal discretization does not formally suffer from the ergodicity problems the exponential discretization suffers.
In Section~\ref{sect:2sites} we provide a simple two-site example where this claim may be directly verified.
If one insists on writing $\det M^d[\phi] = e^{\Phi/2} f[\phi]$, in the $\alpha=0$ case of course $f$ will still be real, but only because the determinant itself is real.
We find no restriction forcing $f[i\phi]$ to be real in the diagonal case.
We later provide an example of $f$ wandering off-axis in the complex plane in Figure~\ref{fig:eigenvalue-evolution}.

As shown, the partitioning of the field space into sectors is a result of the conjugate reciprocity of the bipartite, exponential, $\alpha=1$ case.
It may be that in other cases there are other as-yet formally undemonstrated partitionings.
In Section~\ref{sect:exceptional configurations} we show examples of the determinant flipping sign by crossing zero for both $\alpha=0$ cases.

Let us continue to focus on the $\alpha=1$ case and on the properties of $f[i\phi]$.
Consider now what happens when we increase one of the auxiliary field variables by $2\pi$,
\begin{equation}
    \phi_{x,t} \goesto \phi_{x,t} + 2\pi \delta_{x,x_0}\delta_{t,t_0}
\end{equation}
where $x_0$ and $t_0$ are the space and time coordinates of the field we are changing.
Since $F$ is $2\pi$-periodic in each field variable individually, the determinants must be equal.
Then, we find,
\begin{align}
    f[i\phi_{x,t}] e^{i \Phi/2} = \det M^e[i\phi] &= \det M^e[i\phi+ 2\pi i \delta_{x,x_0}\delta_{t,t_0}] = f[i\phi_{x,t} + 2\pi i\delta_{x,x_0}\delta_{t,t_0}] e^{i (\Phi/2 + \pi)}
    \nonumber\\
    f[i\phi_{x,t}] &= - f[i \phi_{x,t} + 2\pi i \delta_{x,x_0}\delta_{t,t_0}]   \label{eqn:f sign}
\end{align}
so that shifting any field variable by $2\pi$ flips the sign of $f$.
Since this flip is independent of all other field variables, this shows the manifolds of zeroes are codimension 1.

\subsubsection{Summary}\label{sect:eigenvalue summary}

We collect in Table~\ref{tab:eigenvalue constraints} constraints on the eigenvalues in the different discretizations and bases.
Independent of discretization or basis we have, on a half-filled bipartite lattice, charge conjugation symmetry, particle hole symmetry, temporal translation symmetry, time reversal symmetry, and whatever spatial symmetries the lattice exhibits.
In the exponential case we have exact chiral symmetry and when $\alpha=1$ conjugate reciprocity.

\begin{table}[t]
    \begin{center}
    \begin{tabular}{|r|rl|rl|}
            \hline
            &           &   exponential                 &           &   diagonal        \\ \hline
$\alpha=0$  &   $s$     &   $\with s e^{2\pi i/\nt}$    &   $s$     &   $\with -s$      \\
            &   $s$     &   $\with s^*$                 &   $s$     &   $\with s^*$     \\
            &   $s_\up$ &   $\with 1/s_\down$           &           &                   \\ \hline
$\alpha=1$  &   $s$     &   $\with s e^{2\pi i/\nt}$    &   $s$     &   $\with -s$      \\
            &   $s_p$   &   $\with s_h^*$               &   $s_p$   &   $\with s_h^*$   \\
            &   $s_p$   &   $\with 1/s_h$               &           &                   \\ \hline
    \end{tabular}
    \end{center}
    \caption{Demonstrated relationships between the eigenvalues $s$ in the half-filled bipartite case.
    The eigenvalues of the fermion matrix are these eigenvalues plus one.
    When no subscript is attached the relationship is between eigenvalues for the same species.
    When $\alpha=0$ we attach $\up$ and $\down$ subscripts to indicate eigenvalues for the different spin species; for $\alpha=1$ we attach $p$ and $h$ for particles and holes.
    The last two relationships for the exponential $\alpha=1$ case imply conjugate reciprocity and lead to a partitioning of the configuration space into sectors separated by exceptional configurations.}
    \label{tab:eigenvalue constraints}
\end{table}

Earlier we showed the weight $W$ was real.
For a straightforward Monte Carlo method, $W$ should have an interpretation as a probability measure, and should therefore be positive.
When $\alpha=1$, each particle eigenvalue is the complex conjugate of a hole eigenvalue.
This guarantees positivity of the weight $W$.
When $\alpha=0$ the complex conjugate guarantee is not enough.
In the exponential case, we can use~\eqref{eqn:exponential-factorization} to demonstrate positivity, though each determinant individually is not guaranteed to be positive---but the chiral symmetry is enough to guarantee that the two species determinants have the same sign.

Consider the $\alpha=0$ diagonal case.
If $s$ is real it is its own complex conjugate.
If the corresponding eigenvalue of $M$, $\lambda = s+1$, is real and negative it need not have another negative partner in the eigenvalues of either spin species.
Thus, positivity is not, in general, guaranteed in this case.
We later give a simple example in~\eqref{eqn:f two site} and find that positivity can be guaranteed with small enough $\kappatilde^2$.
We observe that positivity can be lost when \nt is odd and $\kappatilde$ is large and have not found an \nt-even example; we conjecture that this is generic and positivity can always be guaranteed by increasing \nt and approaching the continuum limit.
Since we know of no large-scale computational effort using this combination of basis, discretization, and odd \nt we leave a precise determination of how large \nt must be to ensure positivity to future work.
We focus on \nt even.

Knowledge of eigenvalues of $S$ can help us discover eigenvectors of $M$.
Solving the eigenvalue equation using a known eigenvalue $s$, $(S-s\one)v = 0$ yields the eigenvector $v$ of $S$.
Since $S=M-\one$, $S$ and $M$ share all their eigenvectors.
But $S$ may be much better conditioned than $M$ and thus much easier to invert.
It may also provide a numerical speedup to find an additional eigenvector for the cost of a single solve if the associated eigenvalue comes for free.
We are investigating the acceleration of the inversion of $M$ using knowledge of the relationships between eigenvalues in Table~\ref{tab:eigenvalue constraints}.


\subsection{Extreme Limits\label{sect:limits}}

In the general case it is hard to analytically extract features of the probability weight function.
However, in particular limits additional symmetries emerge and yield additional information about the weight.

First, when $U/\kappa$ gets very small, for fixed $U\beta$, the interaction is effectively turned off and $F$ gets close to $\one_\nx$.
Then the path integral nearly factorizes into \nt copies of the non-interacting path integral,
\begin{equation}\label{eqn:weak coupling}
    \lim_{U/\kappa\goesto0} \Z_{\nt} = (\Z_1)^\nt
\end{equation}
where the right-hand side is \nt powers of the $\nt=1$ problem with the same $\Utilde$.
We call this limit the weak coupling limit.

In the opposite limit---the limit of no hopping, $\kappa\goesto0$ with fixed $U\beta$, the partition function on $\nx$ sites factorizes into $\nx$ copies of the one-site the partition function,
\begin{equation}\label{eqn:strong coupling}
    \lim_{\kappa\goesto0} \Z_{\nx} = (\Z_1)^{\nx}.
\end{equation}
where now the right-hand side is \nx powers of the one-site problem.
We call this limit the strong coupling limit, $U/\kappa\goesto\infty$ (with $U\beta$ held fixed implicit).

In the general case we have the charge conjugation symmetry discussed in Section~\ref{sect:charge conjugation},
\begin{equation}
    \phi    \goesto     -\phi
\end{equation}
In the two limits we can make independent negations of $\phi$,
\begin{equation}
    \phi_{x,t}    \goesto
    \begin{cases}
        \text{sign}_t \phi_{x,t}    &   (\text{weak coupling})    \\
        \text{sign}_x \phi_{x,t}    &   (\text{strong coupling})
    \end{cases}.
\end{equation}

In the general case we have invariance under temporal shifts and time reversal, as discussed in Sections~\ref{sect:temporal shifts}~and~\ref{sect:time reversal},
\begin{align}
    \phi_t      &   \goesto     \phi_{t+\tau}
    &
    \phi_{+t}   &   \goesto     \phi_{-t}
\end{align}
In the two limits we again can make a larger set of transformations.
In the weak coupling limit the symmetry is enhanced and we can arbitrarily permute the timeslices,
\begin{equation}
    \phi_t          \goesto     \phi_{\mathcal{T}t}     \quad\quad(\text{weak coupling})
\end{equation}
where $\mathcal{T}$ is a permutation.
In the strong coupling limit we can independently perform these operations on each thread of spatial sites,
\begin{align}
    \phi_{xt}  &   \goesto     \phi_{x(t+\tau_x)}
    &
    \phi_{xt} &   \goesto     \phi_{x(\text{sign}_x t)}.
    &   (\text{strong coupling})
\end{align}

In the general case we have the spatial symmetries discussed in Section~\ref{sect:spatial symmetries},
\begin{equation}
    \phi_{xt}  \goesto \phi_{(\permutation x) t}    \quad\quad(\permutation\text{ a lattice symmetry})
\end{equation}
where the same operation \permutation is a symmetry of the lattice and is applied to every timeslice.
In the weak coupling limit we can apply a different spatial transformation on each timeslice and in the strong coupling limit we can arbitrarily permute the threads of spatial sites,
\begin{equation}
    \phi_{xt}  \goesto \begin{cases}
        \phi_{(\permutation_t x) t}  &   (\text{weak coupling, } \permutation\text{ a lattice symmetry})    \\
        \phi_{(\permutation x)t}     &   (\text{strong coupling, } \permutation\text{ any permutation})
    \end{cases}.
\end{equation}

These operations will provide inspiration for proposal machines which give large field transformations that are still accepted often enough, helping overcome ergodicity problems HMC may encounter.
This strategy is discussed in Section~\ref{sect:jumps}.



\section{Ergodicity Problems}
\label{sect:ergo}

We now turn to the issue of ergodicity, which is required for an accurate, unbiased Markov-Chain Monte-Carlo (MCMC) algorithm.
An \emph{ergodicity problem} arises when the algorithm for updating the state of the Markov chain is unable to visit the neighborhood of every field configuration.
In this case we can introduce bias and find inaccurate results.
We can further delineate between \emph{in-principle} or \emph{formal} ergodicity problems and \emph{in-practice} ergodicity problems.
In a formal ergodicity problem there are regions of configuration space that the update algorithm cannot find, by any means.
An in-practice problem might arise when the update algorithm \emph{could} explore the whole space but is unlikely to find important regions of configuration space in the finite amount of time you are willing to run your computer.
We emphasize that in this context ergodicity is a property of an algorithm, and not of physics itself.

As Hybrid Monte Carlo (HMC) is an MCMC algorithm, it relies on the previous state to propose a new state which is subjected to the Metropolis-Hastings accept/reject step.
This final step is essential for maintaining detailed balance and ultimately corrects (via the ensemble average) for any numerical errors in the evolution of the state~\cite{Duane:1987de}.
The proposed state is obtained by integrating the equations of motion (EoMs) derived from an artificial Hamiltonian $\mathcal{H}$ in a newly-introduced time direction.
In the case of the Hubbard model and $\alpha=0$, this artificial Hamiltonian is the Legendre transform of the action \eqref{eqn:spin basis partition fn}
\begin{equation}
  \mathcal{H}[\pi,\phi] = \frac{1}{2}\sum_{x,t}\left(\pi_{xt}^2+\frac{1}{\tilde U}\phi_{xt}^2\right) -\log\det M[\phi]-\log\det M[-\phi]\ ,\label{eqn:artificial H}
\end{equation}
where $\pi$ are newly-introduced momenta conjugate to the field variables $\phi$.
For $\alpha=1$ given by~\eqref{eqn:ph basis partition fn}, replace $\phi \to i \phi$ in the fermion matrices.
Details of this method can be found in the pioneering Ref.~\cite{Duane:1987de}.

Areas where the integrand of the partition function (\eqref{eqn:spin basis partition fn} and~\eqref{eqn:ph basis partition fn}) has zero weight, for example when $\det M(\phi)=0$, are represented by infinitely tall potential barriers in the Hamiltonian \eqref{eqn:artificial H} which repel the state during the integration of the EoMs.
This, in general, is a wanted feature, since such locations in configuration space contribute nothing to the partition function and should thus be avoided.
Problems arise, however, when such barriers separate regions that \emph{do} contribute to the partition function.
If there are manifolds in configuration space of codimension-1 ($\nx\nt-1$ dimensional, when there is only one degree of freedom per site) the configuration space is partitioned and HMC trajectories cannot propagate to these different sectors, thereby violating ergodicity.
This is an in-principle problem.

It may be that there are extended codimension-1 manifolds in configuration space that terminate on boundaries, in which case it is in principle possible for HMC to find a sequence of updates that visits states on both sides of the manifold.
Whether this is a problem in practice depends on the model of interest; if these manifolds are very big they might take a long time to circumnavigate.
The manifolds in the cases discussed here are boundary-free.

If there are zero-weight manifolds of higher codimension, the configuration space isn't partitioned and HMC can always explore the whole space, though there could still be a problem in practice.

Reference~\cite{Ulybyshev:2017hbs} pointed out that the factorization~\eqref{eqn:exponential-factorization} and reality of $f[i\phi]$ implies a formal ergodicity problem for the exponential discretization when $\alpha=1$.
Since $f[i\phi]$ is real and takes both positive and negative values, by the intermediate value theorem, there must be zeros separating the two regions, as discussed at the end of Section~\ref{sect:chiral symmetry}.
As first suggested in Ref.~\cite{Ulybyshev:2017hbs}, there exist codimension-1 manifolds in $\phi$ where the determinant is zero.
We can understand this result by seeing that $Q$ in~\eqref{eqn:Q} cannot be changed by a continuous change in $f$ unless $f$ passes through zero.
Were the manifolds smaller in dimension, HMC would find its way around, without having to go through the barriers, though there could nevertheless have been an issue in practice, as it might take a long time to circumnavigate the zeros.
Formally, an infinitely precise HMC integrator cannot penetrate these barriers and is therefore not ergodic.

When considering the diagonal discretization in~\eqref{eqn:M3}, the factorization \eqref{eqn:exponential-factorization} does not naturally emerge.
As discussed in Section~\ref{sect:chiral symmetry}, this factorization was the result of the exact chiral symmetry found in the exponential discretization.
When considering the spin basis ($\alpha=0$) with these discretizations the determinant is still real, and may (but need not) still have both negative and positive values and the intermediate zeros may obstruct the exploration of configuration space by HMC.\@

For the particle/hole basis using the diagonal discretization in~\eqref{eqn:M3}, the situation is quite different.
The factorization still does not naturally emerge.
If one insists on writing it that way, the function $f[i\phi]$ is complex and thus we can avoid the conclusion of the intermediate value theorem that would force HMC to go through a zero in order to change the sign of $f$.
The function $f[i\phi]$ may still have zeros, but there no longer need be codimension-1 zero manifolds and thus the space is not partitioned into regions that trap HMC trajectories.
We provide numerical examples of this in the following sections.

We stress that our arguments here do not prove unequivocally that the particle/hole basis using the diagonal discretization~\eqref{eqn:M3} does not suffer from any formal ergodicity issues, only that it does not suffer from those identified in Ref.~\cite{Beyl:2017kwp}.
However, we are not aware of any other formal ergodicity issues this discretization may have.
As one nears the continuum limit, since the two discretizations must agree, very tall potential barriers can rise between the exceptional configurations.
This raises the possibility of an \emph{in practice} problem; how difficult it is to overcome depends on the exact example and how fine a temporal discretization one uses.
This is later demonstrated for a simple problem in Figures~\ref{fig:2sites contours} and~\ref{fig:2sites contours finer}.
When we propose a general solution to the formal ergodicity problem of the exponential $\alpha=1$ case in Section~\ref{sect:jumps}, such a solution can also resolve the in-practice problem in the diagonal $\alpha=1$ case that emerges near the continuum limit.

In the remainder of this section we discuss the dynamics of the eigenvalues and the role they play in formal ergodicity problems, and then present numerical examples that support our discussion above, detailing in-principle and in-practice problems.
We investigate small systems where we can perform direct comparisons with exact solutions.
Though the systems are small in dimension, they capture all the relevant aspects of ergodicity (or lack thereof) that are present in larger simulations, and provide the added benefit that these aspects can be visualized.

In all our HMC simulations, unless otherwise stated, we always target an acceptance rate $\geq 80\%$ by adjusting the accuracy of our numerical integration of the EoMs.
As previously stated, the error in our integration is corrected by the accept/reject step.
A more accurate integration corresponds to a higher acceptance rate, but the configuration space probed by each trajectory is diminished.
Conversely, a less accurate integrator allows HMC to probe more configuration space at the expense of a lower acceptance rate.
Since these examples are so small, we produce an ensemble of fields $\{\phi\}$ by thermalizing for a few thousand trajectories and generating 10,000 to 100,000 HMC trajectories per simulation.
We compute correlators only on every 10th configuration to reduce autocorrelations.
Our uncertainties are given by the standard deviation of bootstrap samples of the particular quantity in question.

The HMC ensembles and correlator data are available online in Ref.~\cite{data2018-12-19}.
This data was generated using Isle~\cite{isle01}, a new library currently in development for HMC calculations in the Hubbard model.

\subsection{Exceptional Configurations And Zero Eigenvalues}\label{sect:exceptional configurations}

An \emph{exceptional configuration} is one with zero weight in the path integral.
The zero weight implies an infinite potential in the classical EoMs used in HMC, and as a trajectory nears an exceptional configuration the force diverges.
In the case of the Hubbard model these zeros arise from a vanishing fermion determinant, which in turn corresponds to a vanishing eigenvalue of the fermion matrix.

For $\alpha=0$, the fact that $\det M$ can be positive and negative, and is always real, implies that manifolds of exceptional configurations partition the configuration space; there is a formal ergodicity problem.
We observe that the frequency with which exceptional configurations are encountered increases as one approaches the continuum limit $\nt\goesto\infinity$.
This can be understood by considering the non-interacting eigenvalues in~\eqref{eqn:M eigenvalues}.
Two factors drive the increased frequency.
First, $\delta$, which vanishes with increasing \nt, controls how close eigenvalues are to the origin.
In the exponential discretization the radii go to 1 with vanishing $\delta$.
In the diagonal discretization the rings' centers converge on 1 with vanishing $\delta$.
The other issue is that the rings of eigenvalues become increasingly dense with \nt, as the Matsubara frequencies come closer together.
This observation also holds for the interacting case with typical auxiliary field configurations.

The frequency of exceptional configurations also depends on the spatial lattice.
Again considering the non-interacting limit and assuming the hopping term $H_0/\kappa$ has a vanishing energy eigenvalue $\epsilon=0$.
In the exponential case there is at least one ring of eigenvalues with unit radius which puts the eigenvalues close to the origin; in the diagonal case one of the rings is centered exactly on 1.

Moreover, as the infinite-volume limit is taken the eigenvalues become more dense.
For regular lattices with $n$ nearest neighbors, the nearest-neighbor hopping Hamiltonian $H_0/\kappa$ has eigenvalues $\epsilon$ bounded by $-n\leq \epsilon \leq n$.
The number of eigenvalues of the non-interacting Hamiltonian is given by the number of sites; going towards the infinite volume limit means more and more rings will appear and can get close to the origin.
This observation, again, is borne out of the interacting case with typical auxiliary field configurations.
Of course, with a nonzero field configuration the eigenvalues move and the observations are only qualitatively true---the eigenvalues are no longer exactly controlled by the noninteracting energies, for example.

Even when the lattice is large, some geometries may be more favorable than others.
The square lattice with periodic boundary conditions always has a zero eigenvalue; the honeycomb lattice only has a zero eigenvalue when the lattice dimensions are congruent to 0 (mod 3).

In \autoref{fig:fermion-matrix-spectra} we show the spectra of fermion matrix eigenvalues in the complex plane.
The interacting eigenvalues for the example exponential $\alpha=0$ configuration lie on rings centered on $1+0i$ at angles determined by the Matsubara frequencies.
Were this generically true for all $\phi$, the determinant would always be positive and there would be no in-principle ergodicity problem.
However, the eigenvalues are only constrained to obey the partnerships in the exponential $\alpha=0$ portion of \Tabref{eigenvalue constraints}.

Another way to satisfy those relationships emerges when two rings of eigenvalues have the same radius.
Then, as $\phi$ changes, the two rings can counter-rotate by the same angle, still obeying the constraints of \Tabref{eigenvalue constraints}.
When the two counter-rotated rings of eigenvalues eventually meet they can then part, moving radially.
Since the counter-rotation is always by the same angle, when the rings part the eigenvalues always lie on the rays determined by the Matsubara frequencies, or exactly halfway between those frequencies---putting eigenvalues on the real axis.
If one ring crosses $r=1$, the determinant changes sign.

We conjecture that once the eigenvalues are off of these rays the only way to maintain all the symmetry properties of the fermion matrix's spectrum is to remain locked to the radius where they first collided, and then once apart the symmetry properties cannot be maintained unless the eigenvalues are locked to the rays.
We have not seen examples where more than two rings all have the same radius and perform an even more complicated dance, though such a dance may be possible.

\begin{figure}[htbp]
	\centering
	\includegraphics[width=\textwidth]{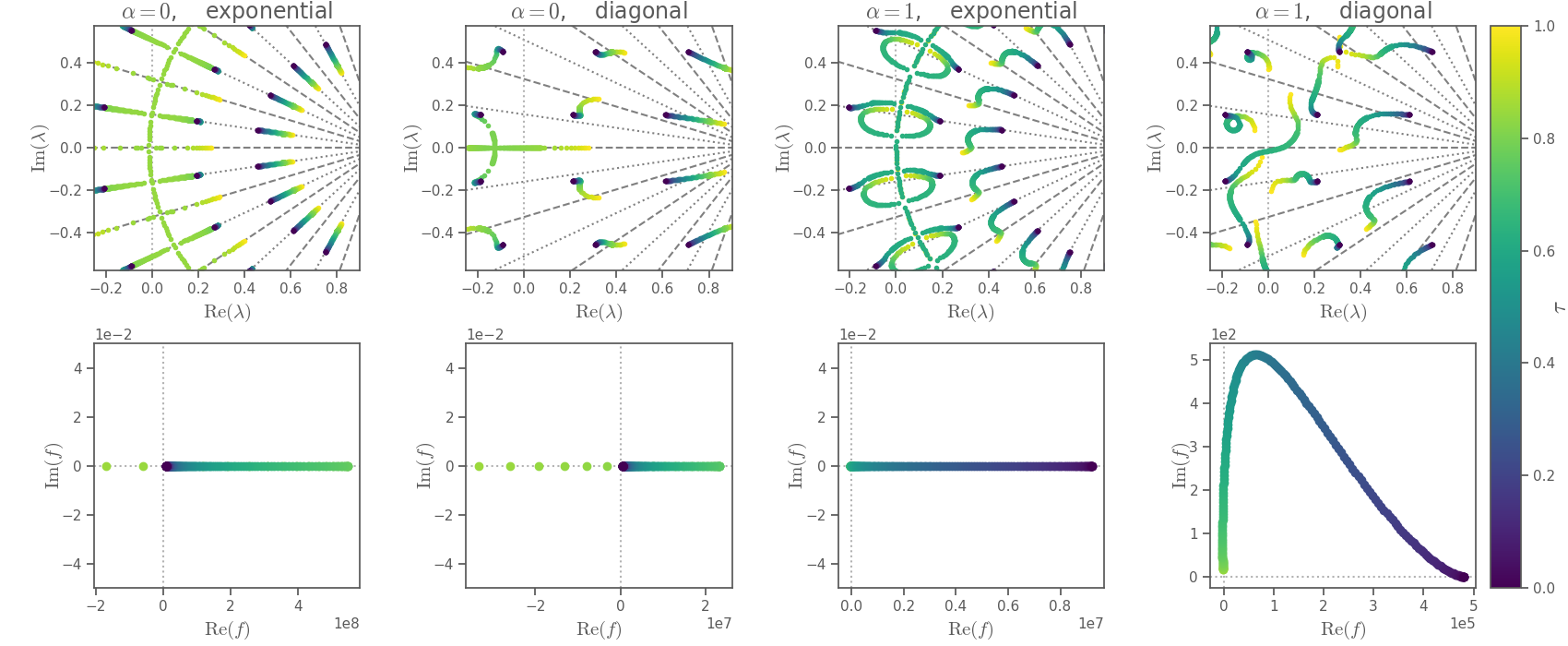}
	\caption{Eigenvalues $\lambda$ and $f$ as in~\eqref{eqn:exponential-factorization} for $\phi = \tau\phi^{-}$, where $\det M[\phi^-] < 0$ in all cases.
		The fictional time $\tau$ runs from 0 (purple), so that $\phi$ is the non-interacting case and the determinant is positive, to 1 (yellow) so that the determinant is negative; intermediate values of $\tau$ are shown in the color bar on the right.
		The eigenvalue trajectories are described in the text.
		The eigenvalue locations were computed on an $N_t=20$ 4-site honeycomb lattice with $\kappa\beta = 4$.
		See the supplementary material~\cite{supplemental} for animations of these eigenvalue dances.
		\label{fig:eigenvalue-evolution}
	}
\end{figure}

To illustrate the above statements, we track eigenvalues of $M$ from a configuration with positive determinant to a configuration with negative determinant in \autoref{fig:eigenvalue-evolution}.
Animations of the processes depicted in the figure are available in the supplementary material~\cite{supplemental}.
In the animation of the $\alpha=0$ exponential case, one can see the inner ring crosses $r=1$ between $\tau=0.8430$ and $\tau=0.8431$ and the determinant flips sign.

We generated a single configuration $\phi^-$ with a real, negative determinant for all four bases and discretization choices.
We show the spectrum of $M[\tau\phi^-]$ where the fictitious time $\tau$ runs from 0 (purple) to 1 (yellow), so that the spectrum goes from the noninteracting spectrum, which has a positive determinant, to the spectrum of $\phi^-$ with negative determinant.
The two leftmost panels shows the eigenvalue trajectories for the exponential $\alpha=0$ case, and $f$, related to the determinant by~\eqref{eqn:exponential-factorization}.
While it looks like the eigenvalues meet and rotate around the point 1 at a radius $r=1$, this is only approximately true.

The other cases are similar to the exponential $\alpha=0$ case, though the constraints on the eigenvalues are different.
In the exponential $\alpha=1$ case, the eigenvalues enjoy conjugate reciprocity.
If one eigenvalue crosses the origin, another eigenvalue must cross in the opposite direction, preserving the sign of the determinant.
When instead two rings meet at $r=1$, the eigenvalues are their own conjugate reciprocals.
Then, the rings can rotate oppositely, though the angles they rotate by need not be equal as there is no complex-conjugation constraint as in the $\alpha=0$ case.
The determinant changes sign while the two rings rotate.
When the two rings again meet, they can part radially, preserving the conjugate reciprocity constraint.

Eigenvalues in the $\alpha=0$ diagonal case are no longer locked to the rays, because the eigenvalues do not come in perfect rings.
However, the complex-conjugate pairing still means a real eigenvalue must cross the origin to flip the sign of the determinant.

In the $\alpha=1$ diagonal case the eigenvalues need not meet at all to traverse from a real and positive determinant to a real and negative determinant.
We show the movement of eigenvalues and $f$ (defined by~\eqref{eqn:exponential-factorization} even for the diagonal case where such a factorization does not naturally emerge) for each case in Figure~\ref{fig:eigenvalue-evolution}.


\subsection{The One-site Problem\label{sect:1site}}

Strictly speaking, the one-site problem is not bipartite.
However, since there is no hopping in this case, the tight-binding Hamiltonian $H_0$ can be ignored ($\kappa=0$) and the exponential, linear, and diagonal discretizations \eqref{eqn:M1},~\eqref{eqn:M2},~and~\eqref{eqn:M3} are all equivalent
\begin{equation}
{M[\phi]}_{x^\prime t^\prime, x t} = \delta_{x^\prime,x} \delta_{t^\prime,t} - e^{\phi_{xt}} \delta_{t^\prime, t+1} \ .
\end{equation}
The determinant of the fermion matrix in this case can be expressed in closed form,
\begin{align}
\det M[\phi] &= 2\cosh\left(\frac{\Phi}{2}\right)\, e^{\Phi/2}\label{eqn:det M1}\\
\det M[i\phi] &= 2 \cos\left(\frac{\Phi}{2}\right)\, e^{i\Phi/2}\ ,\label{eqn:det M2}
\end{align}
where $\Phi=\sum_t^{\nt}\phi_t$ as in \eqref{eqn:polyakov loop}.
We see that
\begin{align}
    f[\phi]     &=2\cosh(\Phi/2) &
    f[i\phi]    &=2\cos(\Phi/2),
\end{align}
that $f$ is even, and that $f[\phi]$ does not go through zero, while $f[i\phi]$ does.
Thus we expect that the particle/hole basis has infinite barriers in the artificial Hamiltonian, and formally has ergodicity issues, while the spin basis has no formal ergodicity issue.

However, as we shall show shortly, calculations in the spin basis exhibit a bimodal distribution that becomes increasingly separated for large $U\beta$.
Though calculations in this basis do not formally suffer from ergodicity issues as $\det M[\phi]$ never vanishes for any $\phi$, in practice the separation of the modes for large $U\beta$ essentially separates two regions that are extremely unlikely to be connected via HMC, regardless of the accuracy of the integration of the EoMs.
This presents an \emph{in practice} ergodicity issue.

Using \eqref{eqn:det M1} with~\eqref{eqn:spin basis partition fn}, one finds that the weight of a field configuration $\phi$ is given, up to an overall normalization, by
\begin{align}
  W[\phi] =
  \det M[\phi] \det M[-\phi] e^{-\frac{1}{2\Utilde} \sum_{t} \phi_{t}^{2}}
  &= 4\cosh^2\left(\frac{\Phi}{2}\right) e^{-\frac{1}{2\Utilde} \sum_{t} \phi_{t}^{2}}\\
  &= 4\cosh^2\left(\frac{\Phi}{2}\right) e^{-\frac{1}{2U\beta}\Phi^2} \exp\left(-\frac{1}{4U\beta} \sum_{t_1,t_2} (\phi_{t_1}-\phi_{t_2})^{2}\right).\label{eqn:kernel1}
\end{align}
Note that after completing the squares in the exponent, we have completely exposed the $\Phi$ dependence of the probability weight.
The other factor,
\begin{equation}\label{eqn:other factor}
    \exp\left(-\frac{1}{4U\beta} \sum_{t_1,t_2} (\phi_{t_1}-\phi_{t_2})^2\right)
\end{equation}
can be shown to be independent of $\Phi$, by directly differentiating with respect to $\Phi$ and using $\partial \Phi / \partial \phi_t = 1$ and $\partial \phi_i / \partial \phi_j = \delta_{ij}$ for simplification.
Thus, the distribution of $\Phi$ is bimodal, as seen in the factor $4\cosh{\left(\frac{\Phi}{2}\right)}^2e^{-\frac{1}{2U\beta}\Phi^2}$ in \eqref{eqn:kernel1}.
This distribution is strongly peaked about $\Phi \approx \pm U\beta$, implying that the peaks of the modes are separated by a distance $2U\beta$, and is exponentially small in between.
This analysis also shows that the modes become further separated when either $U$ is increased (strong coupling limit), or $\beta$ is increased (zero temperature limit), or both.

For the $\alpha=1$ case, using \eqref{eqn:det M2} with~\eqref{eqn:ph basis partition fn} gives
\begin{equation}\label{eqn:kernel2}
    W[\phi] = \det M[i\phi] \det M[-i\phi] e ^ { - \frac { 1} { 2\Utilde} \sum _ { t } \phi _ { t } ^ { 2} }
  =4\cos^2\left(\frac{\Phi}{2}\right)e^{-\frac{1}{2U\beta}\Phi^2} e ^ { - \frac { 1} { 4U\beta} \sum _ { t_1,t_2 } (\phi _ { t_1 }-\phi_{t_2}) ^ { 2} }.
\end{equation}
The last factor is the same as before and is $\Phi$-independent, and thus the manifold of zeroes is indeed codimension 1 so that a precise HMC integrator cannot circumnavigate the zeros and a formal ergodicity problem arises.
The distribution of $\Phi$ is determined by
$4\cos^2\left(\frac{\Phi}{2}\right)e^{-\frac{1}{2U\beta}\Phi^2}$, a product of a gaussian of width $\sqrt{U\beta}$ centered at $\Phi=0$ and a simple, periodic $\cos^2\left(\frac{\Phi}{2}\right)$ function.
The zeros of $\cos\left(\frac{\Phi}{2}\right)$ dictate the zeros of the kernel, and we find zeros at $\Phi= (2n+1)\pi$ for all integers $n$, independent of the value $U\beta$.
This function is multimodal, but the modes remain close together, even as $U$ or $\beta$ are taken large.
Rather than separating two important modes, taking the low-temperature or strong-coupling limit broadens the gaussian and increases the number of important modes.
We can analytically determine the probability distribution for $\Phi$,
\begin{align}
W[\Phi]=&\frac{e^{-\frac{\Phi ^2}{2 U\beta}-\frac{U\beta}{4}}
   \cosh ^2\left(\frac{\Phi
   }{2}\right)}{\sqrt{2 \pi U\beta}\ \cosh\left(\frac{U\beta}{4}\right) } \label{eq:alpha0PPhi}
   & (\alpha&=0)
   \\
   W[i\Phi]=&\frac{e^{-\frac{\Phi ^2}{2 U\beta}+\frac{U\beta}{4}}
   \cos ^2\left(\frac{\Phi
   }{2}\right)}{\sqrt{2 \pi U\beta} \ \cosh\left(\frac{U\beta}{4}\right)} \label{eq:alpha1PPhi}
  & (\alpha&=1).
   \end{align}
Note that these expressions depend only on the product $U\beta$, and in particular are independent of $\nt$.

\begin{figure}
	\resizebox{.8\textwidth}{!}{\input{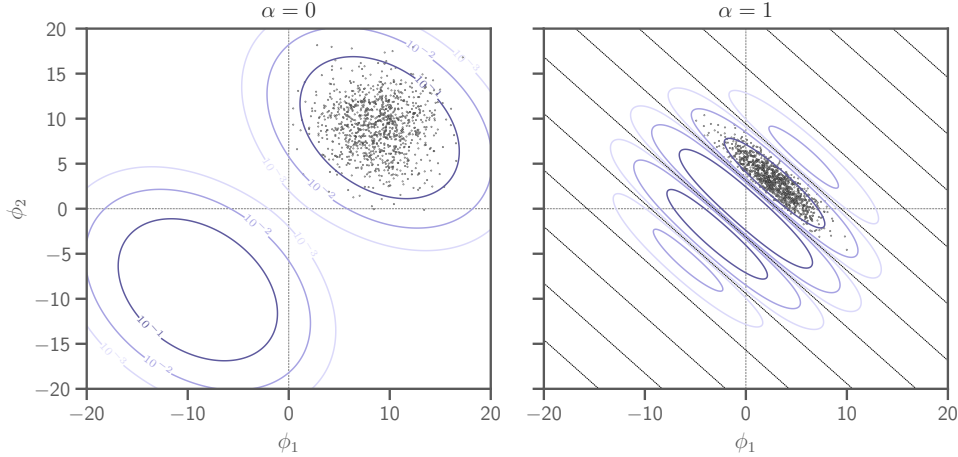}}
	\caption{
		Contours of constant probability, \eqref{eqn:kernel1} (left, $\alpha=0$) and \eqref{eqn:kernel2} (right, $\alpha=1$) with $U\beta=18$ and $\nt=2$ for the 1-site problem, where the subscript on the Hubbard-Stratonovich fields $\phi$ indicates the timeslice.
		The dark contours represent high probability density, which decreases as the contours get lighter.
		The $\Phi$ axis runs from bottom left to top right.
		In the right panel the diagonal (black) lines correspond to locations where the probability weight \eqref{eqn:kernel2} is exactly zero, creating barriers to HMC; their regular spacing is a consequence of the periodicity of the determinant, discussed in \Secref{shifts and periodicity}.
		We show every tenth configuration of 10,000 configurations (generated with a very precise integrator) as points.
		It is evident that the HMC algorithm was trapped in a region of high probability density in the $\alpha=0$ case by the wide separation with low probability density and in the $\alpha=1$ case by the lines of zero weight.
		\label{fig:1site contours}
	}
\end{figure}

As a visual aid, we consider the $\nt=2$ case.
Here there are only two Hubbard-Stratonovich degrees of freedom, $\phi_1$ and $\phi_2$.
We plot contours of the kernels for the two bases (\eqref{eqn:kernel1} and \eqref{eqn:kernel2}) in the case when $U\beta=18$, so that the modes in the $\alpha=0$ case are well-separated, in \autoref{fig:1site contours} and also show the codimension-1 manifolds that produce a formal ergodicity problem in the case of $\alpha=1$.
Note that such lines are absent in the $\alpha=0$ case.
Nevertheless, in both cases we will get a biased result---the $\alpha=0$ case has an in-practice problem caused by the isolation of modes by a region of very small probability density.
For both cases we generated 10,000 configurations using a precise HMC integrator (with 20 leapfrog steps for a unit-length molecular dynamics trajectory, yielding a near-100\% acceptance rate), shown as points.
The $\Phi$ axis in these plots runs along the diagonal from bottom-left to top-right, and the numerical distribution generated by HMC is clearly not symmetric around $\Phi=0$, a feature clear in the analytic expressions \eqref{eq:alpha0PPhi} and \eqref{eq:alpha1PPhi} and guaranteed by the fact that $f$ is even \eqref{eqn:exponential-factorization}.

\begin{figure}
	\centering
	\includegraphics[width=.8\textwidth]{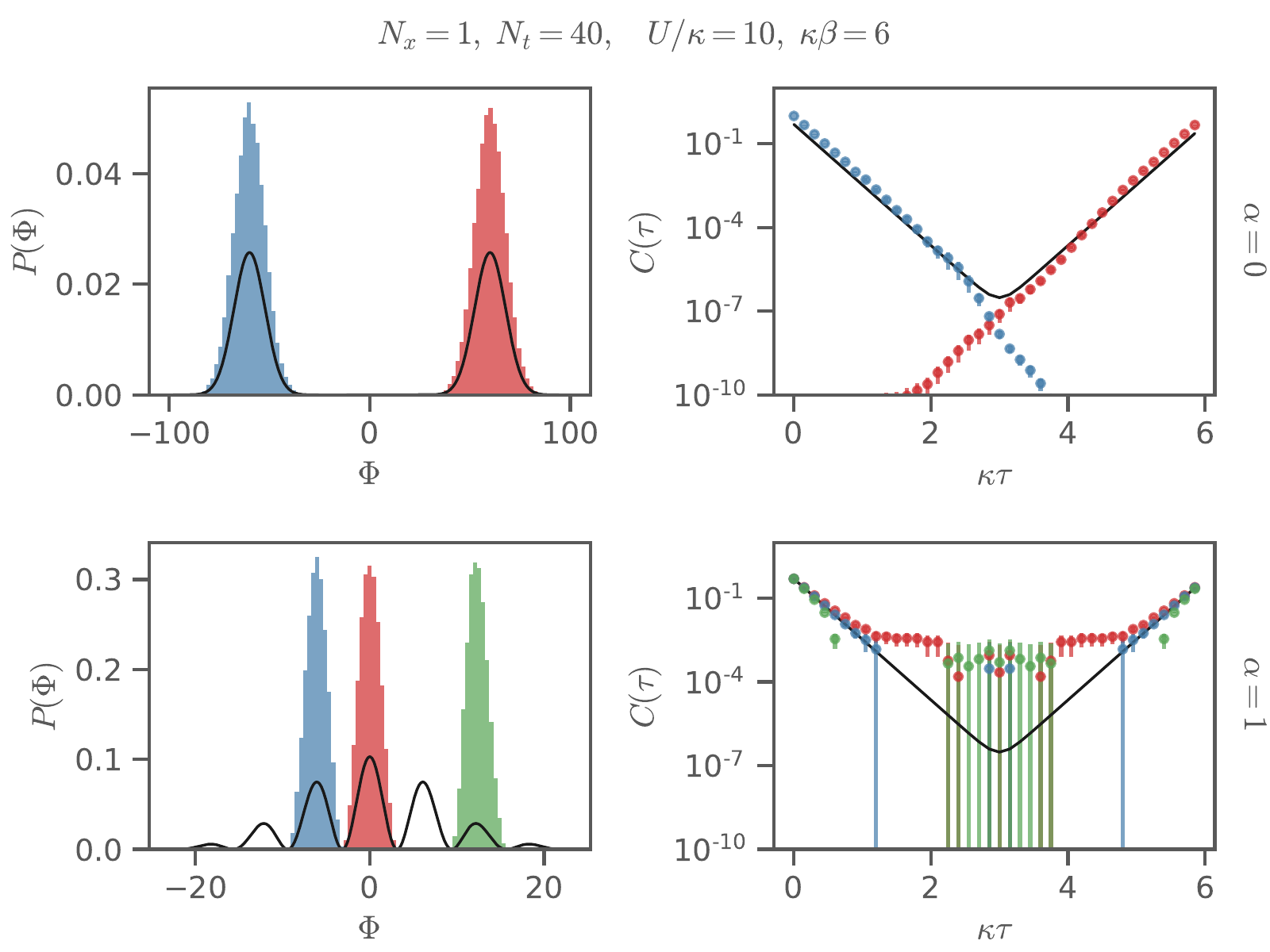}
	\caption{
		Histograms of $\Phi$ and correlators for the one-site problem with very fine integration.
		Monte Carlo evolution was started in different lobes and the trajectories are stuck in those lobes, shown by the different colors.
		All ensembles consist of 50k thermalized configurations.
		Single particle correlators to the right are color coded to match the histograms in the left column.
		Points where $C(\tau) \leq 0$ are not shown because of the log scale.
		The black lines show the exact results.
		The histograms are all normalized but, because HMC was trapped, the numerically generated histograms are much taller and thinner than the exact results.
		\label{fig:alpha01 ergo problem}
	}
\end{figure}

We can also study problems with substantially larger $\nt$, or equivalently, finer discretizations.
One observable that is of interest is the correlation function $C_{ij}(\tau)$, where $i,j$ refer to spatial locations.
For the 1-site problem we are restricted to $i=j=1$.
In the continuum limit we can compute the correlation function exactly
\begin{equation}\label{eqn:1 site correlator}
    C_{11}(\tau) \equiv \left\langle a(\tau)\ a^\dagger(0)\right\rangle
            = \left\langle \sum_t {\left(M_{11}^{-1}\right)}_{t+\tau,t}\right\rangle
            =      \frac{\cosh(U(\beta-2\tau)/4)}{2\cosh(U\beta/4)}
\end{equation}
which, being a continuum-limit quantity, is independent of $\delta$.
We can estimate $C$ as an ensemble average over configurations generated by HMC by
\begin{equation}\label{eqn:correlator}
    C_{ij}(\tau) \approx \frac{1}{N_{c}}\sum_{\phi\in\{\phi\}}\sum_t M_{ij}^{-1}
{\left[\begin{cases}
    \phi    &   \text{if }\alpha=0 \\
    i\phi   &   \text{if }\alpha=1
\end{cases}\right]}_{t+\tau,t}
\end{equation}
where $N_{c}$ is the number of generated Hubbard-Stratonovich field configurations in the ensemble $\{\phi\}$.
Lack of ergodicity in our sampling of $\phi$ will give disagreement between simulated and exact correlators.
To demonstrate this, we consider in \autoref{fig:alpha01 ergo problem} as case using a precise MD integrator (acceptance rate > 99\%) with extreme $U/\kappa=10$ and $\beta\kappa=6$, so that when $\alpha=0$ the modes are widely separated and when $\alpha=1$ many modes contribute.
The top row corresponds to $\alpha=0$, whereas the bottom row is $\alpha=1$.
The left column shows the histogram of $\Phi=\sum_t\phi_t$ for different runs, while the right column shows the calculated correlators.
In all plots the black line is the exact result.
Different HMC runs are differentiated by color.
For the $\alpha=0$ case it is clear that the HMC trajectories are trapped in one of the two modes (top left panel), despite there being no regions with $\det M=0$.
The corresponding correlators calculated with these fields are color-matched and shown in the top right panel.
Clearly the sampling of fields is grossly biased, and this is reflected in the large disagreement between simulated and exact correlators.
For the $\alpha=1$ case three different runs (red, blue, green) were performed with different starting points for $\Phi$, each separated by a point where $\det M=0$.
From the histogram (lower left panel) it is clear that the runs are trapped within their respective sectors, and their corresponding correlators (color matched with the histograms) each differ from the exact result.
Presumably, the correct linear combination of these correlators (with relative weights given by~\eqref{eq:alpha1PPhi}) would give the exact correlator.
Of course, in a more complicated system we do not know such weights \emph{a priori}, and therefore would not know how to combine such correlators to produce the correct result.

These examples, though extreme, demonstrate how both \emph{formal} (for $\alpha=1$) and \emph{in practice} (for $\alpha=0$) ergodicity problems can arise in HMC simulations, even when simulations are of the same physical system.
The choice of basis greatly influences the behavior of the sampled field configurations, and in turn can drastically impact calculated observables such as the correlator.



\subsection{The Two-site Problem\label{sect:2sites}}


\begin{figure}[b]
	\includegraphics[width=.8\columnwidth]{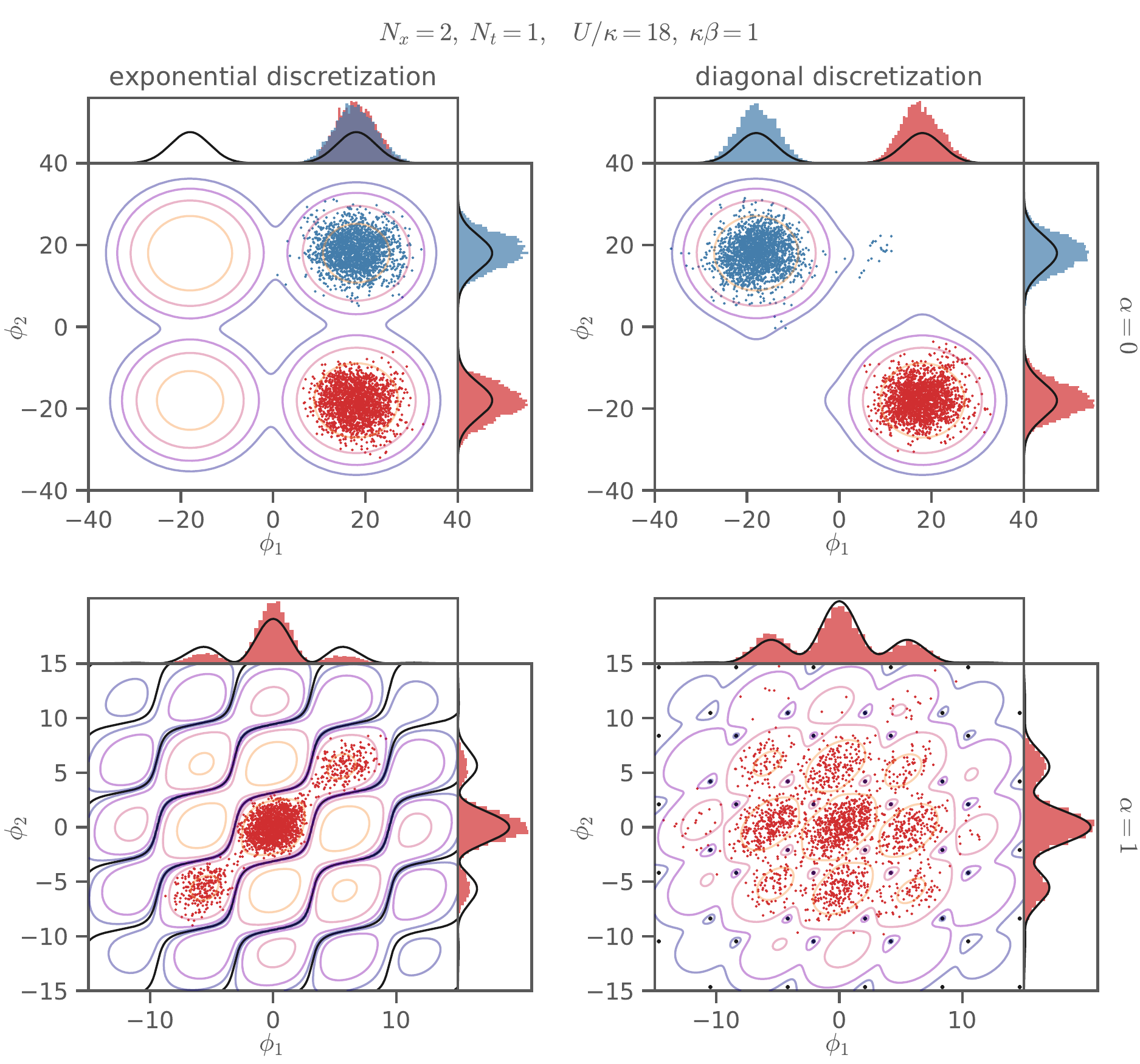}
	\caption{
		Contours of constant probability weight \eqref{eqn:probability} using the exponential discretization (left column) and the diagonal discretization (right column) with $\alpha=0$ (top row) and $\alpha=1$ (bottom row) for the 2-site problem with $U\beta=18$, $\kappa\beta=1$, and $\nt=1$.
		Contour colors range from orange (light) for large to blue (dark) for small values in arbitrary units.
		In the bottom panels, the black diagonal squiggly lines and regularly-spaced black dots correspond to locations where \eqref{eqn:probability} is exactly zero; their regular spacing is a consequence of the periodicity of the determinant, discussed in \Secref{shifts and periodicity}.
		Crossing a black line changes the sign of the fermion determinant.
		The colored points show 10k HMC trajectories.
		For $\alpha=0$, evolution was started in different modes of the field space, shown in different colors.
		For $\alpha=1$, evolution was started at the origin with a small random deviation.
		The outer histograms show the marginal distributions in $\phi_1$ and $\phi_2$ where the black lines are the exact results obtained from \eqref{eqn:probability}.
		\label{fig:2sites contours}
	}
\end{figure}

We now consider the two-site problem to demonstrate how different discretizations can lead to ergodicity issues.
To keep the presentation reasonable, we restrict our analysis to the exponential and diagonal discretizations given by~\eqref{eqn:M1}~and~\eqref{eqn:M3}, respectively, using the $\alpha=0$ and $1$ bases\footnote{In simple cases we have found that the linear discretization~\eqref{eqn:M2} exhibits the same behavior as that of the diagonal discretization~\eqref{eqn:M3} but have not explored the linear case as extensively.
The lack of conjugate reciprocity suggests there ought to be no formal ergodicity problem when $\alpha=1$.}.
In~\eqref{eqn:M3} the term $h$ is given by
\begin{equation}
h_{x',x} =
\kappatilde\delta_{\langle x',x\rangle}=
\begin{pmatrix}
    0               & \kappatilde  \\
    \kappatilde    & 0
\end{pmatrix}\ ,
\end{equation}
while the matrix in~\eqref{eqn:M1} is its exponential, given by
\begin{equation}
{\left[e^h\right]}_{x',x}=
\begin{pmatrix}
    \cosh\kappatilde & \sinh\kappatilde   \\
    \sinh\kappatilde & \cosh\kappatilde
\end{pmatrix}\ .
\end{equation}

For extreme simplicity, we turn to the problem of two sites on a single timeslice where, as in the previous section, there are two degrees of freedom, $\phi_1$ and $\phi_2$, the label now indicates the spatial site.
Adopting the factorization shown in~\eqref{eqn:exponential-factorization} of the determinant of the fermion matrix, we have that
\begin{equation}\label{eqn:f two site}
f[\phi]=
\begin{cases}
2\left[\cosh (\Phi/2)+\cosh\left(\frac{\phi_1-\phi_2}{2}\right)\cosh(\kappatilde)\right] & \text{exponential discretization}\\[0.25cm]
2\left[\cosh (\Phi/2)+\cosh\left(\frac{\phi_1-\phi_2}{2}\right)-\frac{\kappatilde^2}{2}e^{-\Phi/2}\right] & \text{diagonal discretization}
\end{cases}
\end{equation}
where again $\Phi=\phi_1+\phi_2$.
Close inspection of the equations above shows that both $f[\phi]$ and $f[i\phi]$ are always real in the exponential discretization.
In the diagonal discretization, on the other hand, only $f[\phi]$ is real and $f[i\phi]$ is in general complex.

\begin{figure}[b]
	\includegraphics[width=.8\textwidth]{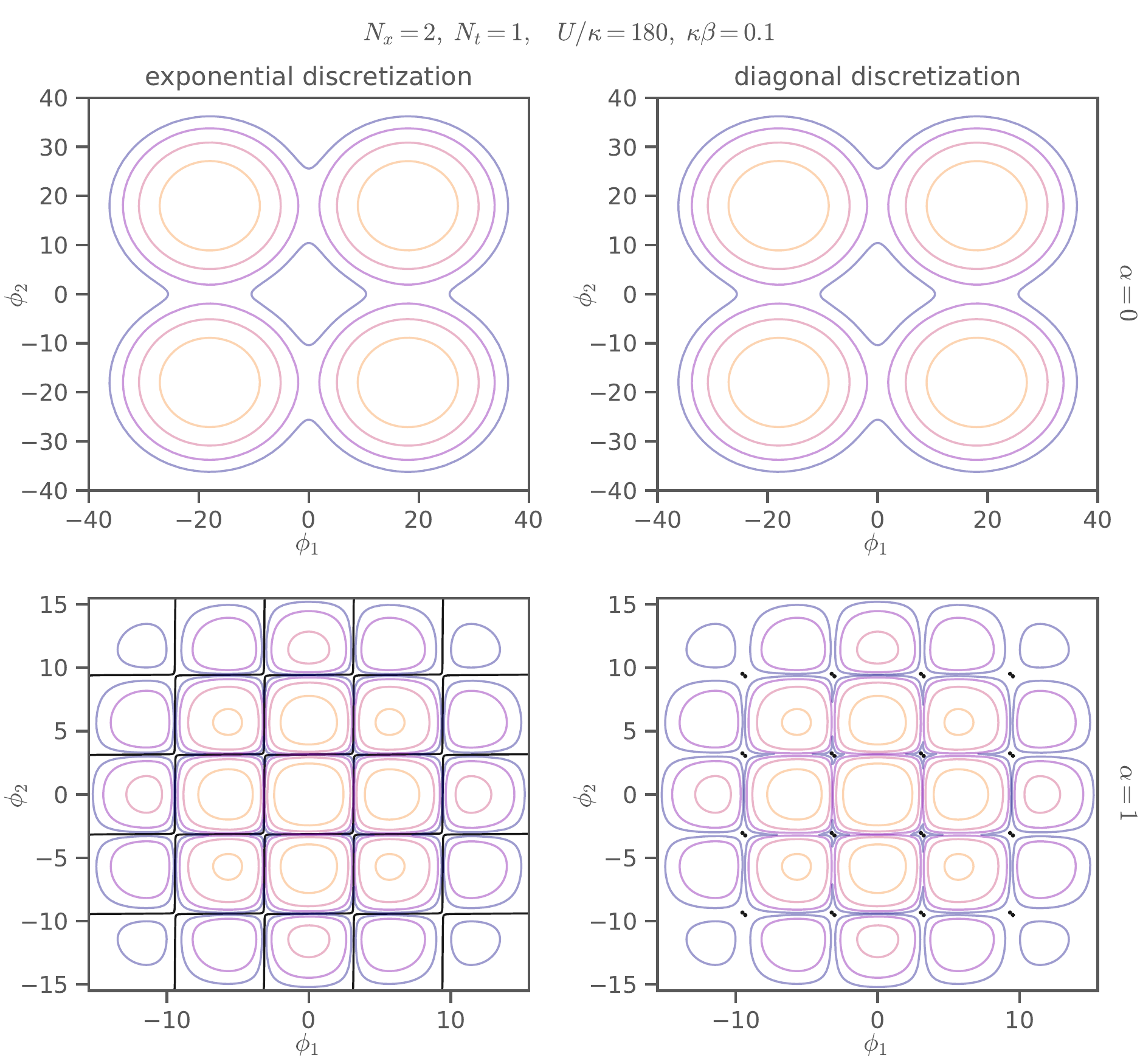}
	\caption{
		Analogous contours to \autoref{fig:2sites contours} but using $\kappa=0.1$ to mimic approaching the continuum limit.
		The exponential and diagonal discretizations now have very similar probability contours.
		For the $\alpha=1$ case, the black lines of zero weight of the exponential discretization  pinch and nearly close, while the zero-weight configurations remain isolated in the diagonal discretization.
		\label{fig:2sites contours finer}
	}
\end{figure}

The probability density
\begin{equation}\label{eqn:probability}
W[\phi] \propto f[\phi]f[-\phi] e^{-\frac{\phi_1^2+\phi_2^2}{2U\beta}}
\end{equation}
is positive semi-definite for both discretizations with $\alpha=1$ because $f[i\phi]f[-i\phi]=|f[i\phi]|^2$.
In the exponential $\alpha=0$ case it is positive definite as well because $f[\phi]$ is.
Positive (semi)-definiteness is however not guaranteed in the diagonal discretization with $\alpha=0$.
For large enough values of $\kappatilde$ the product $f[\phi]f[-\phi]$ can become negative, which means that~\eqref{eqn:probability} cannot be interpreted as a probability distribution.
Normally this is not an issue; we can increase $\nt$ and positivity is eventually assured.
However, for this particular example we must enforce an additional constraint to maintain positivity, namely $\kappatilde^2\le 1$.

\autoref{fig:2sites contours} shows probability contours of this single-timeslice problem in the case when $\tilde{U}=U\beta=18$ and $\kappatilde=\kappa\beta=1$ for these different discretizations and bases as well as field configurations generated by HMC with a fine integrator (acceptance rate $>99\%$).

Regardless of discretization, $\alpha=0$ exhibits no field configurations of zero weight when $\kappatilde\leq1$, so there is no formal ergodicity problem.
HMC nevertheless gets trapped in single lobes of the probability distribution, and there is a problem in practice.
In the case of the diagonal discretization, the two modes are symmetric about the $\phi_1=\phi_2$ line, and the histogram of $\Phi$ will \emph{not} show evidence of this bimodal structure as $\Phi$ is peaked about zero in both modes in the same way.

\begin{figure}[b]
	\includegraphics[width=.8\columnwidth]{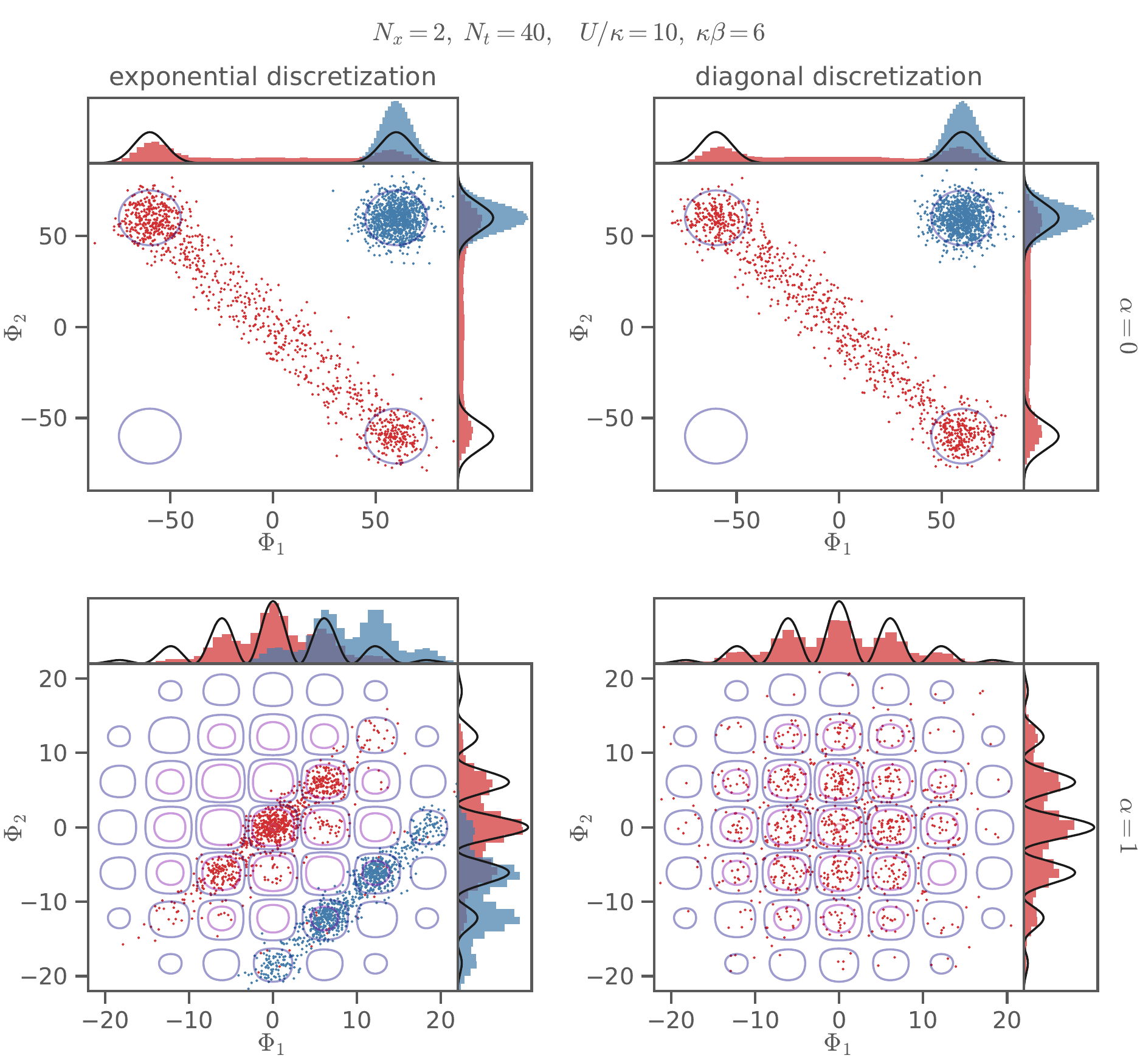}
	\caption{
		HMC history of $\Phi_x = \sum_t \phi_{xt}$ for the 2-site problem with 100k trajectories with $U\beta=60$, $\kappa\beta=6$, and $\nt=40$.
	    Only every 100th trajectory of each ensemble is shown.
	    Several HMC streams for each case were produced by starting in different modes in configuration space as indicated by color.
        Only for $\alpha=1$, diagonal there is no difference, where the run is started, so only one stream is shown.
	    The black lines in the marginal distributions for $\Phi_1$ and $\Phi_2$ are the exact \emph{1-site} distributions which are recovered in the strong coupling limit.
	    The contours show the product of the 1-site distributions and are \emph{not} exact results for this case.
		\label{fig:2site histograms}
	}
\end{figure}

With $\alpha=1$ there are field configurations with zero weight.
With the exponential discretization given by~\eqref{eqn:M1} there are entire lines of zero weight, separating the field space into different sectors, and giving rise to a formal ergodicity problem.
With a fine integrator, HMC gets trapped between the infinite barriers.

The $\alpha=1$ case with the diagonal discretization given by~\eqref{eqn:M3} is particularly interesting since the probability density vanishes only at isolated points, not lines.
Regions of relevant weights are no longer separated by infinite barriers, and therefore ergodicity is formally preserved.
Evidence of this is seen in the distribution of field configurations generated by HMC, shown as red points.
The HMC algorithm successfully reached into nearby basins that would have been unreachable in the case of the other discretization.
The diagonal discretization is clearly less restrictive in the $\alpha=1$ case, compared to the exponential discretization.

The preceding example has only little bearing on a realistic calculation due to the fact that it represents a single timeslice calculation.
Before considering a larger $\nt$ case, we point out that we can mimic a finer time discretization by reducing $\kappa$.
This can also be viewed as taking the strong coupling $U/\kappa\gg1$ limit.
\autoref{fig:2sites contours finer} shows the contours in the case where $\kappa$ has been reduced by an order of magnitude compared to \autoref{fig:2sites contours}.
Note that all contours within their respective $\alpha$ bases are nearly the same for both discretizations, giving credence to our claim that the discretizations become equivalent in the continuum limit.
This is despite the fact that for $\alpha=1$ the allowed values where $\det M[i\phi]=0$ (black lines and dots of bottom row) are topologically different.

\begin{figure}[b]
	\centering
	\includegraphics[width=.8\columnwidth]{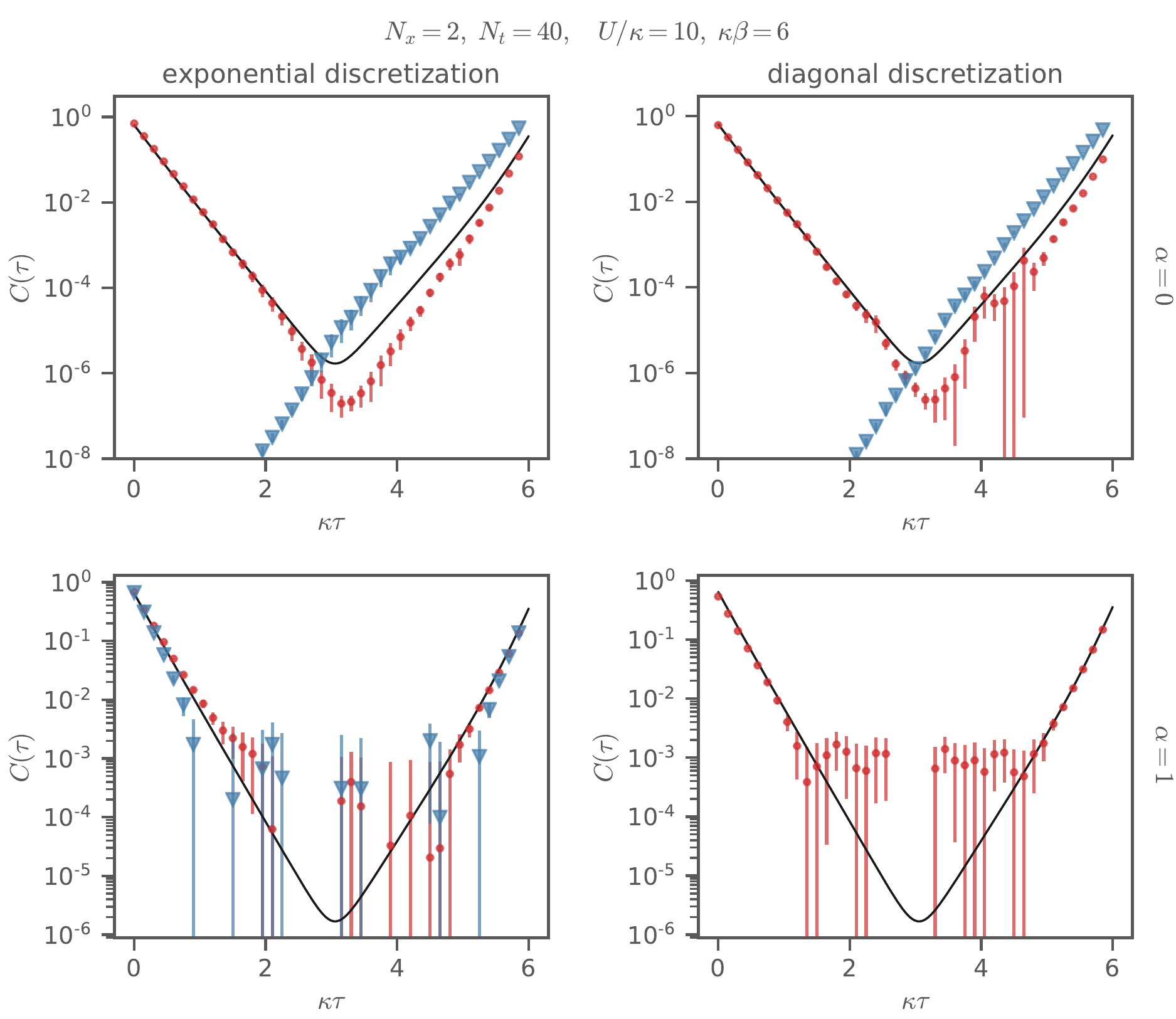}
	\caption{
		Correlators from Eq.~\eqref{eqn:Cpm} color coded to match the ensembles shown in Figure~\ref{fig:2site histograms}.
		Points where $C(\tau) \leq 0$ are not shown because of the log scale.
		The black lines show exact results from direct diagonalization of the 2-site Hubbard model.
		For clarity, the figure shows only $C_+$, $C_-$ behaves in the same way.
		\label{fig:2site corr ergo}
	}
\end{figure}

We now consider $\nt=40$ again with extreme $U/\kappa=10$ and $\beta\kappa=6$.
In \autoref{fig:2site histograms} we show the distribution of $\Phi_x\equiv \sum_t\phi_{xt}$ as in \eqref{eqn:polyakov loop} from different HMC simulations using the different bases and discretizations.
We again use a high precision integrator with acceptance rate $>99\%$.
The left column represents the exponential discretization and the right the diagonal one.
The top row has $\alpha=0$, the bottom $\alpha=1$.
Histograms of $\Phi_1$ and $\Phi_2$ are also shown in the figures.
Recall that in the large $U/\kappa$ limit, the hopping term becomes negligible compared to the on-site interaction and the 2-site problem factors into the product of two 1-site problems, as discussed in Sec.~\ref{sect:limits}.
One expects then that the distributions of $\Phi_x$ are given by~\eqref{eq:alpha0PPhi} and~\eqref{eq:alpha1PPhi}, which are shown as black lines in the marginal histograms.
The fact that our simulated distributions qualitatively agree with these distributions is due to the fact that $U\beta=60$ is in the strong coupling regime.
For the $\alpha=0$ case, it is clear that there is a multi-modal distribution and HMC trajectories are separated into these modes, even though there are no configurations with $\det M=0$, which is similar to the 1-site case.
HMC samplings are again grossly biased.
For the exponential $\alpha=1$ case (bottom left panel), trajectories are also biased because of the separation of regions by $\det M=0$.
The sampling of fields for the diagonal $\alpha=1$ case (bottom right) on the other hand is relatively symmetric and seemingly unbiased.

For the two-site system there are two linearly independent correlators.
If we label one site $A$, and the other $B$, then the two correlators are
\begin{equation}\label{eqn:Cpm}
C_{\pm}(\tau)=\frac{1}{2}\left(C_{AA}(\tau)+C_{BB}(\tau)\pm\left[C_{AB}(\tau)+C_{BA}(\tau)\right]\right)\ ,
\end{equation}
where $C_{ij}(\tau)$ represents the correlator of a quasi-particle starting at site $i$ and propagating to site $j$ and is estimated by \eqref{eqn:correlator}.
In the strong coupling limit these two correlators approach the 1-site correlator solution of \eqref{eqn:1 site correlator}.
In \autoref{fig:2site corr ergo} we show the corresponding calculated correlators from the field distributions in \autoref{fig:2site histograms}, arranged and colored in the same way.
As expected, the correlators for the $\alpha=0$ case agree very poorly with the exact result given by the black lines due to the biased sampling of fields.
In the exponential $\alpha=1$ case, the red points sampled from the diagonal band in \autoref{fig:2site histograms} show only a small deviation from the exact result in regions with small noise.
If a different band is sampled however, deviations can be larger as demonstrated by the blue points which show a clear deviation from the black line.
The impact of an ergodicity problem on correlators depends on the contributing states as demonstrated in the next section.
On the other hand, correlators calculated in the diagonal $\alpha=1$ case have better agreement with the exact results, particularly for early and late times, due to a less biased sampling of fields.

These examples give further evidence of how the choice of bases can impact the sampled fields, as discussed in the previous section.
In addition to this, however, is the fact that different discretizations can also lead to disparate sampling of fields, and ultimately impact the fidelity of observable calculations.
In Appendix~\ref{sect:other-observables} we use these same ensembles to calculate other observables and similarly find that the exponential discretization can suffer from ergodicity problems that are absent in the diagonal case.


\subsection{The Four Site Problem\label{sect:4 and more}}

Up to this point only the exponential $\alpha=1$ discretizaton has exhibited cases where  $\det M$ is always real and can be negative.
With four and more sites, the exponential $\alpha=0$ discretization also exhibits $\det M < 0$ cases, as was originally pointed out in Ref.~\cite{Beyl:2017kwp}.
It would be interesting to understand why the two site problem is protected from negative determinants in the other cases.

Reference~\cite{Beyl:2017kwp} argued that the initial starting point of the HMC evolution can lead to drastically different results due to the separation of $\det M<0$ and $\det M>0$ in the exponential $\alpha=0$ case.
The authors provide an explicit example of the equal-site $\langle C_{ii}(\tau)\rangle$ correlator calculated on a $4\times4$ square lattice, showing a clear dependence of the correlator determined from HMC runs that originated from either a $\det M>0$ or a $\det M<0$ configuration (Figure~1 of Ref.~\cite{Beyl:2017kwp}).
Ergodicity is clearly violated in these extreme cases.
We reproduce these results in Appendix~\ref{sect:sixteen sites}.
However, we find that the diagonal $\alpha=1$ seems not to suffer from this problem (see \autoref{fig:16sites corr} of Appendix~\ref{sect:sixteen sites}).

To substantiate our claim, we consider instead the $2\times2$ square lattice Hubbard model (4 sites) and repeat the exercise that was done for the $4\times4$ case of Ref.~\cite{Beyl:2017kwp}.
The added benefit here is that we can compare directly to exact solutions obtained via direct diagonalization.
We do exactly this by considering correlators in momentum space,
\begin{equation}
C_{\pm}(\vec{k},\tau)=\frac{1}{2}\sum_{\vec{x}}e^{i\vec{k}\cdot \vec{x}}C_{\pm}(x,\tau)\ ,
\end{equation}
 where the sum is over unit cell locations (each unit cell containing one $A$ site and one $B$-site) and $C_{\pm}(\vec{x},\tau)$ is given by Eq.~\eqref{eqn:Cpm} but now with explicit unit cell location $x$ in its argument.
As there are two allowed momenta for this system, $a\vec{k}=(0,0)$ and $(\pi/2,\pi/2)$, there are in principle a total of four possible correlators.
However, at $a\vec{k}=(\pi/2,\pi/2)$ one has $C_+=C_-$, and thus we have only three distinct correlators.
\begin{figure}
	\centering
	\includegraphics[width=.8\columnwidth]{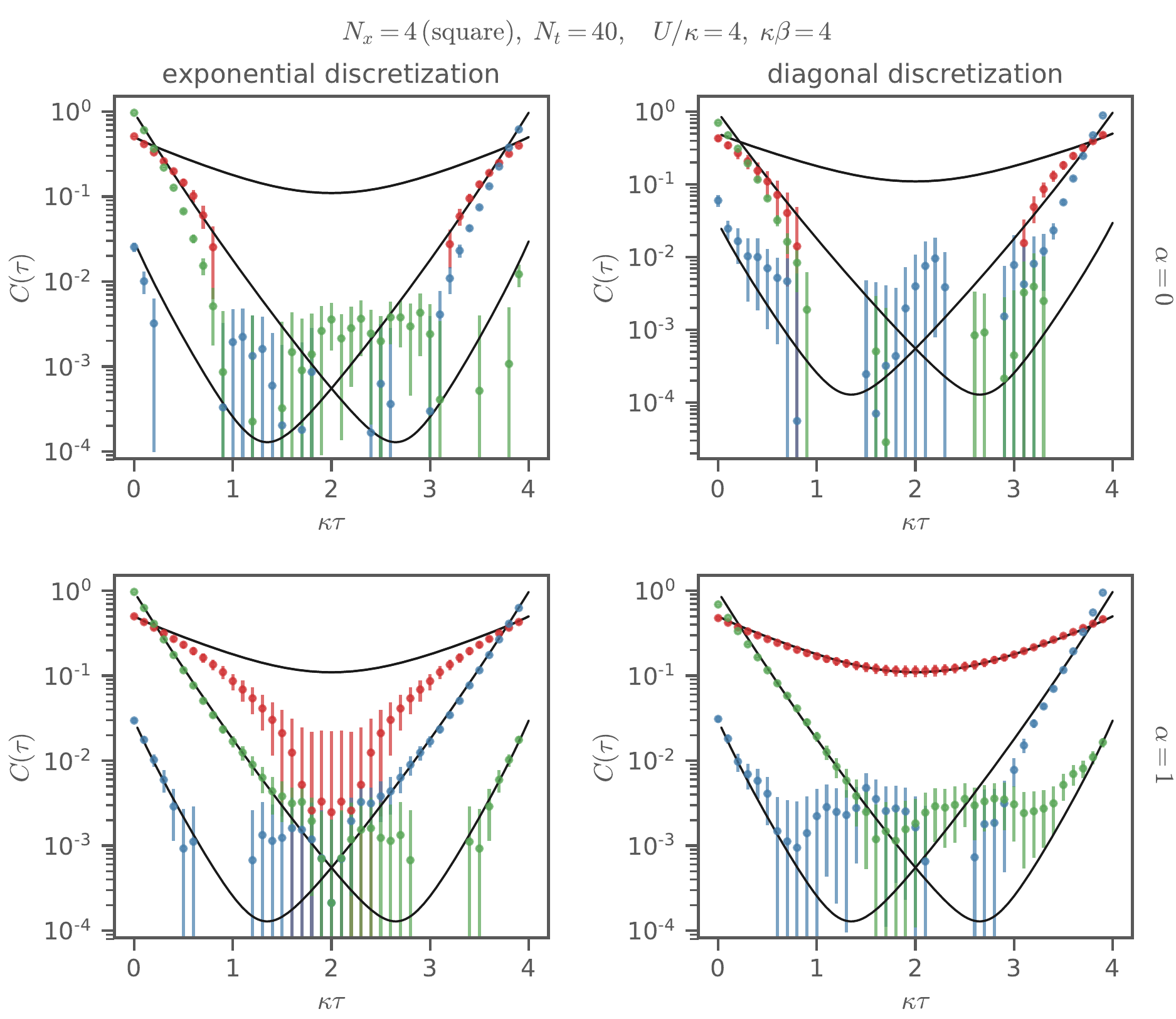}
	\caption{
		Dependence of correlators on initial HMC configuration.
		The red correlators are calculated at $a\vec{k}=(\pi/2,\pi/2)$ which for this case means that $C_+=C_-$.
		The green and blue correlators (lower, off center minima) are calculated at $a\vec{k} = (0,0)$.
		For $\alpha=0$ the initial configuration was chosen such that $\det M[\phi] < 0$.
		For $\alpha=1$, exponential discretization the starting configuration satisfies $0 > e^{-i \Phi / 2} \det M[\phi] \in \mathbb{R}$.
		For $\alpha=1$, diagonal no particular starting criterion was set.
		All plots show data from 2400 trajectories.
		The starting criteria stayed fulfilled for all of those configurations.
		\label{fig:4sites}
	}
\end{figure}

In \autoref{fig:4sites} we show these three correlators for the four different discretization schemes and bases.
For both $\alpha=0$ cases we start the HMC evolution from a $\det M[\phi] <0$ configuration while for the $\alpha=1$, exponential discretization the evolution starts at a configuration such that $0 > e^{-i \Phi / 2} \det M[\phi] \in \Reals$ in accordance with equation~\eqref{eqn:alpha1 exp reality}.
For the remaining diagonal $\alpha=1$ case, no such criterion can be formulated since the determinant is complex and does not factorize as in the exponential case.
In all cases we use a very precise MD integrator.
It is clear from the disagreement with the exact result that the HMC trajectories are \emph{not} properly sampling all the important regions of configuration space.
It is worth noting that not all correlators are affected by the lack of ergodicity in the same way.
For the exponential $\alpha=1$ case, the higher energy correlators (green and blue) are very precise, only the ground state correlator shows a strong deviation.

\begin{figure}
	\centering
	\includegraphics[width=.8\columnwidth]{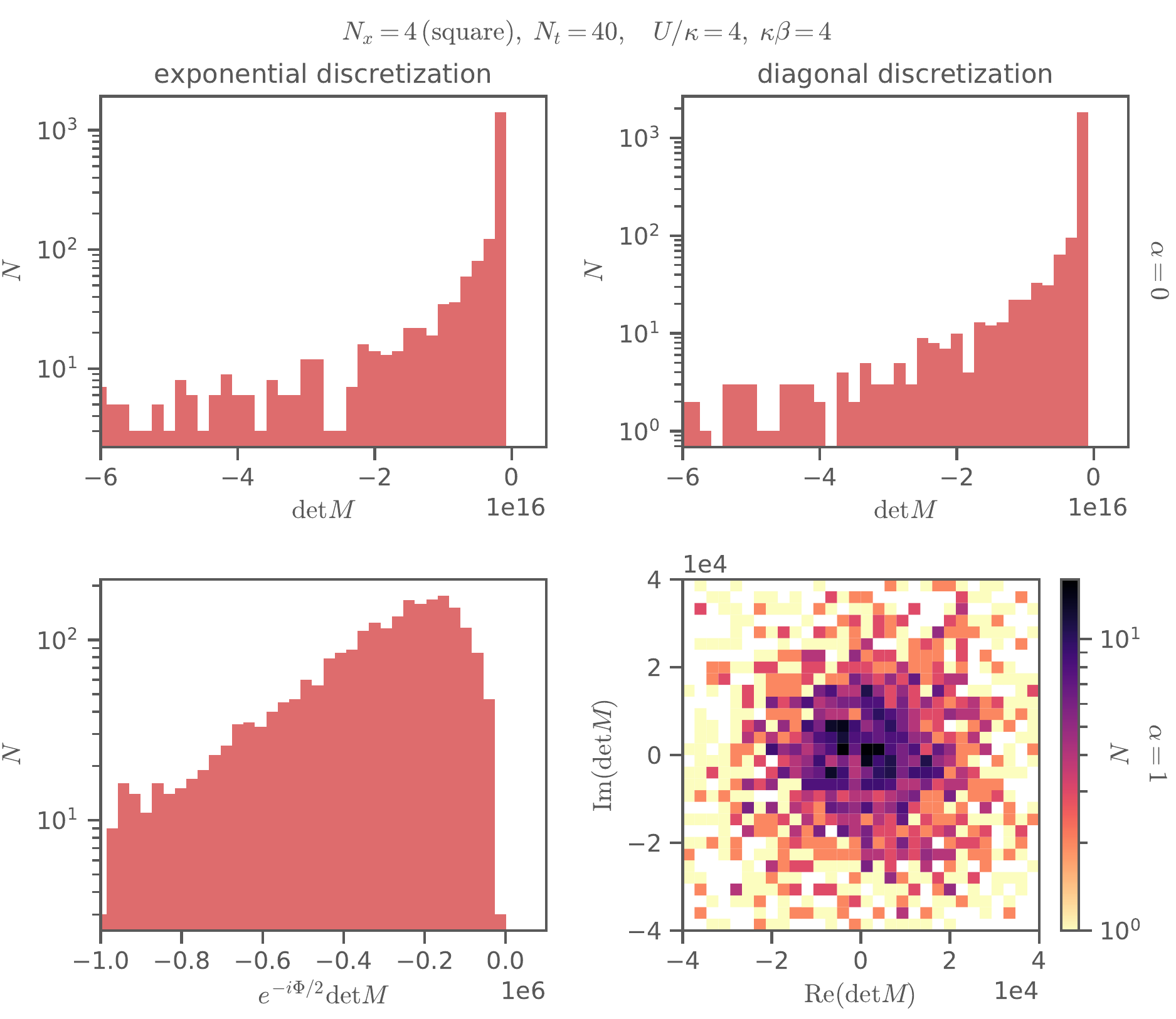}
	\caption{
		Histograms of $\det M$ for the $\alpha=0$ basis (top row) and diagonal $\alpha=1$ case (bottom right), and  $e^{-i \Phi / 2} \det M[\phi] $ for the exponential $\alpha=1$ case (bottom left) for the ensembles used for \autoref{fig:4sites}.
		Except for the diagonal $\alpha=1$ case, all quantities are real.
		In the diagonal $\alpha=1$ case $\det M$ is complex and the domain of the histogram is the complex plane.
		\label{fig:4sites_hist}
	}
\end{figure}

When we consider the histograms of the MC histories of $\det M$ for the $\alpha=0$ and exponential $\alpha=1$ cases, as shown in \autoref{fig:4sites_hist}, we find that $\det M$ is confined to the negative region only.
Ergodicity is indeed violated.
The lone exception is the diagonal $\alpha=1$ case whose correlators agree very well with the exact result (bottom right panel of \autoref{fig:4sites}).
The corresponding histogram of the MC history is shown in the bottom right panel of \autoref{fig:4sites_hist}.
In this case, since $\det M$ is complex, the histogram is shown as a density on the complex plane.
In this case the zero of $\det M$ can be easily circumnavigated and there does not exist an ergodicity problem.




\section{Overcoming Ergodicity Issues}
\label{sect:solution}

Reference \cite{Beyl:2017kwp} already showed that one may avoid ergodicity issues by complexifying the auxilliary field (taking an intermediate value of $\alpha$).
In this section we will examine a variety of other solutions.

\subsection{Coarse Molecular Dynamics Integration}

When the molecular dynamics integrator is not very precise, the Markov Chain can hop over the barriers that separate neighboring basins in configuration space in the particle/hole basis.
This, for all practical purposes, avoids the ergodicity issues as long as the integrator takes sufficiently coarse steps.
We have found that targeting an acceptance rate of around 70\% allows the integrator to readily explore the areas in field space that would be separated by an impenetrable barrier in the case of a very precise integrator.

We emphasize that the numerical ``errors'' introduced by using an imprecise molecular dynamics integrator do not invalidate the stochastic algorithm.
In other words, HMC with a coarse integrator is still a valid proposal generator from a Metropolis-Hastings perspective.
As long as the accept/reject step is maintained and the integration is reversible, the algorithm still faithfully samples the distribution dictated by the action in the path integral.

In the spin basis, the basins of $\Phi$ for the one-site problem grow farther apart when increasing $U\beta$.
Between the two modes the probability is never zero, but is nevertheless exponentially small.
Unless an \emph{extremely} coarse integrator with a long trajectory length is used, it seems unlikely that molecular dynamics can cross such a wide exponentially small region.

With a fixed molecular dynamics time between trajectories, there is a limit to how coarse the integration may be made.
With either a very or moderately coarse integrator combined with a long molecular dynamics time between configurations, we expect the acceptance rate to plummet.
There thus may be an in-practice problem of rejecting too frequently, not evolving the configuration enough, or an explosion of computational cost.

\begin{figure}[h]
	\centering
	\includegraphics[width=.8\columnwidth]{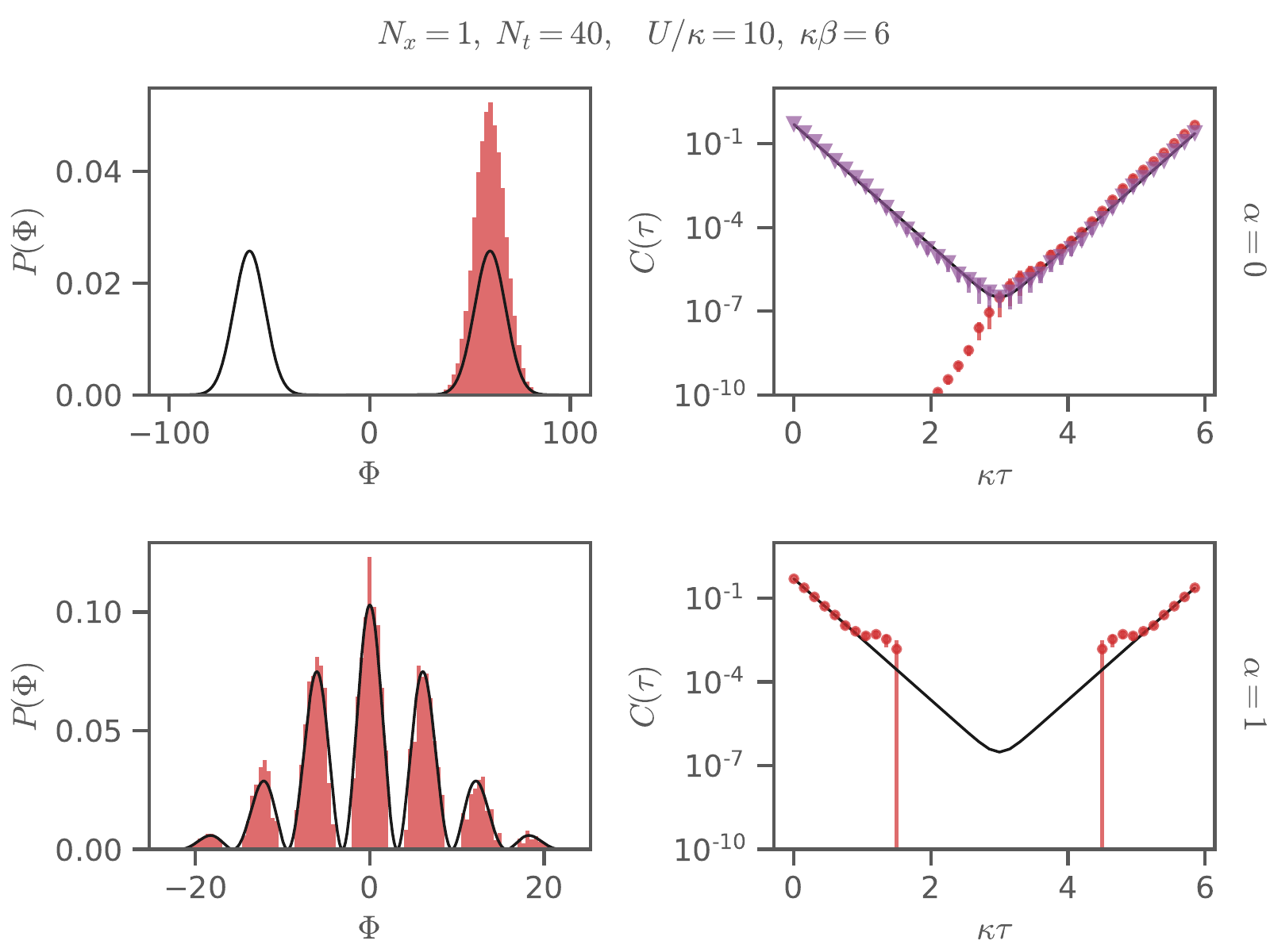}
	\caption{
		Same as \autoref{fig:alpha01 ergo problem} but with coarser integration with an acceptance rate of about $60\%$.
		For $\alpha=0$ Monte Carlo evolution was started in the right hand lobe and got stuck there.
		Both ensembles consist of 50k thermalized configurations.
		The purple correlator is $C_{\purple\blacktriangledown}(\phi) = (C_{\red\bullet}(\phi) + C_{\red\bullet}(-\phi)) / 2$, the result of the symmetrization discussed in Section~\ref{sect:symmetrization}.
		\label{fig:alpha01 coarse}
	}
\end{figure}

In Figure~\ref{fig:alpha01 coarse} we see that coarse integration allows the $\alpha=1$ case to jump over the exact zeros, whereas the fine integration was trapped in the initial mode, as seen in Figure~\ref{fig:alpha01 ergo problem}.
In contrast, we can see that coarse integration does not alleviate the in-practice problem seen in the $\alpha=0$ case because the modes are so widely separated.

\begin{figure}
	\centering
	\includegraphics[width=.8\columnwidth]{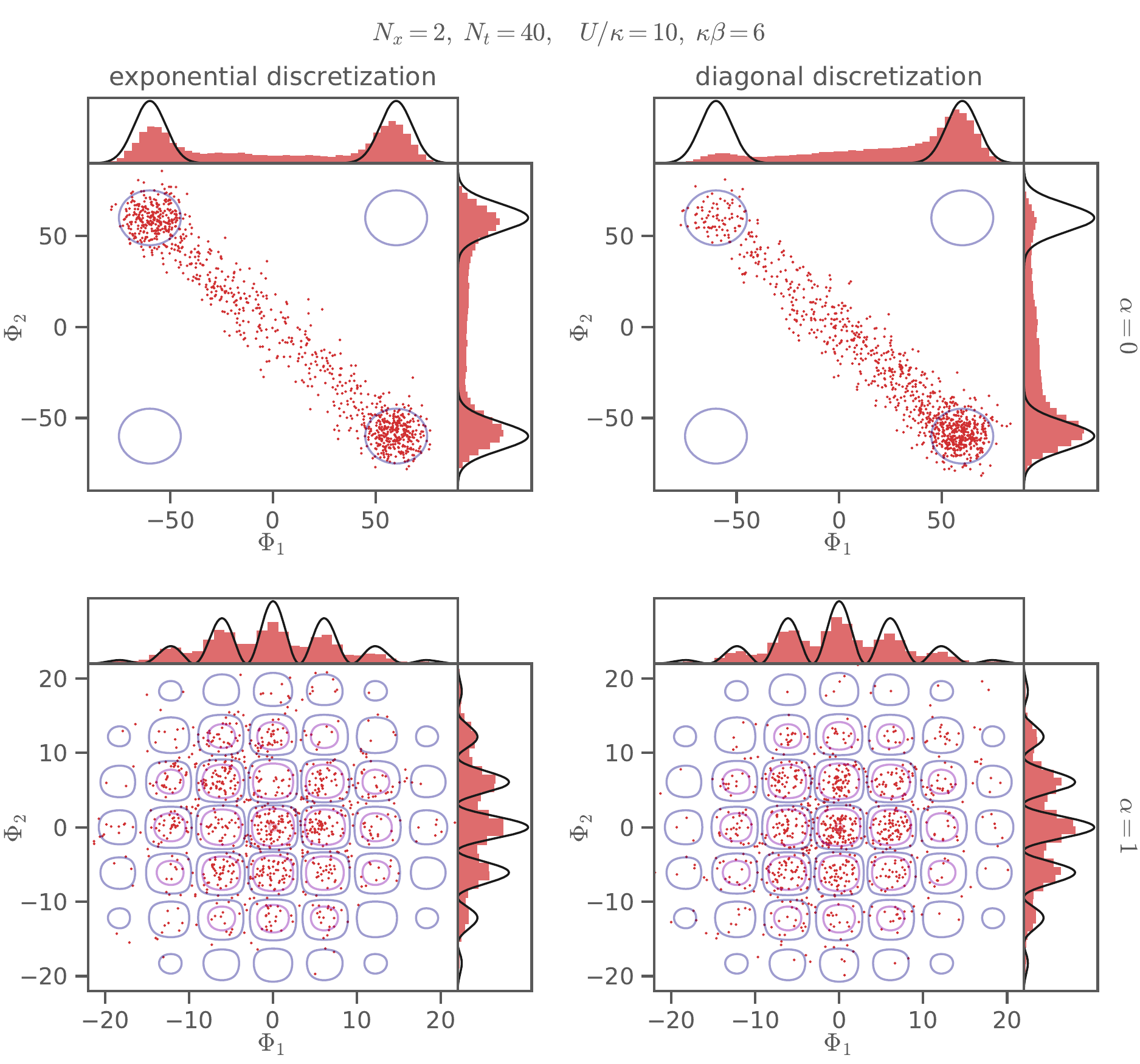}
	\caption{
		HMC history of $\Phi_x$ for the 2-site problem with 100k trajectories and a coarse integrator.
    	Only every 100th trajectory of each ensemble is shown.
    	The black lines in the marginal distributions for $\Phi_1$ and $\Phi_2$ are the exact \emph{1-site} distributions which are recovered in the strong coupling limit.
    	The contours show the product of the 1-site distributions and are \emph{not} exact results for this case.
		\label{fig:2site histograms coarse}
	}
\end{figure}

A similar improvement can be observed in two-site problem when using a coarse integrator with $\approx 60\%$ acceptance rate.
Figure~\ref{fig:2site histograms coarse} shows that configurations in this case are more evenly distributed for $\alpha=1$, especially in the exponential discretization; compare with Figure~\ref{fig:2site histograms}.
For $\alpha=0$, however, there is no change compared to the fine integrator as expected.

\begin{figure}
	\centering
	\includegraphics[width=0.8\columnwidth]{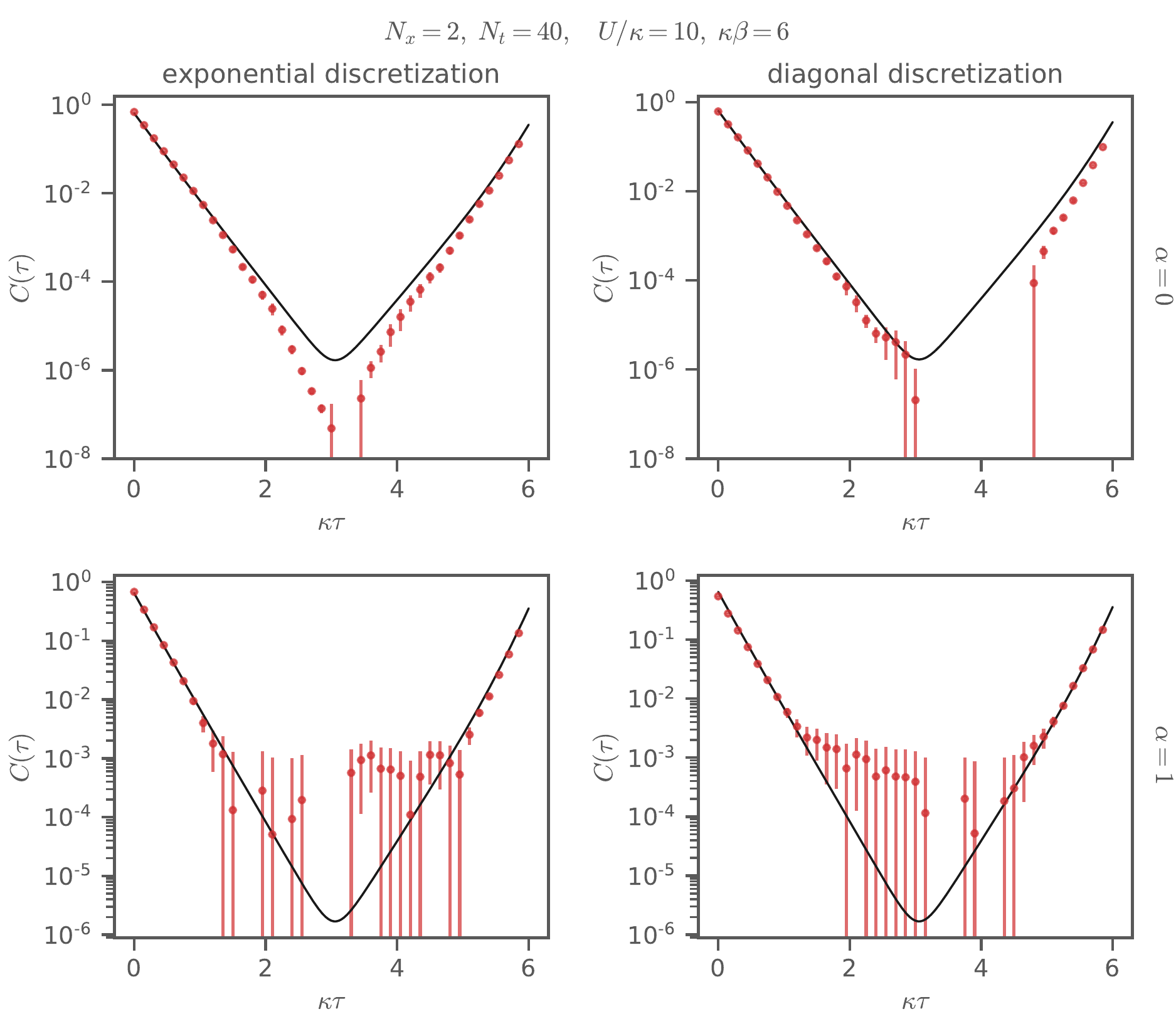}
	\caption{
		Correlators from Equation~\eqref{eqn:Cpm} for the ensembles shown in \autoref{fig:2site histograms coarse}.
		Points where $C(\tau) \leq 0$ are not shown because of the log scale.
		The black lines show exact results from direct diagonalization of the 2-site Hubbard model.
		For clarity, the figure shows only $C_+$, $C_-$ behaves in the same way.
		\label{fig:2site corr ergo coarse}
	}
\end{figure}

The correlators in Figure~\ref{fig:2site corr ergo coarse} show no significant improvement over those in Figure~\ref{fig:2site corr ergo} for a fine integrator.
It is however no longer possible to construct an ensemble that is stuck in a specific region of configuration space and leads to a systematically biased correlator in the exponential $\alpha=1$ case, c.f.~\autoref{fig:2site histograms} and~\ref{fig:2site corr ergo}.
Noise in the systematically wrong medium $\tau$ range has increased for $\alpha=0$, particularly strongly in the diagonal discretization.

Once many sites are included, it may be infeasible to rely on a coarse integrator for crossing zeroes, so it is worthwhile to consider additional techniques to aid ergodic exploration of the configuration space\cite{Beyl:2017kwp,Ulybyshev:2017hbs}.

\subsection{Symmetrization}\label{sect:symmetrization}

For the one-site problem in the spin basis, the in-practice ergodicity problem arises entirely because of the large separation of the modes in $\Phi$ in \eqref{eq:alpha0PPhi} and a possible solution immediately presents itself.
If the configuration $\phi$ corresponds to $\Phi$, the field configuration $-\phi$ corresponds to $-\Phi$.

Then, we can perform Monte Carlo on equivalence classes of $\phi$---for each step in the Markov Chain, we receive two (entirely correlated) field configurations, $\phi$ and $-\phi$, perform measurements on both configurations and average the result.
Since their weights are exactly equal, we know they appear equally often in the original method of simply sampling individual field configurations---we simply ensure the different $\Phi$ lobes are sampled equally without the need to cross over the wide, exponentially small barrier.

This approach is not restricted to the one-site problem.
Recall that, as discussed in Section~\ref{sect:charge conjugation}, $W[\phi] = W[-\phi]$ independent of basis and $\alpha$ at half-filling on a bipartite lattice.

In Figure~\ref{fig:alpha01 coarse} the $\alpha=0$ case has coarse HMC integration that is trapped in one of the two modes.
The purple correlator in the top-right panel is the average of the correlation function measured on $\phi$ and $-\phi$ for each configuration.
The improvement is obvious.

Looking at the basins and boundaries in the lower-left panels of \autoref{fig:2sites contours} and \autoref{fig:2site histograms} makes it clear that this symmetrization is not a complete solution for the exponential $\alpha=1$ case.
For example, if HMC winds up in a basin that does not include $\phi=0$, no symmetrization will ever get it there.
For the $\alpha=0$ cases in those figures, it is clear that this symmetrization still fails to join different regions of nonzero probability.

Generalizing from just the charge conjugation operation $\phi\goesto-\phi$, we can also immediately symmetrize in terms of the other operational symmetries discussed in Section~\ref{sect:eigenvalues}: temporal shifts, time reversal, and the spatial symmetries of the lattice.
Put another way, if observables are formulated that are invariant under these operations, the ergodicity problem is reduced from the full configuration space to only the configuration space with these symmetries modded out.

This symmetrization may be performed stochastically, as well.
For example, one often finds that in lattice QCD literature the chiral condensate is evaluated on stochastic noise sources or that correlation functions are measured from sources on different randomly chosen sites in the lattice for each field configuration.

Unfortunately, this is not a cure-all.
Since the determinant is invariant under many of these symmetries, it cannot repair the in-principle problem we might encounter in, for example, the exponential $\alpha=0$ many-site problem where we know, from \eqref{eqn:exponential-factorization}, a formal problem that requires flipping the determinant's sign can arise.


\subsection{Large Jumps}\label{sect:jumps}

Rather than restricting Markov chain updates to HMC only, one can use a mix of proposal generators followed by the Metropolis accept/reject step.
Those proposal machines must be statistically balanced so that the first criterion discussed in Section~\ref{sect:hmc} is fulfilled.
In addition, selecting which proposer to use based on the current field configuration is not allowed.
One approach is that of tempered transitions~\cite{neal1996sampling,2014arXiv1405.3489B} which have found recent application in, for example, the study of the 0+1-dimensional Thirring model~\cite{Alexandru:2017oyw}.

Another approach is to interleave large mode jumps~\cite{besag:1995} with normal HMC.\@
By making a large change to the field configuration that is not produced by integrating equations of motion, the barriers that repel HMC trajectories, or the uncrossable valleys separating modes, can be bypassed.
However, as shown by the inefficacy of simple random updates, it can be a challenge to make a proposal with a large change to the field configuration such that the new configuration contributes meaningfully to the partition function and the proposal is thus likely to be accepted.

By taking advantage of some of the features of the problem we can propose large jumps to new regions in configuration space where the weight of the new field configuration is non-negligible.
In the general case it is hard to analytically extract features of the probability weight function.
In the previous section we saw that we could use symmetries to better cover the configuration space.

In fact, the symmetries that emerge in the strong or weak coupling limits can also be used to generate proposals.
However, while a true symmetry operation is guaranteed to produce an accepted configuration, these other operations need not be accepted by the Metropolis step.
The symmetries that emerge in the strong and weak coupling regime were reviewed in Section~\ref{sect:limits}.

It may be possible to craft proposals based on knowledge of the probability distribution in the weak-coupling limit.
However, the failure of perturbation theory for problems of interest suggests these proposals will hardly ever be accepted in a calculation.
We therefore leave the detailed description of these proposals to other work and focus on the approximate factorization of~\eqref{eqn:strong coupling}.

In the strong coupling limit we can flip the sign of $\phi$ on a thread of spatial sites, temporally shift or time-reverse threads individually, and arbitrarily permute the spatial threads rather than remain restricted to symmetries of the lattice.
Proposals adjusting $\phi$ according to those manipulations are likeley to be accepted if the system of interest is sufficiently close to the appropriate limit.

\subsubsection{Spin Basis}\label{sect:spin jumps}

As an example, in the strong coupling limit we can take advantage of our knowledge of the probability distribution for the one-site field configurations.
For convenience we reproduce~\eqref{eq:alpha0PPhi} here,
\begin{align}
    W_1[\phi]
    &=
        \frac{  e^{-\frac{\Phi ^2}{2 U\beta}-\frac{U\beta}{4}}
                \cosh ^2\left(\frac{\Phi}{2}\right)}
             {  \sqrt{2 \pi U\beta}\ \cosh\left(\frac{U\beta}{4}\right) }
    & (\alpha&=0)
\end{align}
where we attach a $1$ subscript to emphasize it is for the one-site problem.

The one-site probability distribution is symmetric under $\Phi\goesto-\Phi$.
In the strong coupling limit the probability distribution for a problem with \nx spatial sites is simply the product of the one-site distribution for each site.
The probability is concentrated on the corners of an $\nx$-dimensional hypercube whose $2^{\nx}$ vertices have value $\Phi_x \simeq \pm U\beta$.

That distribution is invariant if we send $\Phi\goesto-\Phi$ independently for each single spatial site across all time,
\begin{equation}\label{eqn:ergo sign flip}
    \Phi_x \goesto \text{sign}_x\Phi_x
\end{equation}
where $\Phi_x = \sum_t \phi_{x,t}$, the auxiliary field summed across all time but only on a single site, as in \eqref{eqn:polyakov loop} (and so $\Phi = \sum_x \Phi_x$).
We can pick any subset of the spatial lattice, and negate all the auxiliary fields for all time for those sites.

For the one-site problem with $\alpha=0$ in the strong-coupling limit, this operation allows us to propose a new configuration with the same weight but in another mode that will always be accepted.
Away from strong coupling the guarantee of equal weight no longer holds as the factorization of the path integral fails.
Assuming the probability weight does not deform too much, we can visit all the modes with proposals of this type.
Regardless of $U\beta$, this eliminates the in-practice sampling problem one might encounter due to widely-separated modes, but does not resolve the possible formal problems.

\begin{figure}
	\centering
	\includegraphics[width=.8\columnwidth]{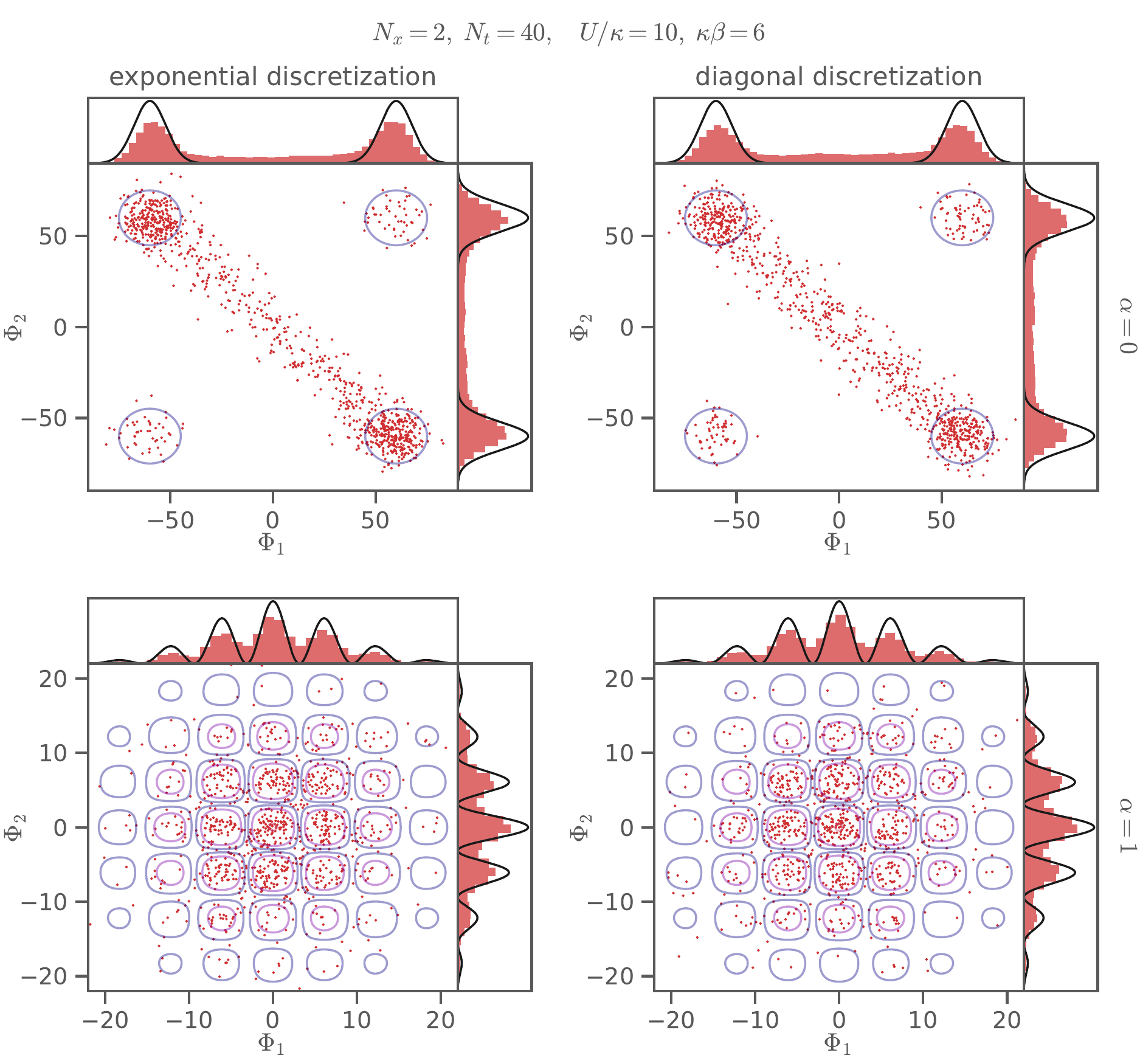}
	\caption{
		HMC history of $\Phi_x = \sum_t \phi_{xt}$ for the 2-site problem with 100k trajectories and a \textbf{sign flip} every 100 trajectories according to~\eqref{eqn:ergo sign flip}.
    	Only every 100th trajectory of each ensemble is shown.
    	The black lines in the marginal distributions for $\Phi_1$ and $\Phi_2$ are the exact \emph{1-site} distributions which are recovered in the strong coupling limit.
    	The contours show the product of the 1-site distributions and are \emph{not} exact results for this case.
		\label{fig:2site histograms ergo jumps}
	}
\end{figure}

\begin{figure}
	\centering
	\includegraphics[width=.8\columnwidth]{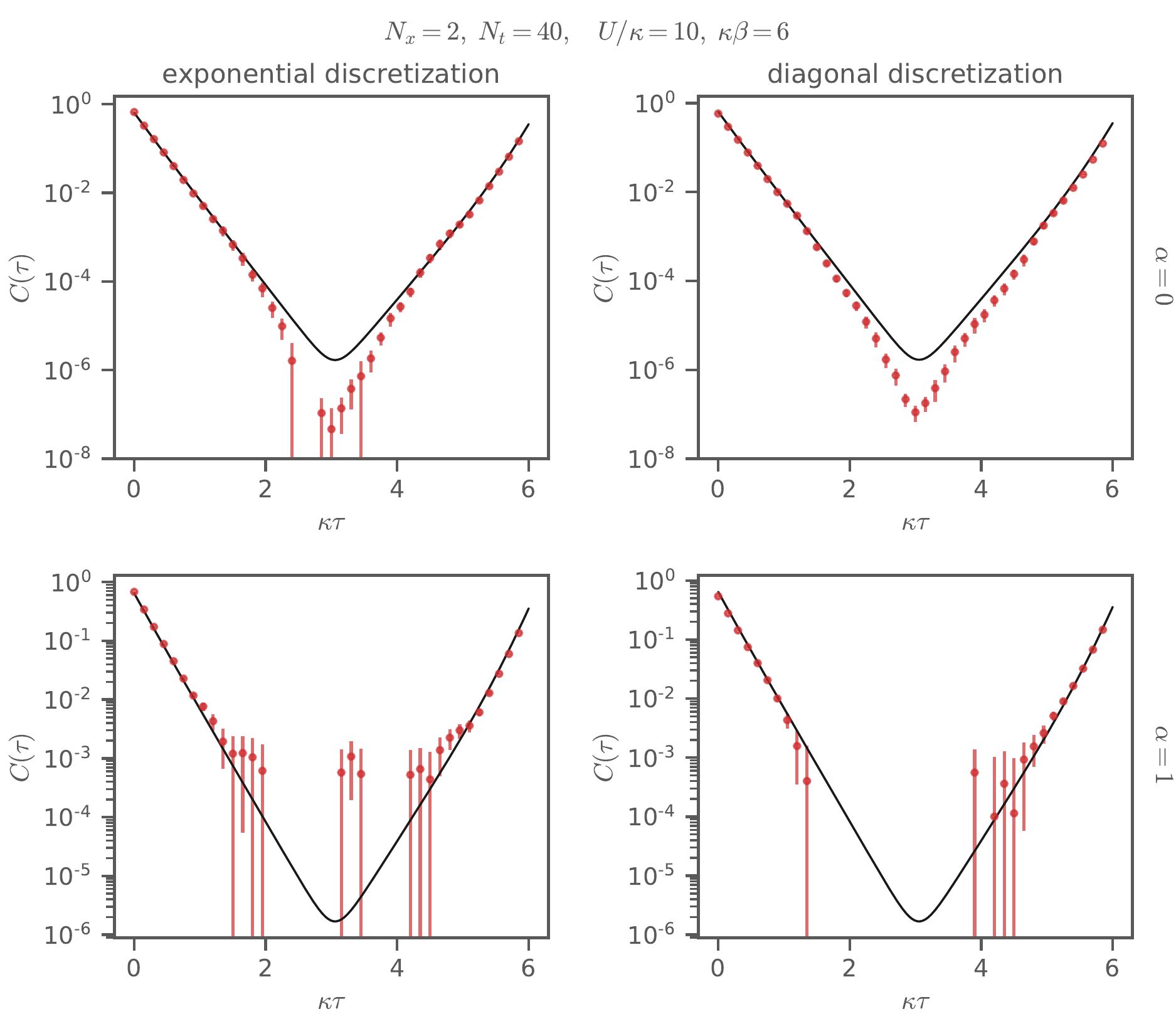}
	\caption{
		Correlators from Eq.~\eqref{eqn:Cpm} from the ensembles with \textbf{sign flips} shown in Figure~\ref{fig:2site histograms ergo jumps}.
		Points where $C(\tau) \leq 0$ are not shown because of the log scale.
		The black lines show exact results from direct diagonalization of the 2-site Hubbard model.
		For clarity, the figure shows only $C_+$, $C_-$ behaves in the same way.
		\label{fig:2site corr ergo jumps}
	}
\end{figure}

We demonstrate the success of this method in Figures~\ref{fig:2site histograms ergo jumps} and~\ref{fig:2site corr ergo jumps}.
They show results from a run using a coarse integrator with an acceptance rate of $\approx 60\%$ augmented by random sign flips based on~\eqref{eqn:ergo sign flip} every 100 trajectories.
The sign is chosen for each spatial lattice site separately and with equal probability for either sign.
Evidently, these jumps allow sampling the far apart lobes for $\alpha=0$ which is not possible with just the fine or coarse integrator (compare to Figures~\ref{fig:2site histograms} and~\ref{fig:2site histograms coarse}).

The correlators in \autoref{fig:2site corr ergo jumps} are similar to those in \autoref{fig:2site corr ergo}.
The most notable difference for is that for $\alpha=0$ the minimum is now in the correct place whereas it was shifted to the right for the fine integrator without sign flips.
Another unknown systematic remains however.

\subsubsection{Particle-Hole Basis}\label{sect:ph jumps}

For the one-site problem with $\alpha=1$ in the strong-coupling limit, the sign-flipping operation in~\eqref{eqn:ergo sign flip} yields a proposed configuration in a different domain from where HMC might be currently stuck.
However, as mentioned in Section~\ref{sect:symmetrization}, this operation helps to symmetrize the distribution around zero, but it does not allow us to access all the modes and thus does not offer a complete solution.

We nonetheless use those sign flips in Figures~\ref{fig:2site histograms ergo jumps} and~\ref{fig:2site corr ergo jumps}.
In this simple case this operation (paired with a coarse MD integrator) is sufficient to sample the relevant region of configuration space properly.
This has no discernible impact on the correlators however which is likely caused by the large noise for medium $\tau$.

When $\alpha=1$ the one-site result is~\eqref{eq:alpha1PPhi}
\begin{align}
   W_1[i\phi]
   &=
       \frac{   e^{-\frac{\Phi ^2}{2 U\beta}+\frac{U\beta}{4}}
                \cos ^2\left(\frac{\Phi}{2}\right)}
            {\sqrt{2 \pi U\beta} \ \cosh\left(\frac{U\beta}{4}\right)}
    & (\alpha&=1).
\end{align}
The full probability distribution in the strong coupling limit is simply the product of this distribution for each site, and is concentrated around the origin $\Phi_x=0$.

In this case we can also propose a new configuration by increasing or decreasing all the auxiliary fields on a single spatial thread by $2\pi/\nt$,
\begin{equation}
    \phi_{x,t} \goesto \phi_{x,t} \pm \frac{2\pi}{\nt}\delta_{x,x_0}
\end{equation}which changes $\Phi_{x_0}$ by $2\pi$, putting it on the other side of an exact zero and thus into a different mode.
Such a proposed update will be accepted according to the ratio of probabilities\footnote{This identity holds in general and not only in the strong coupling limit.}
\begin{equation}
    \frac{W[i(\phi_{x,t} \pm \frac{2\pi}{\nt}\delta_{x,x_0})]}{W[i \phi]} = (\text{ratios of determinants that cancel at strong coupling})\times e^{-\frac{2\pi}{U\beta}(\pi \pm \Phi_{x_0})},
\end{equation}
meaning that if the proposal drives $\Phi_{x_0}$ towards zero it will always be accepted and will occasionally be accepted if it drives $\Phi_{x_0}$ away from zero.
This update will allow us to access all the modes of $\Phi_{x_0}$ without encumbrance from exact zeros in the probability distribution.
In the strong coupling limit we may make this proposed update to each thread of spatial sites independently.
Away from strong coupling this proposal may nevertheless prove beneficial, assuming the probability weight doesn't deform too much.
In fact, the $2\pi$ change in $\Phi_{x_0}$ need not be evenly spread across all the timeslices, at the cost of reducing the likelihood according to a change in \eqref{eqn:other factor}.

In a similar spirit, we can use the field-shift transformation \eqref{eqn:field shift}, shifting each timeslice by its own $\theta_t$, and as long as the shifts sum to 0 modulo $2\pi$, according to \eqref{eqn:periodicity condition}, we can make the cancellation of the determinant exact,
\begin{equation}
    \frac{W[i(\phi_{x,t} + \theta_t)]}{W[i\phi]} = (\text{ratios of determinants that exactly cancel}) \times e^{-\frac{1}{2\Utilde} \sum_t 2 \phi_t \theta_t + \nx \theta_t^2}.
\end{equation}
If we pick a constant $\theta_t = \pm2\pi/\nt$ we can simplify further,
\begin{equation}\label{eqn:coordinated-jump-acceptance}
        \frac{W[i(\phi_{x,t} \pm 2\pi/\nt)]}{W[i\phi]} = e^{-\frac{2\pi}{U\beta}\left(\pi \nx \pm \Phi\right)}.
\end{equation}
The average acceptance rate is analyzed in \Appref{acceptance}.
In \Figref{1site contours}, such a coordinated jump is a jump in the diagonal $\Phi$ direction, orthogonal to the lines of zero determinant.

Most generally, we know that when $\alpha=1$ the determinants are $2\pi$ periodic in each field variable.
For a $\pm 2\pi$ change on site $x_0$ at time $t_0$ the ratio of the weights is simply given by the ratio of the gaussian factors,
\begin{equation}\label{eqn:jump-acceptance}
    \frac{W[i(\phi_{x,t}\pm2\pi \delta_{x,x_0}\delta_{t,t_0})]}{W[i \phi]} = (\text{ratios of determinants that exactly cancel}) \times e^{-\frac{2\pi}{\Utilde}(\pi \pm \phi_{x_0,t_0})}.
\end{equation}
We can independently propose and accept or reject a change by $\pm2\pi$ on each site, knowing the determinant will remain invariant.
Again, if this drives the auxiliary field on a site towards zero it will be accepted and it will occasionally be accepted when the field is driven away from zero.
This proposal is entirely local and extremely speedy, not requiring an evaluation of the determinant.
Moreover, in the exponential $\alpha=1$ case, each accepted change hops over a zero manifold, as can be seen from~\eqref{eqn:f sign}.
This alleviates the formal ergodicity problem arising from conjugate reciprocity.
In the diagonal $\alpha=1$ case, while there is no formal ergodicity problem, interleaving such a proposal can help in practice, especially as barriers are raised as one approaches the continuum limit.

Reconsidering Figure~\ref{fig:1site contours}, it is clear that an evenly-distributed change of $\Phi$ by $2\pi$ is more likely to be accepted than a $2\pi$ shift in either auxiliary field variable alone, because a coordinated jump tends to keep you close to the middle of the gaussian.
As discussed in \Appref{acceptance}, the acceptance rate of coordinated jumps remains finite in the continuum limit, while the acceptance rate for jumps in individual field components vanishes in the continuum limit.
Whether such coordinated changes are guaranteed to cross every ergodicity barrier remains obscure to us, while it seems apparent that independent changes provide such assurance.

\begin{figure}
  \centering
  \includegraphics[width=.65\columnwidth]{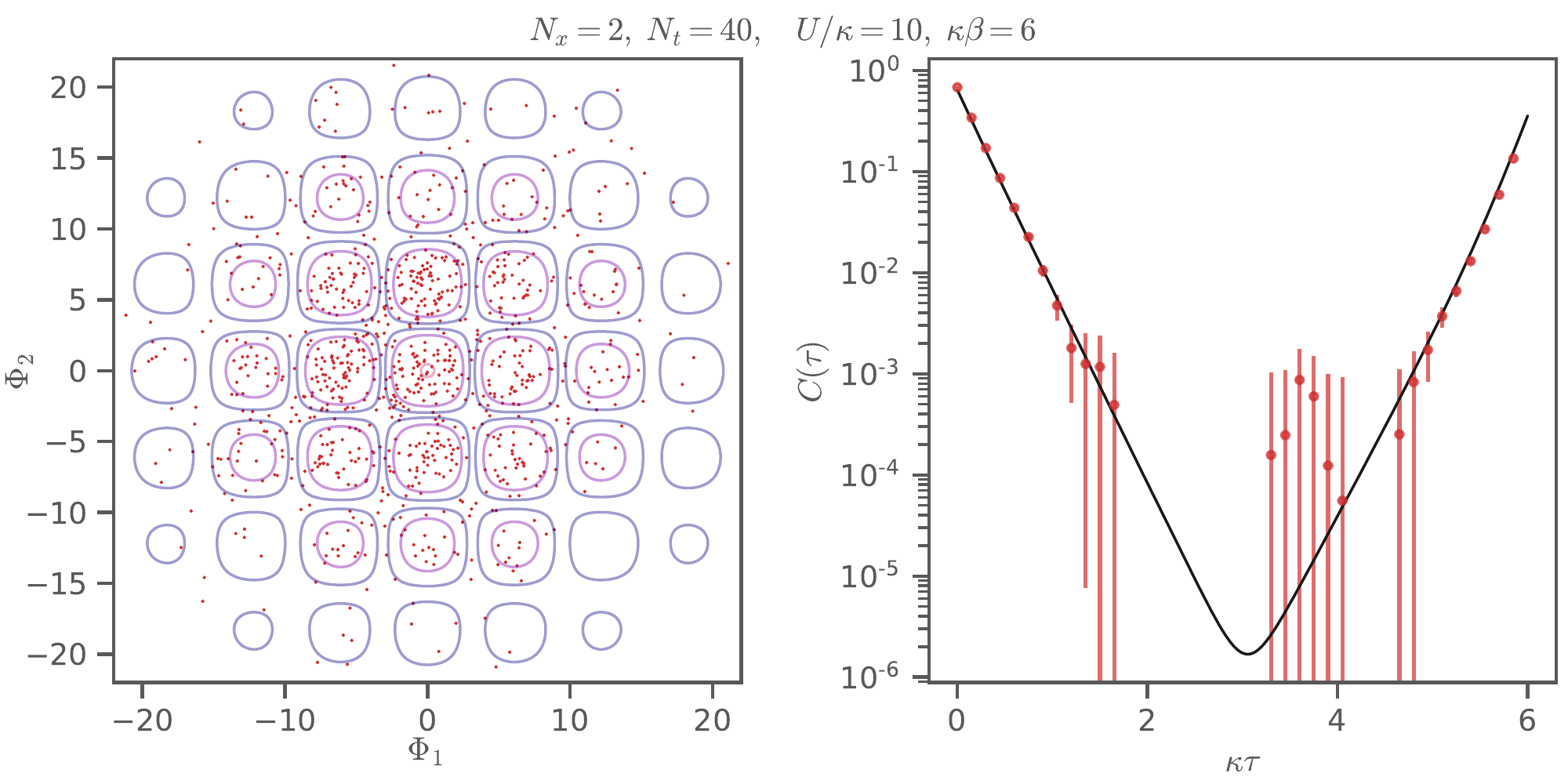}
  \caption{HMC trajectories and correlator for the exponential $\alpha=1$ case with the same parameters as Figures~\ref{fig:2site histograms}, and~\ref{fig:2site corr ergo} but with large jumps as in equations~\eqref{eqn:coordinated-jump-acceptance} and~\eqref{eqn:jump-acceptance} every 100 trajectories.
    Only every 100th trajectory is shown in the left panel and the ensemble consists of 100k trajectories in total.\label{fig:2site 2pi jumps}}
\end{figure}

In Figure~\ref{fig:2site 2pi jumps} we show an example run in the two site system with the exponential discretization.
The integrator is as fine as in Figures~\ref{fig:2site histograms} and~\ref{fig:2site corr ergo} but every 100 trajectories two jumps are performed.
First, $N_t N_x$ single site jumps by $2\pi$ are performed on random sites and each is accepted or rejected according to~\eqref{eqn:jump-acceptance}.
Then, all field variables are shifted by $\pm 2\pi/N_t$ in a coordinated jump and the change is accepted or rejected according to~\eqref{eqn:coordinated-jump-acceptance}.
The result shows a similar improvement to the sign flips in Figures~\ref{fig:2site histograms ergo jumps} and~\ref{fig:2site corr ergo jumps}.
The histogram shows marked improvement, showing no hint of trapping, and the correlation function is no longer incompatible with the exact result, especially visible around $\kappa\tau\sim1.5-2$.

\section{Conclusion\label{sect:conclusion}}

The issue of ergodicity is of great significance for the validity of any stochastic calculation which relies on an MCMC process.
Throughout this work, we have emphasized that ergodicity (or the lack thereof) has a \emph{formal} as well as a \emph{practical} meaning.
In the former case, regions in the configuration space exist which are \emph{formally} disconnected due to appearance of boundaries of co-dimension~$1$, at which $\det M[\phi]=0$.
In the latter case, such boundaries do not exist, yet MCMC cannot sample all relevant regions of configuration space in a practical amount of computing time.
In other words, thermalization and decorrelation would be exponentially slow.
In such cases, we speak of an \emph{in practice} ergodicity issue.
We have investigated both situations, in terms of the Hubbard model in 2D.
In particular, we have shown that the exponential discretization in the particle-hole basis exhibits a \emph{formal} ergodicity problem arising from conjugate reciprocity, while the diagonal discretization in the particle-hole basis does not.\footnote{In Ref.~\cite{Ulybyshev:2017ped} the authors use a fermion operator not considered here, where transport in time alternates between a linearized hopping operator and interaction through the auxiliary field.  They claim that the simulations are free from ergodicity problems because the determinant is complex, in line with our detailed argument.}
Furthermore, for the case of $\alpha=0$ (spin basis) at large $U\beta$, an \emph{in practice} ergodicity problem also appears, as the field distributions fragment into widely separated multi-modal lobes.
HMC algorithms become trapped in one of these lobes and require an exponential simulation time to traverse to any other lobe, despite there being no boundaries where the fermion determinant changes sign.

We note that ergodicity issues have been encountered numerous times in MCMC simulations of lattice gauge theories.
In cases where the problem is \emph{formal}, it is often observed that it is sufficient (at least for small systems) to reduce the accuracy of the MD integrator so that the $\det M[\phi] = 0$ barriers can be traversed.
We have observed such behavior for the $\alpha=1$ exponential discretization which has $\det M[\phi]=0$ boundaries, which nevertheless could be traversed with sufficiently coarse MD integration.
However, this comes at the price of reducing the acceptance rate of the HMC algorithm, and for cases where $U$ and $\beta$ are very large, ergodicity can no longer be restored by such brute force methods.
For the $\alpha=0$ basis (both exponential and diagonal), the separation of the field distributions into widely spaced lobes at large $U\beta$ presents a more serious challenge for HMC, despite it being an \emph{in practice} rather than \emph{formal} issue.
For such cases, we have proposed a new algorithm which takes advantage of the symmetries of the action and the fermion matrix $M$.
Effectively, we augment the standard HMC algorithm with large, carefully crafted jumps between regions of large probability.
The accept/reject step of the algorithm ensures that each region is sampled with the correct relative probability.
We have found that such jumps fix this particular type of \emph{in practice} ergodicity problem.

Our studies were concentrated on histograms of the HS field and single particle correlators.
We stress that this does not present a full proof that the algorithm is ergodic.
It is, of course, still possible that there is a problem caused by a barrier that has escaped notice.
It is therefore useful to carefully verify that observables match known results that might be calculated.
If the algorithm is ergodic one expects a faithful ensemble and that every observable match its true value.

In the $\alpha = 0$ basis, we have observed the appearance of configurations with $\det M[\phi] \le 0$ for lattices with $4$~or more sites, as was also found in Ref.~\cite{Beyl:2017kwp} for the $16$-site problem.
We observe that the frequency with which negative determinants occur increases dramatically with system size, which we have attributed to an accumulation of states with nearly zero energy.
This is highly detrimental to the HMC algorithm and its evolution, as the frequency with which exceptional configurations are encountered also increases.

Our studies show how the \emph{reduction} of symmetries removed formal ergodicity issues.
Missing from this discussion are the implications of reduced symmetry for observables.
The reduction of symmetries had little to no effect on the two-point correlators we have concentrated on (which give access to quasi-single-particle energies). We note that this is by no means a general statement. Still, a similar behavior for the spin-density wave (SDW) order parameter was found in Ref.~\cite{Beyl:2017kwp}.
On the other hand, in Ref.~\cite{Buividovich:2016tgo} the
diagonal discretization led to chiral symmetry violations that impacted the spin-density wave order parameter, which can only be restored in the continuum limit. Thus the restoration of ergodicity in this case made the calculation of the SDW order parameter more difficult, and emphasizes the need for a robust continuum limit extrapolation.
For additional evidence that the diagonal discretization compromises on exact symmetries in favor of better ergodic properties, the reader is pointed to Appendix~\ref{sect:other-observables}.

In the examples we have considered here, we have taken values of $U$ and $\beta$ which are deliberately chosen to be large (such as $U\beta \simeq 60$), in order to ensure that ergodicity issues are not unduly suppressed.
However, we note that the $\alpha = 1$ diagonal discretization was found to suffer from neither \emph{formal} nor \emph{practical} ergodicity issues, even for extreme values of $U$ and $\beta$.
When comparisons with exact results are feasible (such as for $1$-site, $2$-site, and $4$-site square and hexagonal lattices), we have found that this discretization performed equally well as the other ones (after application of the remedies mentioned above).
For cases with large $U$ and $\beta$, the diagonal $\alpha = 1$ discretization was found to be superior, as it requires no modification to the HMC algorithm, nor monitoring for exceptional configurations.
For the $16$-site calculation, we were able to reproduce published results from the BSS algorithm and never encountered any exceptional configurations.
For these reasons, we conclude that the diagonal discretization with $\alpha = 1$ is optimal for our purposes, even though it only recovers the exact chiral symmetry of the exponential discretization in the continuum limit.



\section*{Acknowledgements}
We thank Stefan Krieg, Johann Ostmeyer, Carsten Urbach and Enrico Rinaldi for insightful discussions. 
We are indebted to Stefan Beyl for his assistance in comparing our results with the BSS formalism.
C.K. gratefully acknowledges funding through the Alexander von Humboldt Foundation through a Feodor Lynen Research Fellowship.
This work was done in part through financial support from the Deutsche Forschungsgemeinschaft (Sino-German CRC 110).
The authors gratefully acknowledge the computing time granted through JARA-HPC on the supercomputer JURECA~\cite{jureca} at Forschungszentrum J\"ulich.


\FloatBarrier
\appendix

\section{Determinants}\label{sect:determinants}

\subsection{Cyclic Lower Block Bidiagonal Matrices}
In this section we derive the central equations for the determinants of the fermion matrices.
We look at matrices of the form
\begin{align}
  M_{x't',xt} &= D_{x',x}\delta_{t',t} - {(T_{t'})}_{x',x}\mathcal{B}_{t'}\delta_{t',t+1}\ ,
\end{align}
where
\begin{align}
  \mathcal{B}_t \equiv
  \begin{cases}
    +1,\quad 0 < t < \nt \\
    -1,\quad t = 0
  \end{cases}
\end{align}
encodes anti-periodic boundary conditions in time and is factored out to simplify representing $M=\Mexp$ or $M=\Mdia$.
In time-major layout $M$ is a cyclic lower block bidiagonal $\nt\times\nt$ matrix with blocks of size $\nx\times\nx$.
Written in matrix form this is
\begin{align}
  M =
  \begin{pmatrix}
    D    &      &        &        & T_0 \\
    -T_1 & D    &        &        &     \\
         & -T_2 & D      &        &     \\
         &      & \ddots & \ddots &     \\
         &      &        &-T_{\nt-1}&D   \\
  \end{pmatrix}.
\end{align}

We can compute the determinant of $M$ by means of an LU-decomposition in terms of the matrices $D$ and $T_t$.
This decomposition can be performed by hand thanks to the sparsity of $M$.
We use the following ansatz which is an adaptation of the ansatz presented in Ref.~\cite{zivkovic:2013}
\begin{align}
  L =
  \begin{pmatrix}
    1   &     &    &        &        &\\
    l_0 & 1   &    &        &        &\\
        & l_1 & \ddots &        &        &\\
        &     & \ddots & 1      &        &\\
        &     &   & l_{n-3} & 1      &\\
        &     &   &        & l_{n-2} & 1
  \end{pmatrix},
  \; U =
  \begin{pmatrix}
    d_0 &     &      &   &        & v_0\\
        & d_1 &     &    &        & v_1\\
        &     & d_2 &    &        & \vdots \\
        &     &     & \ddots &        & v_{n-3} \\
        &     &     &    & d_{n-2} & v_{n-2} \\
        &     &     &    &        & d_{n-1}
  \end{pmatrix}
\end{align}
Multiplying out $LU = M$ and solving the straightforward recursive equations gives
\begin{align}
  d_i &= D \quad \text{for} \quad 0 \le i < \nt-1\\
  d_{\nt-1} &= D (1 + A)\\
  v_i &= D A_{0,i}\\
  l_i &= - T_{i+1} D^{-1}\ .
\end{align}
Here we have used
\begin{align}
  A_{t,t'} &\equiv D^{-1} T_{t'} D^{-1} T_{i-1} \cdots D^{-1} T_t\ ,\label{eqn:def partial A}\\
  A &\equiv A_{0,\nt-1}\ .\label{eqn:def A}
\end{align}
The determinant can be computed from the determinants of the blocks on the diagonals using
\begin{align}
  \det M &= \det L \det U
             \nonumber\\
             &= \left(\prod_{i=0}^{\nt-1}\,\det 1\right) \left(\prod_{i=0}^{\nt-1}\, \det d_i\right)
             \nonumber\\
             &= {(\det D)}^{\nt-1} \det D(1+A)
             \nonumber\\
             &= {(\det D)}^{\nt} \det (1+D^{-1} T_{\nt-1} D^{-1} T_{\nt-2} \cdots D^{-1} T_{1} D^{-1} T_{0})\ .\label{eqn:det M 1}
\end{align}
This result can alternatively be written as
\begin{align}
  \det M &= \det (T_{\nt-1}T_{\nt-2} \cdots T_0) \det (1 + T^{-1}_{0} D T^{-1}_{1} D \cdots T^{-1}_{\nt-2} D T^{-1}_{\nt-1} D)\label{eqn:det M 2}\\
         &= \det (T_{\nt-1}T_{\nt-2} \cdots T_0) \det (1 + A^{-1})
\end{align}
which can be beneficial depending on the forms of $D$ and $T$.
Expression~\eqref{eqn:det M 2} can be obtained from~\eqref{eqn:det M 1} by factoring in one $D$ and factoring out one $T$ at a time.

\subsection{Fermion Determinants}
The determinant of $\Mexp$~(\ref{eqn:M1}) can be calculated from~(\ref{eqn:det M 1}) by inserting
\begin{align}
  D \mapsto \one\ , \qquad T_{t'} \mapsto e^{h} F_{t'}\ ,
\end{align}
with
\begin{align}
  {F_{t'}[\phi]}_{x',x} &\equiv e^{\phi_{x(t'-1)}}\delta_{x',x}\ .\label{eqn:def F}
\end{align}
Thus
\begin{align}
  \det \Mexp[\phi, h] &= \det (1 + e^{h}F_{\nt-1}[\phi] e^{h}F_{\nt-2}[\phi] \cdots e^{h}F_{1}[\phi] e^{h}F_{0}[\phi])\label{eqn:det Mexp}\\
                   &= \det (1+B[\phi, h])\ .
\end{align}

We can proceed in a similar way for the matrix in the diagonal discretization $\Mdia$~(\ref{eqn:M3}).
This time we insert the following into the alternate form~(\ref{eqn:det M 2})
\begin{align}
  D \mapsto \one - h\ , \qquad T_{t'} \mapsto F_{t'}\ ,
\end{align}
with the same $F$ as before~\eqref{eqn:def F}.
Thus
\begin{align}
  \det \Mdia[\phi, h] &= e^{\Phi} \det (1 + F^{-1}_{0}[\phi] (\one-h) F^{-1}_{1}[\phi] (\one-h) \cdots F^{-1}_{\nt-2}[\phi] (\one-h) F^{-1}_{\nt-1}[\phi] (\one-h))\label{eqn:det Mdia}\\
                   &= e^{\Phi} \det (1+A^{-1}[\phi, h])\ .
\end{align}

At this point, a note on numerical stability is in order.
The calculation of determinants of dense $\nx\times\nx$ matrices via a standard LU-decomposition should not be a problem.
Especially so since we are ultimately interested in $\log \det M$ which is more stable for large matrices.
The spatial matrices $A$ and $B$ are however constructed from a product of $2\nt$ matrices.
Such a product can incur large round-off errors when the involved matrices have significantly different scales.
Both $\one - h$ and $e^{h}$ have elements $\lesssim 1$ and present no problem.

For the spin basis, the matrices $F \sim e^{\phi}$ on the other hand have elements of widely varying size which can be significantly larger than one.
We have observed instabilities of the action and therefore HMC evolution as well as the solver for systems of linear equations based on equations~\eqref{eqn:det Mexp} and~\eqref{eqn:det Mdia}.

For the particle/hole basis, the problematic matrices are replaced by $F \sim e^{i\phi}$ whose elements are on the unit circle.
All matrices that contribute to $A$ and $B$ are therefore of the same order and floating point errors are greatly reduced.
Our tests show that both action and solver are precise to 12 or more digits for lattices of size $\nx \approx 100$ and greater.

We use algorithms based on~\eqref{eqn:det Mexp} and~\eqref{eqn:det Mdia} for $\alpha=1$.
For $\alpha=0$ we use dense $L-U$ based algorithms for the full space-time matrices $M$.
Other algorithms have been used to solve this problem, including a stabilization of this algorithm using singular value decompositions~\cite{Bulgac:2008zz} and a different solver based on Schur complements~\cite{Ulybyshev:2018dal}.


\section{Sixteen-site Problem}\label{sect:sixteen sites}

\begin{figure}[h]
	\centering
	\includegraphics[height=.55\columnwidth]{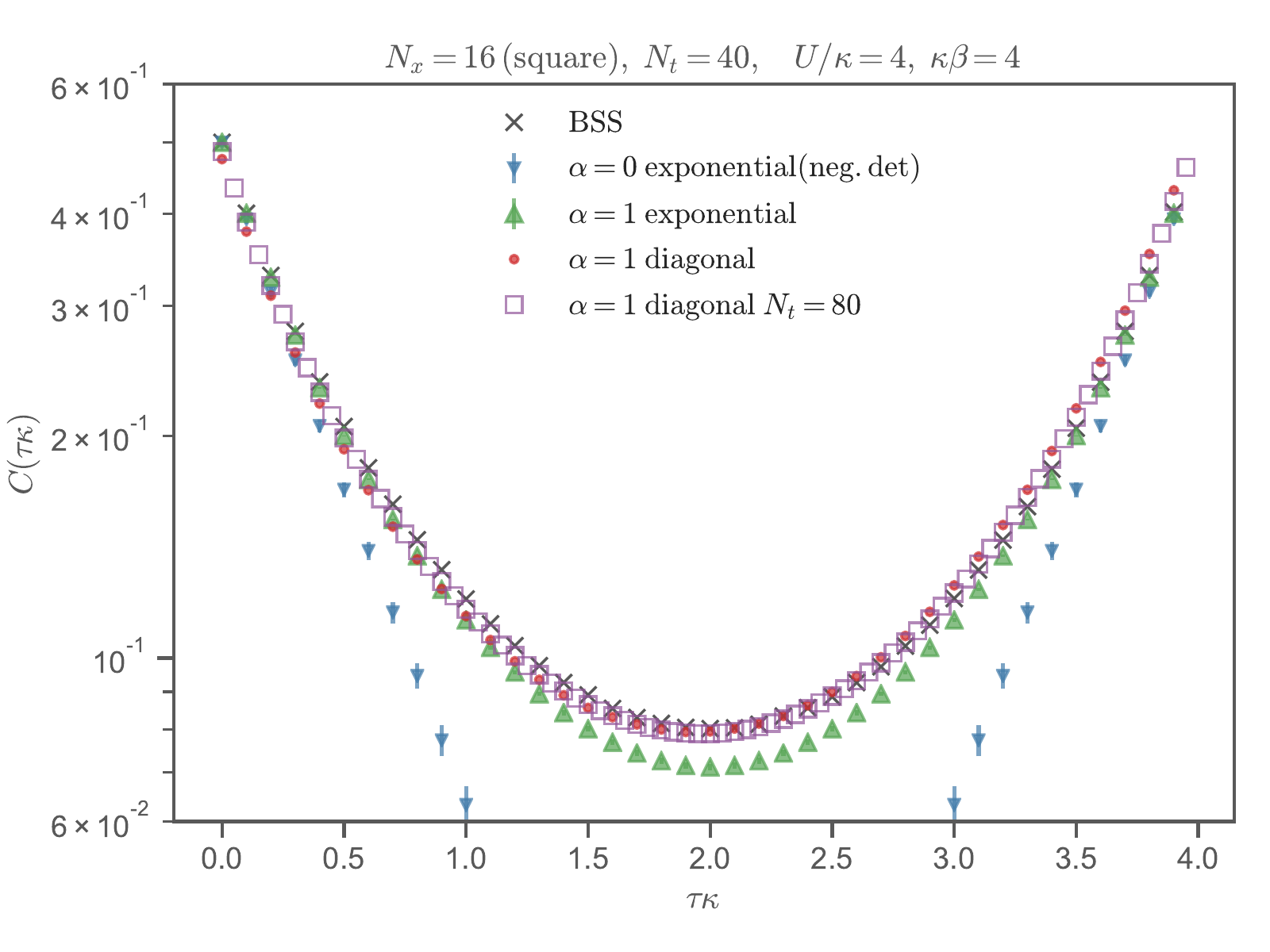}
	\caption{
		Comparison of correlators on a $4\times4$ square lattice.
		The data points for BSS are taken from Ref.~\cite{Beyl:2017kwp}.
		The exponential $\alpha=0$ HMC run has negative $\det M[\phi]$ for all configurations.
		Started from a positive determinant this case produces results consistent with BSS and $\alpha=1$.
		\label{fig:16sites corr}
	}
\end{figure}

For completeness we reproduce Figure~1 of Ref.~\cite{Beyl:2017kwp} which outlines the susceptibility of correlators to the starting configuration in case HMC is not ergodic.
The correlators examined in this case are the equal-site correlators,
\begin{displaymath}
C(\tau)=\frac{1}{N_x}\sum_i\langle C_{ii}(\tau)\rangle\ ,
\end{displaymath}
where the sum is over all lattice sites.
We show three different HMC runs in Figure~\ref{fig:16sites corr} as well as the result obtained from the Blankenbecler-Scalapino-Sugar method (BSS) taken from Ref.~\cite{Beyl:2017kwp}.
The latter is claimed to be free of any ergodicity problems and thus provides a benchmark for our HMC results.

One HMC run is in the exponential discretization in the spin basis and starts with a configuration $\phi_0$ such that $\det M[\phi_0] < 0$.
The MD integrator is very fine (acceptance rate $>98\%$) in this particular example, to ensure no barriers are accidentally jumped by coarse integration.
This run clearly deviates from the BSS result and matches the (green) correlator shown in Figure~1 of Ref.~\cite{Beyl:2017kwp} (shown as black $\times$s in \Figref{16sites corr}).
Because of the very fine MD integration, HMC gets trapped in a part of the integration domain where the determinant is negative.
For a coarser integrator HMC is able to traverse to $\det M>0$ regions and in this case we find good agreement with the BSS points, though we do not show these results here.

The second HMC run is again in the exponential discretization but in the particle-hole basis, again with a very fine MD integrate to ensure that the ergodicity barriers are impenetrable.
While matching the BSS results at early times, in the middle of the temporal extent it differs markedly from the BSS result.

The third HMC run is in the diagonal discretization in the particle/hole basis.
Since the determinant is complex in this case, there is no particular starting criterion to be chosen and we stress that we do not find any dependence on the initial configuration.
The correlator agrees well with the BSS result at intermediate times; at early times there is a systematic discrepancy; recall that different discretizations need only agree in the continuum.

To emphasize this point, we also show a third HMC run in the diagonal discretization in the particle/hole basis, but with twice as many time slices.
On each common timeslice, the finer discretization is closer to the BSS result than the coarser.

Also note that at early times the trend towards the BSS result is upward, while at later time it is downward.
Averaging the correlator and its time-reversed partner results in a correlator indistinguishable from the BSS result.
In future work we will detail methods for reducing discretization artifacts of correlator measurements.
We emphasize that one \emph{still must take a controlled continuum limit}, even if results are formally closer to the continuum.

\section{Jump Acceptance}\label{sect:acceptance}

In this section we analyze the acceptance rate of the jumps proposed in \Secref{jumps} that take advantage of the periodicity of the determinant, a coordinated jump uniform across all sites
\begin{equation}
    \phi_{x,t} \goesto \phi_{x,t} + \frac{2\pi}{\nt}
\end{equation}
and an individual jump
\begin{equation}
    \phi_{x,t} \goesto \phi_{x,t} + 2\pi \delta_{x,x_0}\delta_{t,t_0}
\end{equation}
in the field component on site $(x_0,t_0)$.

For the coordinated proposal, the acceptance rate is given by \eqref{eqn:coordinated-jump-acceptance},
\begin{equation}
    \frac{W[i(\phi_{x,t} \pm 2\pi/\nt)]}{W[i\phi]} = e^{-\frac{2\pi}{U\beta}\left(\pi \nx \pm \Phi\right)}
\end{equation}
To calculate the average acceptance rate, we should average that weight change in such a way that $\Phi$ is generally representative of a field configuration.
Were the action entirely controlled by the gaussian part, $\phi$ would be distributed normally with mean $0$ and variance $\sigma^2 = \Utilde$.
Then, summing $\nx\nt$ such random variables, we expect $\Phi$ to be drawn from a zero-mean gaussian with variance $\nx\nt\Utilde = \nx U \beta$.
So, we find that the acceptance rate is independent of \nt, but decreases markedly with spatial volume.

When implementing the ergodicity jumps of individual field components it is natural to ask: should each site's jump be accepted or rejected independently?
Or should they be amalgamated into one proposal and considered as a whole?
Here we will argue that it is better to consider each decision independently.

The change in the weight \eqref{eqn:jump-acceptance} for a $2\pi$ shift on site at location $x_0$ and time  $t_0$ is given by
\begin{equation}
    \frac{W[i(\phi_{xt}\pm2\pi \delta_{x,x_0}\delta_{t,t_0})]}{W[i \phi]} = (\text{ratios of determinants that exactly cancel}) \times \exp\left(-\frac{2\pi}{\Utilde}(\pi \pm \phi_{x_0,t_0}) \right),
\end{equation}
while if we did many $2\pi$ jumps, the change in the action would be summed, so that the acceptance would be controlled by
\begin{equation}
	\exp\left(-\frac{2\pi^2}{\Utilde} \sum_x \left(1 \pm_x \frac{\phi}{\pi}\right)\right)
\end{equation}
where $x$ summarizes the space and time coordinates and the subscript on $\pm_x$ is there to remind us that the sign choice is $x$-dependent and can be zero for some sites.

To calculate the average acceptance rate we should again average that weight change in such a way supposes we start in a verly likely configuration.
As above, we assume each $\phi$ is a normal, random variable with variance $\Utilde$.

If, in a single proposal, each site is given a probability $p$ of undergoing a jump, on average we have to sum the action changes from $p\nx\nt$ jumps.
We would like to find the best possible $p$, deciding whether we should completely amalgamate the proposals into one, consider each individually, or something in between.
We can estimate
\begin{equation}
	\sum_x \left(1\pm_x \frac{\phi}{\pi}\right) = p \nx\nt + \frac{\varphi}{\pi}
\end{equation}
where $\varphi$ is drawn from a zero-mean gaussian with variance $p\nx\nt \Utilde$.
On average $\varphi$ is 0 and we get an acceptance probability of
\begin{equation}
    \exp\left(-\frac{2\pi^2}{U\beta/\nt} p \nx \nt\right).
\end{equation}
That is, the acceptance rate is reduced with increasing $p\nx\nt$.
The best choice is to set $p=1/\nx\nt$---that is, to accept or reject a jump in one auxiliary field at a time.
Still, in that case, the acceptance rate exponentially decreases with \nt, so that these jumps are much less important as one takes the continuum limit, though they increase in importance with $U\beta$.
Moreover, when making proposals of jumps in individual components, there is no suppression with the spatial volume \nx, unlike the case of coordinated jumps.

\section{Other Observables\label{sect:other-observables}}

In this appendix we consider other observables.  In particular, we consider correlations between local bilinear operators.

Letting $x$ denote a spatial index, we define the bilinear operators
\begin{align}
    S^0_x &= \frac{1}{2} \left[ a_x a_x^\dagger - b_x b_x^\dagger +1 \right]
    &
    S^1_x &= \frac{1}{2} (-1)^x \left[ b_x^\dagger a_x^\dagger + a_x b_x \right]
    \nonumber\\
    S^2_x &= \frac{i}{2} (-1)^x \left[ b_x^\dagger a_x^\dagger - a_x b_x \right]
    &
    S^3_x &= \frac{1}{2} \left[ a_x a_x^\dagger + b_x b_x^\dagger -1 \right],
\end{align}
raising and lowering operators,
\begin{equation}
    S^\pm_x = S^1_x \pm i S^2_x
\end{equation}
and the charge-density operator $\rho = 1-2S^0$ so that at half-filling, the expectation value $\langle \rho \rangle$ vanishes.
The correlation functions
\begin{equation}
    C^{ij}_{xy}(\tau) = \frac{1}{\nt}\sum_t \left\langle S^i_{x,t+\tau}S^{j\dagger}_{y,t} \right\rangle
\end{equation}
carry two spatial indices $x$ and $y$ which can be projected to definite irreps of the lattice (the dagger is just to provide a consistent notation with \eqref{eqn:1 site correlator}; the $S$ and $\rho$ operators are hermitian).
The two irreps we will consider here are the spatially uniform irrep and its doubler, the alternating irrep.
Let $A$ and $B$ be the two sublattices and $\hat{A}$ and $\hat{B}$ be idempotent orthogonal projection operators that sum to the identity.
Denoting the uniform irrep $+$ and the alternating, staggered irrep $-$, projection to those irreps requires performing the sums
\begin{align}
    +:\phantom{xx}& \nx^{-1/2} \left(\sum_{x \in A} + \sum_{x \in B} \right) = \nx^{-1/2}\sum_x (\hat{A} + \hat{B})
    \\
    -:\phantom{xx}& \nx^{-1/2} \left(\sum_{x \in A} - \sum_{x \in B}\right) = \nx^{-1/2}\sum_x (\hat{A} - \hat{B})
\end{align}
where $A$ and $B$ indicate the set of sites belonging to the two different sublattices.
Once projected to definite irrep, the correlation functions are diagonal.
We denote irrep-projected correlators with the $+$ and $-$ subscripts, rather than spatial subscripts.

We can study these correlators themselves or boil them down into order parameters.
One possible charge density wave (CDW) order parameter is given by\footnote{At equal time this definition matches the definition in \Reference{Meng2010}.}
\begin{equation}
    \text{CDW} = V\inverse\;\lim_{\tau\goesto0} C^{\rho\rho}_{--}(\tau)
\end{equation}
where the $\rho$ index indicates that we should use the charge operator, rather than one of the spin operators\footnote{This definition matches, for example, equations (2) and (3) of Ref.~\cite{PhysRevLett.122.077602} which was also studied in Ref.~\cite{Chen:2018tbe}}, and we can also define the (extensive) antiferromagnetic structure factor\cite{Lang2010}
\begin{equation}
    S_{AF}    = \lim_{\tau\goesto0} C^{33}_{--}(\tau)
\end{equation}
and the antiferromagnetic susceptibility\cite{Lang2010}
\begin{equation}
    \chi_{AF} = V\inverse\;\lim_{\tau\goesto0} C^{+-}_{--}(\tau).
\end{equation}
With this formalism we can also characterize spatially uniform order parameters, such as the mean squared magnetization\cite{Buividovich:2018yar}
\begin{equation}
    \left\langle m_i^2 \right\rangle = V\inverse\;\lim_{\tau\goesto0} C^{ii}_{++}(\tau).
\end{equation}

The antiferromagnetic order parameters in Ref.~\cite{Buividovich:2018yar} can similarly be extracted from same-time information in the correlators.
There, the authors explain that the true indication of a staggered order is a spin (or charge) separation between the two sublattices, but that in a finite volume this order parameter will exactly vanish unless a bias is introduced, simulated with, and ultimately taken to zero.
Rather than simply squaring this difference, which corresponds to considering the above correlators projected to the alternating irrep on both indices at vanishing temporal separation, they define additional quadratic observables.
The total-per-sublattice order parameters $\left\langle q^2 \right\rangle$ and $\left\langle S_i^2\right\rangle$ in Ref.~\cite{Buividovich:2018yar} are given by
\begin{align}
    \left\langle q^2 \right\rangle &= V\inverse\;\lim_{\tau\goesto0} (\hat{A}C^{\rho\rho}\hat{A}+\hat{B}C^{\rho\rho}\hat{B})_{--}
    &
    \left\langle S_i^2 \right\rangle &= V\inverse\;\lim_{\tau\goesto0} (\hat{A}C^{ii}\hat{A}+\hat{B}C^{ii}\hat{B})_{--}
\end{align}
which differ from those parameters defined above by the absence of cross-terms $\hat{A}C\hat{B}$ and $\hat{B}C\hat{A}$.
Note that the signs these cross terms carry are what distinguish the $+$ irrep from the $-$ irrep, in the sense that dropping them from the $++$ correlators yields the same numerical data as dropping them from the $--$ correlators,
\begin{equation}
    (\hat{A}C^{\rho\rho}\hat{A}+\hat{B}C^{\rho\rho}\hat{B})_{--}
    = (\hat{A}C^{\rho\rho}\hat{A}+\hat{B}C^{\rho\rho}\hat{B})_{++},
\end{equation}
for example.
Hence, once the cross terms are dropped, the distinction between staggered and uniform order is lost.
In order to distinguish a uniform order from a staggered order, it makes physical sense to consider
\begin{align}
    \left(\sum_{\substack{x\ \in A \\ y \in B}} + \sum_{\substack{x\ \in B \\ y \in A}}\right)&C^{\rho\rho}_{xy}(\tau)
    &
    \mathrm{or}&
    &
    \sum_{\langle x, y \rangle} &C^{\rho\rho}_{xy}(\tau)
\end{align}
where the sum can be taken over $x \in A$ and $y \in B$ (and vice-versa) or, if contributions from wide spatial separations contribute too much noise, over nearest neighbors $\langle x,y \rangle$.  In the two-site example considered below we simply do irrep projection.

The uniform irreps yield correlations between the operators summed over space.
These operators commute with the Hamiltonian and are therefore conserved.
Commuting with the Hamiltonian also implies the time-independence of the spectral decomposition of the respective correlator.
We may still calculate the correlation function as a function of time and average over time to get an estimator for the same-time correlator,
\begin{equation}\label{eqn:uniform irrep time independence}
    \left\langle \rho_+ \rho_+ \right\rangle = \frac{1}{\nt}\sum_\tau C^{\rho\rho}_{++}(\tau),
\end{equation}
for example.

\subsection{Numerical Results}

In \Figref{spin-spin-correlators} we reuse the ensembles of field configurations from the two-site problem of \Secref{2sites} (in particular, the $\alpha=1$ ensembles used in Figures~\ref{fig:2site histograms} and \ref{fig:2site corr ergo}, produced with a very fine molecular dynamics integration), measuring a variety of correlation functions, which are shown in different colors.
We show the exponential and diagonal discretizations in the left and right columns, respectively, and the uniform irrep and the staggered irrep in the top and bottom rows, respectively.
The results from an exact diagonalization are shown as thin lines, while the numerical results are shown as error bars without markers.
In the exponential case, we show two data sets---the darker corresponds to the red ensemble of the mentioned figures and the lighter to the blue ensemble.
For visibility, $C^{11}$ and $C^{33}$ are slightly offset horizontally, as are $C^{-+}$ and $C^{+-}$.

As mentioned, the uniform irrep has time-independent correlators.  The exact uniform charge-charge correlator is not shown because it is approximately $5.8 \times 10^{-12}$.  The darker exponential ensemble is clearly incompatible with this result, while the lighter is entirely uncertain 0.007(10); the diagonal ensemble is similarly uncertain, 0.001(5), except for the first timeslice, which is incompatible with the exact result.

Additionally, in the diagonal case we see that $C^{-+}$ and $C^{+-}$ are not time independent (we average them, shown as black points).
We expect this behavior to vanish in the continuum limit, though we emphasize that according to \eqref{eqn:uniform irrep time independence} the time average is a good estimator of the true value.
The reason for this non-constant behavior is presumably related to the violations of chiral symmetry in the diagonal case.
We see the further effects of chiral symmetry breaking---$C^{11}$ and $C^{33}$ differ, as do $C^{-+}$ and $C^{+-}$, while those pairs are the same in the exponential case, as in Reference~\cite{Buividovich:2016tgo}.  Further, in the uniform irrep we see that the diagonal correlators are not perfectly equal from timeslice to timeslice (the first timeslice, in particular, differs markedly), in contrast to the exponential case.

This figure also emphasizes the need for a controlled continuum limit, even in the exponential case where, remarkably, the $S_{AF}$ and $\chi_{AF}$ order parameters are very near to their continuum values that come from direct diagonalization, though the corresponding intermediate-time behaviors differ from the exact result.
Absent a continuum limit, a formal demonstration that the equal-time correlator is improved is needed to justify not performing a controlled limit.
Interestingly, the diagonal $C^{33}_{++}$ correlator gets very close to the continuum (though the $C^{11}_{++}$ differs---it should converge to the exact answer in the continuum limit).
At intermediate times the diagonal discretization seems to give results closer to the exact result; this observation is in line with what was found in Appendix~\ref{sect:sixteen sites}.
Nevertheless, we stress that in either discretization a continuum limit is required for reliable, systematically controlled results.

The CDW order parameter, which is known to be susceptible to ergodicity problems when using the exponential discretization\cite{Buividovich:2018yar}, can be seen to be incompatible with the exact result using that discretization, even in this small example---and the two different ensembles give different values, indicating a dependence on the ensemble, and therefore, sensitivity to a violation of ergodicity.
In contrast, the CDW order parameter comes out correct for the diagonal discretization, bolstering our claim that the diagonal discretization can be advantageous for large-scale simulations.
The exact $C^{\rho\rho}_{--}$ correlator falls extremely quickly for these values of $U\beta$ and $\kappa\beta$---and neither exponential ensemble gives much indication of decay, while the diagonal discretization is statistically compatible with zero after the third timeslice.

\afterpage{
    \begin{sidewaysfigure}[p!]
        \includegraphics[width=\textheight]{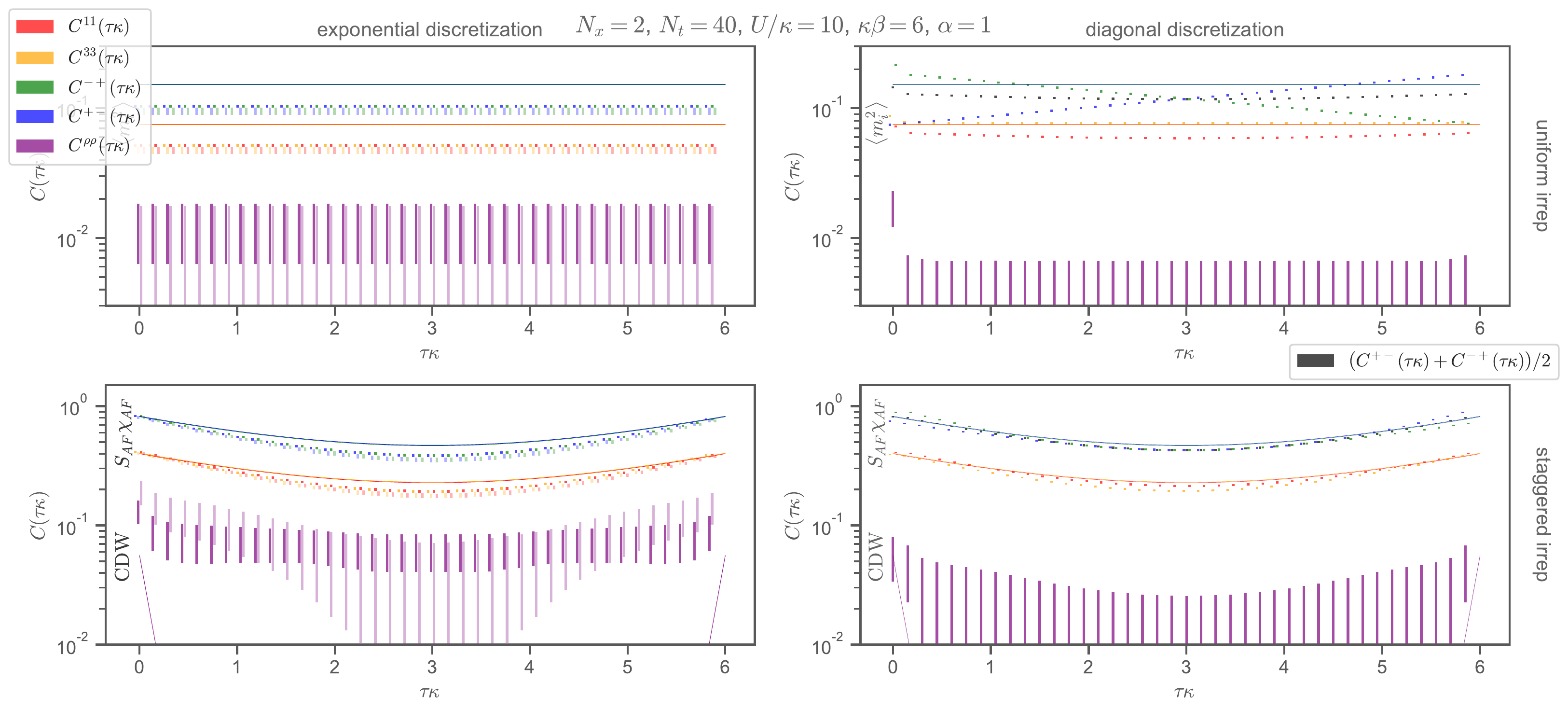}
        \caption{A variety of correlation functions (shown in different colors) of local bilinear operators as a function of Euclidean time, for the exponential and diagonal discretizations, projected to the uniform and staggered irreps on both spatial indices.  Exact results are shown as thin solid lines, while numerical results are shown as markerless error bars, and the respective equal-time order parameters are indicated in text, though the volume factors are omitted for clarity.  In the left panel, the dark and light results correspond to different ensembles, as described in the text.  For additional details and discussion, see the text.}
        \label{fig:spin-spin-correlators}
    \end{sidewaysfigure}
    \clearpage
}


\FloatBarrier
\bibliography{cns}

\end{document}